\documentclass[sigconf]{acmart}

\AtBeginDocument{%
  \providecommand\BibTeX{{%
    \normalfont B\kern-0.5em{\scshape i\kern-0.25em b}\kern-0.8em\TeX}}}

\setcopyright{acmlicensed}
\copyrightyear{2018}
\acmYear{2018}
\acmDOI{XXXXXXX.XXXXXXX}

\acmConference[Conference acronym 'XX]{Make sure to enter the correct
  conference title from your rights confirmation email}{June 03--05,
  2018}{Woodstock, NY}
\acmBooktitle{Woodstock '18: ACM Symposium on Neural Gaze Detection,
 June 03--05, 2018, Woodstock, NY} 
\acmISBN{978-1-4503-XXXX-X/18/06}

\usepackage{multirow}
\usepackage{mdframed}
\usepackage{framed,enumitem} 
\usepackage{graphicx}
\usepackage{xspace}
\usepackage{makecell}
\usepackage{soul}
\usepackage{textcomp}
\usepackage{array}
\usepackage{longtable}
\usepackage{tabularx}
\usepackage[table,xcdraw]{xcolor}
\usepackage{caption}
\usepackage{wrapfig}
\usepackage{subcaption}
\usepackage[normalem]{ulem}
\usepackage{fancyvrb}
\usepackage{fvextra}
\usepackage{tikz}
\usepackage[most]{tcolorbox}
\tcbset{enhanced}

\definecolor{mygreen}{RGB}{0,150,0}
\definecolor{myblue}{RGB}{30,100,200}
\definecolor{myred}{RGB}{180,0,0}
\definecolor{myyellow}{RGB}{200,150,0}

\definecolor{articulation_color}{HTML}{8BD5CE}
\definecolor{exploration_color}{HTML}{7DABF1}
\definecolor{management_color}{HTML}{8376F2}
\definecolor{synchronization_color}{HTML}{D47EFF}

\setul{1pt}{1pt}
\newcommand{\sysname}{\textsc{IntentFlow}}
\newcommand{\add}{\setulcolor{mygreen}\ul{\textsf{\textbf{Add}}}}
\newcommand{\delete}{\setulcolor{myblue}\ul{\textsf{\textbf{Delete}}}}
\newcommand{\correct}{\setulcolor{myred}\ul{\textsf{\textbf{Correct}}}}
\newcommand{\adjust}{\setulcolor{myyellow}\ul{\textsf{\textbf{Adjust}}}}
\newcommand{\rollback}{\textsf{\textbf{\textbullet{} Rollback}}}

\newcommand{\articulation}{\textcolor{articulation_color}{\Large\textbullet} Articulation}
\newcommand{\exploration}{\textcolor{exploration_color}{\Large\textbullet} Exploration}
\newcommand{\management}{\textcolor{management_color}{\Large\textbullet} Management}
\newcommand{\synchronization}{\textcolor{synchronization_color}{\Large\textbullet} Synchronization}

\newcommand*{\inlineimage}[1]{%
    \raisebox{-.3\baselineskip}{%
        \includegraphics[
        height=\baselineskip,
        keepaspectratio,
        ]{#1}%
    }%
}
\definecolor{blockcolor}{HTML}{000000}
\definecolor{blockrule}{gray}{0.6}
\newmdenv[
  topline=false,
  bottomline=false,
  rightline=false,
  leftline=true,
  linecolor=blockrule,
  linewidth=1pt,
  innertopmargin=2pt,
  innerbottommargin=2pt,
  innerleftmargin=10pt,
  innerrightmargin=10pt,
  skipabove=10pt,
  skipbelow=10pt,
  backgroundcolor=white,
  font=\small\color{blockcolor}
]{block}

\newmdenv[
  topline=true,
  bottomline=true,
  rightline=true,
  leftline=true,
  linecolor=blockrule,
  linewidth=1pt,
  innertopmargin=5pt,
  innerbottommargin=5pt,
  innerleftmargin=5pt,
  innerrightmargin=5pt,
  skipabove=5pt,
  skipbelow=10pt,
  backgroundcolor=white,
  font=\sffamily\color{blockcolor},
]{task}

\newtcolorbox{promptbox}{
  breakable,
  colback=gray!8, colframe=black!30,
  arc=4pt,            
  boxrule=0.4pt,
  left=8pt, right=8pt, top=8pt, bottom=6pt
}
\fvset{breaklines, breakanywhere, breaksymbolleft={}, breaksymbolright={}, breaksymbolindent=0pt}
\begin{document}


\title[\sysname{}]{\sysname{}: Investigating Fluid Dynamics of Intent Communication in Generative AI}

\author{Yoonsu Kim}
\orcid{0000-0002-9782-086X}
\affiliation{\institution{School of Computing \\ KAIST}
\city{Daejeon}
\country{Republic of Korea}}
\email{yoonsu16@kaist.ac.kr}

\author{Kihoon Son}
\orcid{0000-0001-7224-2947}
\affiliation{\institution{School of Computing \\ KAIST}
\city{Daejeon}
\country{Republic of Korea}}
\email{kihoon.son@kaist.ac.kr}

\author{Seoyoung Kim}
\orcid{0000-0002-0680-0856}
\affiliation{\institution{School of Computing \\ KAIST}
\city{Daejeon}
\country{Republic of Korea}}
\email{youthskim@kaist.ac.kr}

\author{Brandon Chin}
\orcid{0009-0001-2003-5580}
\affiliation{\institution{College of Engineering \\ University of California Berkeley}
\city{Berkeley}
\state{California}
\country{USA}}
\email{brandoncjw@hkn.eecs.berkeley.edu}

\author{Juho Kim}
\orcid{0000-0001-6348-4127}
\affiliation{\institution{School of Computing \\ KAIST}
\city{Daejeon}
\country{Republic of Korea}}
\email{juhokim@kaist.ac.kr}

\renewcommand{\shortauthors}{Yoonsu Kim et al.}



\begin{teaserfigure}
  \centering
  \includegraphics[width=1\textwidth]{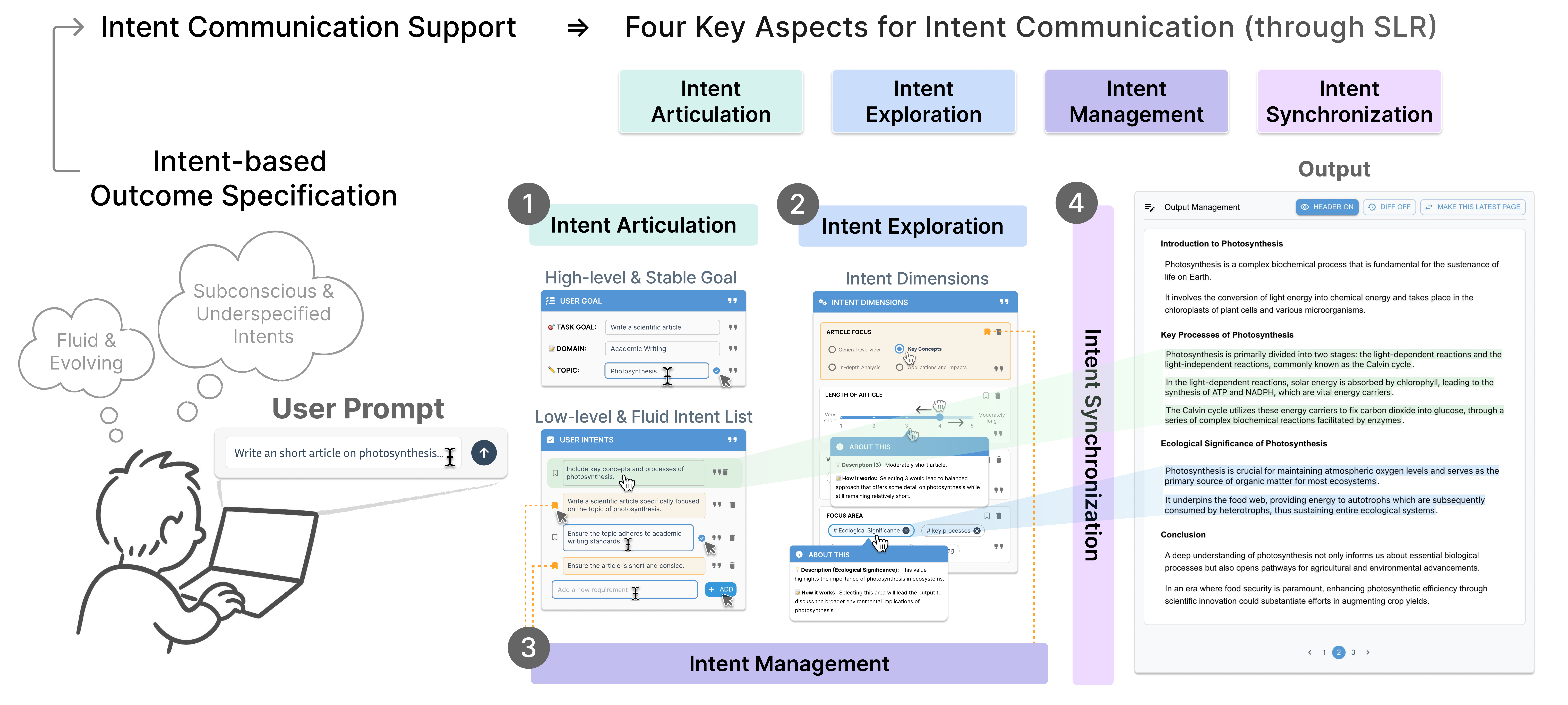}
  \caption{Our work addresses the challenges of intent-based outcome specification in human–LLM interaction, where user intents are often subconscious and underspecified, and tend to be fluid and evolving. Through a systematic literature review, we identified four aspects of intent communication support: articulation, exploration, management, and synchronization. We instantiate these aspects in \sysname{}, an LLM system for writing tasks that (1) helps users articulate vague or subconscious intents, (2) supports exploration of alternative or emerging directions, (3) manages evolving intents over time, and (4) synchronizes intents with generated outputs through linking.}
  \Description{This figure illustrates the motivation and design of IntentFlow. On the left, a user provides a natural language prompt, reflecting the paradigm of intent-based outcome specification but also highlighting its limitations when intents are subconscious and underspecified, and tend to be fluid and evolving. The center depicts four aspects of intent communication support derived from our systematic literature review: articulation, exploration, management, and synchronization. On the right, these aspects are instantiated in the LLM-based writing system IntentFlow, that (1) provides interactive scaffolds to articulate vague or subconscious intents, (2) offers adjustable dimensions to explore alternative or emerging directions, (3) keeps track of and organizes intents to manage their evolution, and (4) links each intent transparently to the generated output for synchronization.}
  \label{fig:teaser}
\end{teaserfigure}

\begin{abstract}
Generative AI shifts interaction toward intent-based outcome specification, despite inherently vague, fluid, and evolving intents. While a growing body of HCI research has proposed diverse interaction techniques to support this process, there is limited understanding of what the key aspects of intent communication are and how they interplay to shape users’ workflows. To bridge this gap, we first conduct a systematic literature review of 46 HCI papers and identify four core aspects of intent communication support: intent articulation, exploration, management, and synchronization. To investigate how these aspects interplay in practice, we developed \sysname{}, a research probe that embodies all four aspects for a writing task, and conducted a comparative study (N=12). 
Our action-level behavioral analysis reveals that comprehensive support enables verification-driven refinement and progressive intent curation, reduces cognitive effort, and improves users’ sense of control and understanding of intent–output alignment. We conclude with design implications for building generative AI systems that support intent communication as a dynamic, iterative process.
\end{abstract} 


\maketitle
\section{Introduction}\label{section:intro}
Advances in generative AI, such as large language models (LLMs) and vision-language models (VLMs), allow people to easily obtain high-quality outputs, ranging from polished text~\cite{laban2024beyondthechat, zhang2023visar, yeh2024ghostwriter, kim2023cells, reza2024abscribe} to functional code~\cite{liu2023whatitwantsmetosay, liang2024largescale} to complete visual designs~\cite{riche2025ai-instruments}, simply by describing what they want in a natural language prompt. 
This represents what Jakob Nielsen calls a new UI paradigm in computing history---\textit{intent-based outcome specification}---where users tell the computer \textit{what they want}, rather than \textit{how to do it}~\cite{AIFirstN88:online}.

While prompting may appear straightforward, communicating intent through natural language is inherently complex. Drawing from communication theory~\cite{stamp1990construct, knapp2011sage, gussow2023language}, we can understand prompting as conveying two key types of information: the user’s \textbf{goal} and their \textbf{intent}. A \textbf{goal} denotes the high-level objective (e.g., writing a cover letter), which tends to be explicit and stable throughout the process. In contrast, \textbf{intents} captures low-level strategies, preferences, or constraints for achieving that goal, often emerging or shifting during the process (e.g., deciding whether to introduce one’s background before the motivation, adopting a polite tone, or keeping certain sections concise). Communication scholars emphasize that intent is inherently fluid and often subconscious~\cite{gussow2023language, stamp1990construct, knapp2011sage, Kraljic2024Collaborating}. People may not be fully aware of their own intents at the outset; these may evolve, surface gradually, or even fade as they reflect on outcomes and iterate on their actions. 

HCI researchers have similarly noted that users’ intents are often not fully articulated at the beginning of interaction~\cite{schon1992designing, buxton2010sketching, gmeiner2025intenttagging, Tankelevitch2024metacognitivedemands, Vaithilingam2024Imagining}. Because intent is not static or fully specified upfront, effective interaction with generative AI requires careful support for intent communication as an ongoing process. Across HCI research on generative AI systems, a wide range of interaction techniques have been proposed to support how users communicate their intents over time~\cite{Vaithilingam2025Semantic, Zhang2025NeuroSync, Wang2025IntentPrism}. However, these efforts are scattered across systems that address different aspects of intent communication in isolation, leaving open questions about how such support mechanisms interact and reinforce one another in practice.

To synthesize existing insights, we conducted a systematic literature review (SLR) of 46 HCI papers on generative AI systems related to intent communication. From this review, we identified four key aspects that support intent communication: (1) \textbf{Intent Articulation}---helping users express vague intents, (2) \textbf{Intent Exploration}---supporting users in discovering and expanding their intents, (3) \textbf{Intent Management}---providing a structured way to manage evolving intents, and (4) \textbf{Intent Synchronization}---ensuring a mutual understanding between the user and LLMs regarging communicated intents. 
At the system feature level, our analysis revealed common design patterns across these aspects (Table~\ref{tab:slr-feature}). 
While prior work has contributed valuable techniques for individual aspects, there is a limited understanding of how they interplay with each other to shape users' dynamic intent communication process.

This gap motivates the following research questions, which focus on how the four aspects interplay in practice within a single workflow and jointly shape the dynamic intent communication process: 
\begin{itemize}
    \item \textbf{RQ1}: How do \textbf{users communicate their intents} when interacting with a system that integrates support for \articulation{}, \exploration{}, \management{}, and \synchronization{}?
    \item \textbf{RQ2}: How does the interplay among different \textbf{intent communication support features} shape users' intent communication behaviors?
    \item \textbf{RQ3}: How do intent communication support features affect users' cognitive effort, sense of control, and understanding of intent--output alignment during interaction?
\end{itemize}
To address these questions, we need a probe that integrates all four aspects of intent communication support to make them operate together within a single interaction workflow. To this end, we developed \sysname{}, a research probe that instantiates all four aspects---\articulation{}, \exploration{}, \management{}, and \synchronization{}---through representative and dominant design features identified in our SLR. We situate \sysname{} in the context of LLM-based writing tasks, which provide a representative testbed due to their iterative nature and the need to balance multiple, evolving intents such as content structure, tone, and emphasis~\cite{reza2024abscribe, flower1981cognitive}. 

Using this research probe, we conducted a within-subjects study (N=12) comparing \sysname{} with a conventional chat-based LLM interface representative of widely-used commercial systems such as ChatGPT Canvas~\footnote{https://openai.com/index/introducing-canvas/} and Claude Artifact~\footnote{https://support.anthropic.com/en/articles/9487310-what-are-artifacts-and-how-do-i-use-them}. We include this baseline to contrast users' intent communication process when supported by all aspects versus current practice (RQ1), which primarily relies on conversational turn-taking with free-form prompts.

Our behavioral analysis revealed distinct interaction patterns when all four aspects of intent communication were jointly supported. Users engaged in iterative cycles of \articulation{} and \synchronization{}, using system feedback to verify how their intents were reflected in the output and refining them accordingly---demonstrating \textit{verification-driven refinement}. We also observed progressive \textit{intent curation}, in which users moved from \articulation{} and \exploration{} to \synchronization{} and \management{}, gradually consolidating intents from tentative expressions into stable configurations. These patterns emerged specifically from the interplay between the aspects: \synchronization{} served as a critical mediating mechanism, grounding tentative intents during \exploration{}, verifying articulated intents, and informing \management{} decisions about which intents to retain; \management{} provided stability that allowed \articulation{} and \exploration{} to build incrementally rather than restart from scratch; \exploration{} fed back into \articulation{} by revealing gaps in users' current intent space. Users also reported significantly easier intent expression and refinement, improved understanding of intent--output alignment, and reduced cognitive effort throughout the intent communication process. 

Taken together, we derive four design implications that supporting intent communication as a dynamic process requires more than assembling known features; it requires designing how these features interact across different stages of the process. Rather than treating intent expression as a linear process, systems should (1) establish immediate bidirectional traceability, enabling users to verify how their intents are interpreted and trace outputs back to intents, (2) scaffold exploration within synchronized views to enable fluid iteration between probing variations and seeing current realizations rather than restarting articulation, (3) support progressive commitment that enables iterative comparison and refinement during exploration and articulation before committing to management, (4) surface existing intent configurations during new articulation to inform users of conflicts, redundancies, and gaps in their current intent space.

In summary, this work makes the following contributions:
\begin{itemize}
    \item A conceptualization of four key aspects that support intent communication---articulation, exploration, synchronization, and management---derived from a systematic literature review of 46 HCI papers on generative AI systems.
    \item Findings from a comparative user study demonstrating how comprehensive intent communication support reshapes user behavior and perception, reducing effort while improving control and intent--output alignment.
    \item Design implications for generative AI systems that support dynamic intent communication, emphasizing iterative support, transparent intent--output alignment, and mechanisms for stabilizing and reusing evolving intents.
\end{itemize}

\section{Related Work}
\subsection{Challenges of Aligning User Intent in Natural Language Interaction}
Despite the remarkable capabilities of LLMs, several studies have consistently highlighted their limitations in understanding and aligning with user intent. Prior studies have emphasized that LLMs frequently generate responses that misalign with user expectations, sometimes leading to unintended consequences due to inherent biases or prompt sensitivity~\cite{kaddour2023challengesapplicationslargelanguage, kim2023understanding, Zamfirescu2023whyjohnny, anwar2024foundationalchallengesassuringalignment, wu2025collabllmpassiverespondersactive, deng2023surveyproactivedialoguesystems}. 
One contributing factor lies in the interaction setup itself: natural language, the primary medium for prompting, provides important advantages by lowering the barrier to entry, enabling flexible self-expression without requiring specialized syntax~\cite{Zamfirescu2023whyjohnny, Khurana2024whyandwhen}. Yet the very qualities that make natural language accessible also introduce ambiguity and instability, making it difficult for users to articulate their intent with the precise phrasing for LLM and forcing them into cognitively demanding trial-and-error reformulation~\cite{subramonyam2024bridging, liu2023whatitwantsmetosay, masson2024directgpt, gmeiner2025intenttagging, Zamfirescu2023whyjohnny, kim2024evallm}. In other words, natural language both empowers users to communicate their intent and, paradoxically, complicates the alignment of that intent with model behavior.  
The opacity of the intent-to-output connection further worsens this issue: prompts may embed multiple intents, yet users are not informed which parts influenced the output, and small wording changes can yield unpredictable shifts~\cite{Zhao2024Explainability, Khurana2024whyandwhen, masson2024directgpt}.
Moreover, as user intent often evolves during interaction, current chat-based interfaces provide limited support for managing this fluid process. Intents expressed across multiple turns become fragmented within linear conversation histories, making it difficult for users to track, refine, and maintain alignment with model behavior~\cite{kim2023cells, zamfirescu2023herding}.
These challenges collectively highlight the need for new interaction mechanisms that help users articulate their intents, understand how their inputs influence the model, and manage them over time. Our work aims to address these issues by helping users more clearly articulate, refine, and adjust their intents within an LLM system.

\subsection{Subconscious and Fluid Characteristics of User Intents}
While intent alignment itself has been recognized as one of the biggest challenges for LLMs, this issue becomes even more complex when considering the characteristics of human intent during communication. Drawing from interpersonal communication research, we can distinguish user prompts into two types of information: \textbf{goals}, which are explicit and relatively stable, and \textbf{intents}, which refer to more fine-grained, sometimes subconscious strategies or actions that evolve dynamically throughout a conversation~\cite{stamp1990construct, knapp2011sage}. This distinction is particularly relevant as users often interact with LLMs in ways that resemble human-to-human communication~\cite{Zamfirescu2023whyjohnny}. 
Research in cognitive task analysis reveals that intent is not spontaneously generated; rather, it emerges from a foundation of knowledge, cognitive processes, and goal configuration~\cite{clark2008cognitive}. 
Consequently, user intents often shift during communication~\cite{stamp1990construct, knapp2011sage}, and these characteristics of intent complicate how users interact with LLMs, making them a crucial consideration in the design of such interactions.
Furthermore, they can be particularly more apparent in creation tasks, such as writing or drawing, due to their iterative nature, requiring continuous reflection and refinement~\cite{buxton2010sketching, schon1992designing, shneiderman2007creativity, terry2002recognizing, gmeiner2025intenttagging}. Recognizing intent as both subconscious and fluid thus highlights the need for interaction designs that can better accommodate the dynamic and situated nature of human communication with LLMs.

\subsection{Supporting Intent-Aligned Interaction with Generative AI}
There are diverse attempts to support users in interacting with generative AIs in ways that better align with their intent. 
Prior work has explored supporting users to more effectively express their intent, such as providing prompt suggestions to scaffold ideation for text-to-image models~\cite{brade2023promptify}, allowing easier prompting through direct manipulation~\cite{masson2024directgpt}. Furthermore, to allow users to better attain the results aligning with their intents, there are attempts to help users understand LLMs’ behaviors through a visual programming environment for hypothesis testing~\cite{arawjo2024chainforge} or breaking down prompts into smaller subtasks and then aggregating the results~\cite{wu2022aichains}. 

For the cases where user intent is unclear or underspecified, there are attempts to go beyond passively responding to user prompts: proactively retrieving missing information when tasks are ambiguous~\cite{qian2024tell} or asking for follow-up questions to better align with users' overarching goals~\cite{wu2025collabllmpassiverespondersactive}. For another approach to better understand user intents, IntentGPT identifies intent within user utterances, enhancing the system’s ability to respond to varying goals with greater precision~\cite{rodriguez2024intentgptfewshotintentdiscovery}. Moreover, to support changes in user intent, dynamic prompt middleware~\cite{drosos2024dynamicpromptmiddlewarecontextual}, which provides context-specific UI elements to better refine the user prompt, allows users to change their preferred options.

More recently, work has begun to reify intent as manipulable units to support more flexible and reflective workflows. IntentTagger introduces ``intent tag''---atomic conceptual units that enable granular and non-linear micro-prompting, supporting intent elicitation and flexible workflows~\cite{gmeiner2025intenttagging}. In a similar vein, AI Instruments embody prompts as reusable interface objects, reflecting multiple interpretations of ambiguous user intents and enabling iterative, non-linear exploration across creative tasks~\cite{riche2025ai-instruments}. These approaches suggest new paradigms for structuring intent that go beyond linear prompting.

These efforts also exist in the context specific to writing with LLM as it involves cognitive content-generation that requires intensive and precise intent formulation and expression~\cite{goldi2024intelligentwriters}.
One line of research has proposed LLM-based writing-support systems that help users explore and prototype their writing while deciding on their writing intent~\cite{suh2024Luminate, zhang2023visar, kim2023cells}. 
There also exist approaches to help users better express their intent by allowing direct manipulation within the text to match their specific writing intent or style~\cite {masson2024directgpt, yeh2024ghostwriter}. 
Moreover, there are attempts to make AI-generated suggestions more explicit and manageable, for instance by structuring multiple variations side by side for rapid comparison or surfacing executable edits directly in the document with safeguards for accuracy~\cite{reza2024abscribe, laban2024beyondthechat}.

While prior work has contributed to improving how users express, reify, and align their intents with generative AI, it has rarely considered the subconscious and fluid nature of user intent as a basis for interaction. Building on this body of work, we conduct a systematic literature review to provide a comprehensive understanding of intent communication with generative AI and present a system that operationalizes these insights by enabling more flexible and reflective interaction with users' evolving intents in writing tasks. 


\section {Systematic Literature Review: Intent Communication Support}
In this section, we present the process and results of our systematic literature review (SLR) of prior HCI research regarding intent communication support with generative AI systems.
Our goal was to identify recurring \emph{feature-level design patterns} for intent communication and to examine how these features have been combined---or separated---in existing systems.


\subsection{Search and Filtering Process}
We conducted our SLR following the PRISMA guideline~\cite{page2021prisma}, which is a well-known framework for SLR. 
\subsubsection{Venue.}
To focus on how the HCI community has addressed interaction support for intent communication with generative AI,  we limited our scope to the following venues: \textit{the ACM CHI Conference on Human Factors in Computing Systems} (CHI), \textit{the ACM Symposium on User Interface Software and Technology} (UIST), \textit{the ACM Conference on Intelligent User Interfaces} (IUI), \textit{the ACM SIGCHI Conference on Designing Interactive Systems} (DIS), \textit{the ACM Conference on Computer-Supported Cooperative Work and Social Computing} (CSCW), \textit{the ACM Conference on Creativity and Cognition} (C\&C), \textit{Proceedings of the ACM on Human-Computer Interaction} (PACMHCI), \textit{the ACM Conference on Conversational User Interfaces} (CUI), and\textit{ the ACM Conference on User Modeling, Adaptation and Personalization} (UMAP). All papers were retrieved from the ACM Digital Library.

\subsubsection{Search Keywords and Exclusion Criteria.}
We collected papers using the following keyword combination in their title or abstract: `Generative AI (GenAI)' \texttt{OR} `Large Language Model (LLM)' \texttt{AND} `Intent' \texttt{OR} `Intention'. To capture the period when GenAI and LLM systems became widely adopted as intent-specification interfaces, we restricted the search timeframe to December 2022 (the release of ChatGPT) through December 2025.
After conducting an exhaustive search, we applied four exclusion criteria (EC) to filter papers that did not align with our SLR goal:
\begin{itemize}
    \item EC1: We excluded papers where `intent' was used as an adverb (e.g., `intentionally'), as this does not relate to user intent communication.
    \item EC2: We focused on papers where users directly communicate their intent with GenAI or LLM systems. Studies where GenAI mediated intent communication between a user and a third party were excluded.
    \item EC3: We included only system or framework papers that proposed concrete interaction features, including prototypes. We excluded survey and workshop papers.
    \item EC4: We focused on papers involving co-creation or generative tasks (e.g., writing, coding, design) where users expressed their wants to the system and received outputs. 
\end{itemize}
\subsubsection{Round 1-3.}
\begin{table}
\small
\begin{tabularx}{\linewidth}{l *{3}{>{\raggedleft\arraybackslash}X}}
\toprule
Venue & Round 1 & Round 2 & Round 3 \\
\midrule
CHI & 219 & 52 & 21 \\
UIST & 62 & 37 & 18 \\
IUI & 43 & 8 & 2 \\
DIS & 41 & 8 & 4 \\
CSCW & 21 & 2 & 0 \\
C\&C & 24 & 3 & 0 \\
PACMHCI & 12 & 3 & 0 \\
CUI & 19 & 8 & 0 \\
UMAP & 28 & 6 & 0 \\
CHIWORK & - & - & 1 \\
\midrule
Sum & 469 & 127 & 46 \\
\bottomrule
\end{tabularx}
\caption{The number of papers selected in each round of our systematic literature review across different venues}
\Description{This table summarizes the number of papers selected in each round of the systematic literature review, broken down by venue (e.g., CHI, UIST, IUI, DIS). It shows how the paper counts decreased from Round 1 to Round 3, with a total of 469 papers in Round 1, 127 in Round 2, and 46 in Round 3.}
\label{tab:slr_round}
\end{table}
We first conducted an extensive search on the ACM Digital Library using the keywords and search period above, which yielded 2,381 papers. We then filtered this set to only include papers from our selected venues, resulting in 469 papers (Round 1). The first author read the abstracts to apply EC1-EC4, reducing the set to 127 papers (Round 2). Subsequently, the first and second authors independently reviewed the full papers based on EC1-EC4, resulting in 45 papers. In addition, we included one relevant study from \textit{the Symposium on Human-Computer Interaction for Work} (CHIWORK) that closely aligned with our criteria, bringing the final set to 46 papers (Round 3). Finally, three authors meticulously read these 46 papers and extracted the interaction features designed to support intent communication, using these findings as the basis for our analysis. The number of papers at each round is summarized in \autoref{tab:slr_round}. 

\par
\subsection{Analysis}
We conducted an open coding analysis of the extracted interaction features. Three authors independently documented how each system supported users’ intent communication, then iteratively grouped similar features through discussion. Through this process, we identified recurring patterns of intent-related interaction support and synthesized them into higher-level categories. We then revisited all papers to examine how these patterns were instantiated and combined across systems.

\subsection{Findings}
\begin{table*}[]
\resizebox{\linewidth}{!}{%
\begin{tabular}{lllrr}

\toprule
\textbf{Theme} &
  \textbf{Feature} &
  \textbf{Paper} &
  \multicolumn{1}{l}{\textbf{N}} &
  \multicolumn{1}{l}{\textbf{\%}} \\
\midrule
\multirow{5}{*}{Articulation} &

\begin{tabular}[c]{@{}l@{}}\textbf{A1. Allowing users to specify and refine intent by directly referencing} \\ \textbf{specific areas of the output.}\end{tabular}  &
  \cite{masson2024directgpt, Wang2025IntentPrism, Wang2024PromptCharm, laban2024beyondthechat, Ding2025Towards, zhang2023visar, Vaithilingam2025Semantic, Liu2024WeNeedStructured, Kim2025ShoeGenAI, Lim2024Co-Creating, Zhou2024GlassMail, Chung2024Patchview, Zhang2024ProtoDreamer, Zhang2025NeuroSync, Leung2025SQUIRE, subramonyam2024bridging}&
  16 &
  34.78\% 
 \\
 &
 \textbf{A2. Decomposing users' vague input into granular sub-components} &
  \cite{riche2025ai-instruments, Gmeiner2024Evidence-based, kim2025applying, chen2025dango, wang2025harmonycut, suh2024Luminate, Kim2025ShoeGenAI, lee2025veriplan, laban2024beyondthechat, Kazemitabaar2024Improving, Zhang2025NeuroSync, Huang2025SketchGPT, Zhang2025ChainBuddy, Lee2025ThematicPlane} &
  14 &
  30.43\%
 \\
 &
  \begin{tabular}[c]{@{}l@{}}A3. Supporting diverse modalities (e.g., sketch, image, metadata) \\ for expressing user intent beyond text according to the task context\end{tabular} &
  \cite{Liu2024WeNeedStructured, gmeiner2025intenttagging, drosos2024dynamicpromptmiddlewarecontextual, Peng2024DesignPrompt, Zhou2024GlassMail, Tilekbay2024ExpressEdit, Chung2024Patchview, Zhang2024ProtoDreamer, Yen2024Memolet} &
  9 &
  19.57\% \\
 &
  A4. Elaborating intent using users’ vague inputs as seeds&
  \cite{chen2025dango, kim2024DiaryMate, suh2024Luminate, Wang2024PromptCharm, Lim2024Co-Creating, Yen2024CoLadder, Zhang2025ChainBuddy} &
  7 &
  15.22\% \\ 
\midrule
\multirow{4}{*}{Exploration} &
 \textbf{E1. Supporting navigation of intent variation spaces through output spectra}&
  \cite{reza2024abscribe, gmeiner2025intenttagging, suh2024Luminate, lee2025veriplan, Chen2024AutoSpark, Zhang2024ProtoDreamer, brade2023promptify, Leung2025SQUIRE, Sun2025Creative, Marquardt2025ImaginationVellum, Choi2025IdeaBlocks, Lee2025ThematicPlane} &
  12 &
  26.09\% 
  \\
 &
  \textbf{E2. Suggesting alternative intents} &
  \cite{riche2025ai-instruments, Gmeiner2024Evidence-based, choi2025Expandora, gmeiner2025intenttagging, Shanmugarasa2025Privacy, Lim2024Co-Creating, laban2024beyondthechat, Leung2025SQUIRE, subramonyam2024bridging, Choi2025IdeaBlocks} &
  11 &
  23.91\% 
  \\
 &
  E3. Supporting to remix intents or intermediate output &
  \cite{wang2025harmonycut, Siddiqui2025ScriptShift, Kim2025ShoeGenAI, Peng2024DesignPrompt, Sun2025Creative} &
  5 &
  10.87\% \\
 &
  E4. Providing exploratory nudges through prompts and questions &
  \cite{Gmeiner2024Evidence-based, Lee2024Closer, brade2023promptify, subramonyam2024bridging} &
  4 &
  8.70\% \\
\midrule
\multirow{3}{*}{Management} &
  \textbf{M1. Structuring intents into manageable representations} &
  \cite{reza2024abscribe, riche2025ai-instruments, kim2025applying, gmeiner2025intenttagging, Wang2025IntentPrism, Siddiqui2025ScriptShift, drosos2024dynamicpromptmiddlewarecontextual, peng2025navigating, brade2023promptify, zhang2023visar, Yen2024Memolet} &
  10 &
  21.74\% \\
 &
  \textbf{M2. Revisiting and curating past intents for independent editing or reuse} &
  \cite{drosos2024dynamicpromptmiddlewarecontextual, Yen2024CoLadder, Yen2024Memolet, Sun2025Creative, Choi2025IdeaBlocks} &
  5 &
  10.87\% \\
 &
  M3. Managing Relationships among Multiple Intents &
  \cite{Chung2024Patchview, Vaithilingam2025Semantic} &
  2 &
  4.35\% \\
\midrule
\multirow{3}{*}{Synchronization} &
 \textbf{S1. Showing how intents are reflected in the output} &
  \cite{liu2023whatitwantsmetosay, chen2025dango, Peng2024DesignPrompt,  Yen2024CoLadder, Yen2024Memolet, subramonyam2024bridging} &
  6 &
  13.04\% 
   \\
 &
  \textbf{S2. Previewing the effects of intent changes} &
  \cite{masson2024directgpt, gmeiner2025intenttagging, Shanmugarasa2025Privacy, Vaithilingam2025Semantic, Marquardt2025ImaginationVellum, Leung2025SQUIRE} &
  6 &
  13.04\% \\
 &
  S3. Exposing the system’s intent interpretation &
  \cite{Liu2024WeNeedStructured, liu2023whatitwantsmetosay, kim2025applying, chen2025dango, Zhou2024GlassMail, Tilekbay2024ExpressEdit} &
  6 &
  13.04\% \\
\bottomrule
\end{tabular}%
}
\caption{Interaction features for intent communication identified through our SLR, organized into four themes. For each feature, we report the number and percentage of reviewed papers that incorporated it.}
\Description{This table summarizes interaction features for intent communication identified through our SLR. Features are grouped into four themes—Articulation, Exploration, Management, and Synchronization—and each row lists a feature description, the papers that include it, and the corresponding number and percentage of papers.}
\label{tab:slr-feature}
\end{table*}
\subsubsection{Four Key Aspects of Supporting Intent Communication in Human-GenAI Interaction}
Our analysis revealed four central aspects of how prior HCI systems supported intent communication (\autoref{tab:slr-feature}):
\begin{itemize} 
    \item[\color{articulation_color}\Large\textbullet] \textbf{Intent Articulation}: Helping users externalize their underspecified and vague intents into more concrete and actionable forms. This is a convergent process focused on transforming users' vague intent into specific instructions.
    \item[\color{exploration_color}\Large\textbullet] \textbf{Intent Exploration}: Supporting users in discovering new, emerging intents that they may not have been initially aware of. This is a divergent process that encourages users to explore and expand their initial scope.
    \item[\color{management_color}\Large\textbullet]\textbf{Intent Management}: Supporting users in organizing, revisiting, and curating their intents as they evolve over time. This is a stabilizing process focused on maintaining continuity across iterations by structuring intents into persistent and manageable representations.
    \item[\color{synchronization_color}\Large\textbullet]\textbf{Intent Synchronization}: Aligning the user's communicated intents with LLM's output by making transparent how each intent is reflected in the generated output and how modifications of intents trigger corresponding updates. This process enables users to verify whether their intents are realized by the LLM as intended. 
\end{itemize}
These aspects encompass the different yet interdependent ways in which systems can facilitate users in externalizing, expanding, managing, and aligning their intent during interactions with generative AI systems.
\begin{figure*}[t!]
    \includegraphics[width=.7\textwidth]{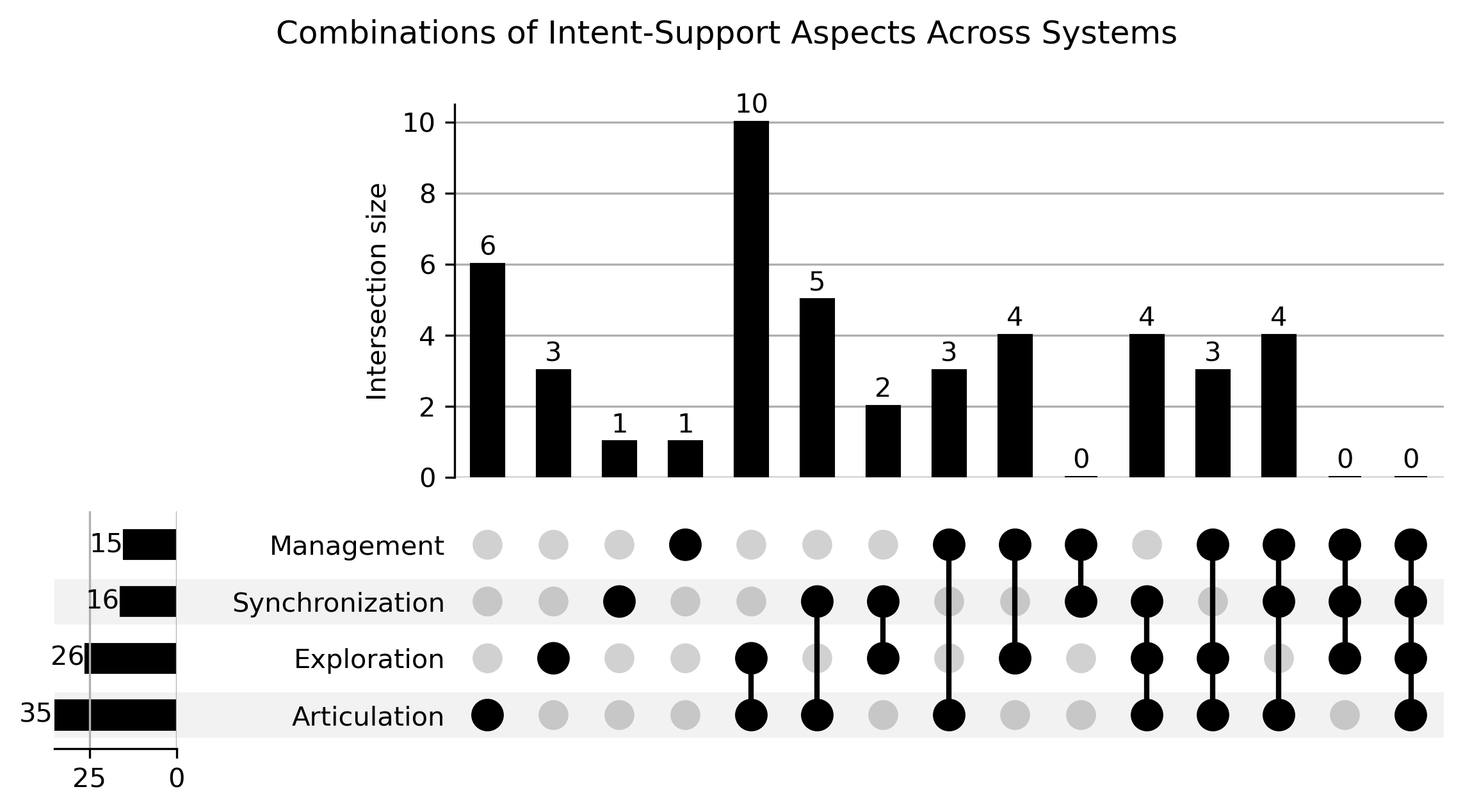}
    \vspace{-1em}
    \caption{Weighted UpSet plot showing how existing systems combine four intent-support aspects---articulation, exploration, management, and synchronization. Each column represents a unique combination of supported aspects, and bar heights indicate the aggregated frequency of occurrences for each combination.}
    \Description{An UpSet plot visualizing combinations of four intent-support aspects across systems. The top bar chart shows the total frequency of each aspect combination. Below, a matrix of dots indicates which aspects are included in each combination, with rows corresponding to articulation, exploration, management, and synchronization, and columns representing different combinations.}
    \label{fig:slr-upset}
    \vspace{-\baselineskip}
\end{figure*}
\subsubsection{Aspect-level Coverage and Limited Integration Across Aspects.}
Prior work has predominantly focused on \articulation{} and \exploration{}, with many systems providing features that help users externalize intents more explicitly or explore alternative intents through variations, suggestions, or prompts (\autoref{tab:slr-feature}). These supports primarily target early-stage intent formulation and ideation, helping users clarify what they want or consider possible alternatives. In contrast, \management{} and \synchronization{} are less consistently supported: relatively few systems offer explicit mechanisms for revisiting, curating, or relating multiple intents over time, or for helping users understand how their intents are interpreted and reflected in system outputs. 

Examining how these aspects are combined reveals further scope for exploration. As shown in the weighted UpSet plot (\autoref{fig:slr-upset}), \articulation{} and \exploration{} frequently co-occur, whereas combinations involving \management{} and \synchronization{} are comparatively rare, and no existing system supports all four aspects simultaneously. This coverage suggests that prior systems emphasize localized support for intent articulation or exploration, but do not provide integrated support for maintaining, coordinating, and aligning intents as interaction unfolds — often leaving users to manage these dynamics on their own.

\subsubsection{Motivation for a Research Probe.}
While prior work has proposed a rich set of features within each aspect, there is a space to explore and understand how these aspects interact to shape users’ intent communication behaviors during end-to-end interaction workflows by integrating four aspects. 
Addressing these gaps requires moving beyond retrospective analysis of existing systems toward designing a probe system that deliberately embeds all four aspects into a unified interaction setting. In the following section, we describe how we construct such a probe system, focusing on how representative features were selected for each aspect and how they are implemented to support intent communication in practice.


\section{Research Probe: \sysname{}}
We designed a system, \sysname{}, as a research probe~\cite{boehner2007probes} that integrates support for articulation, exploration, management, and synchronization within a single interaction workflow, enabling these aspects to interplay during intent communication.
Grounded in representative features from prior work identified through our SLR (\autoref{tab:slr-feature}), \sysname{} brings together these aspects in a unified setting. In this section, we describe the design rationale behind \sysname{}, focusing on how features were selected for each aspect and how they are realized through the system’s interface and underlying pipeline.

\subsection{Design Rationale and Feature Selection}
We selected dominant and representative features for each intent-support aspect to instantiate the probe, which are highlighted in bold in ~\autoref{tab:slr-feature}. Our goal was not to exhaustively implement all prior features, but to capture the core and dominant interaction affordances that characterize each aspect.

\begin{figure}
    \centering
        \includegraphics[width=\linewidth]{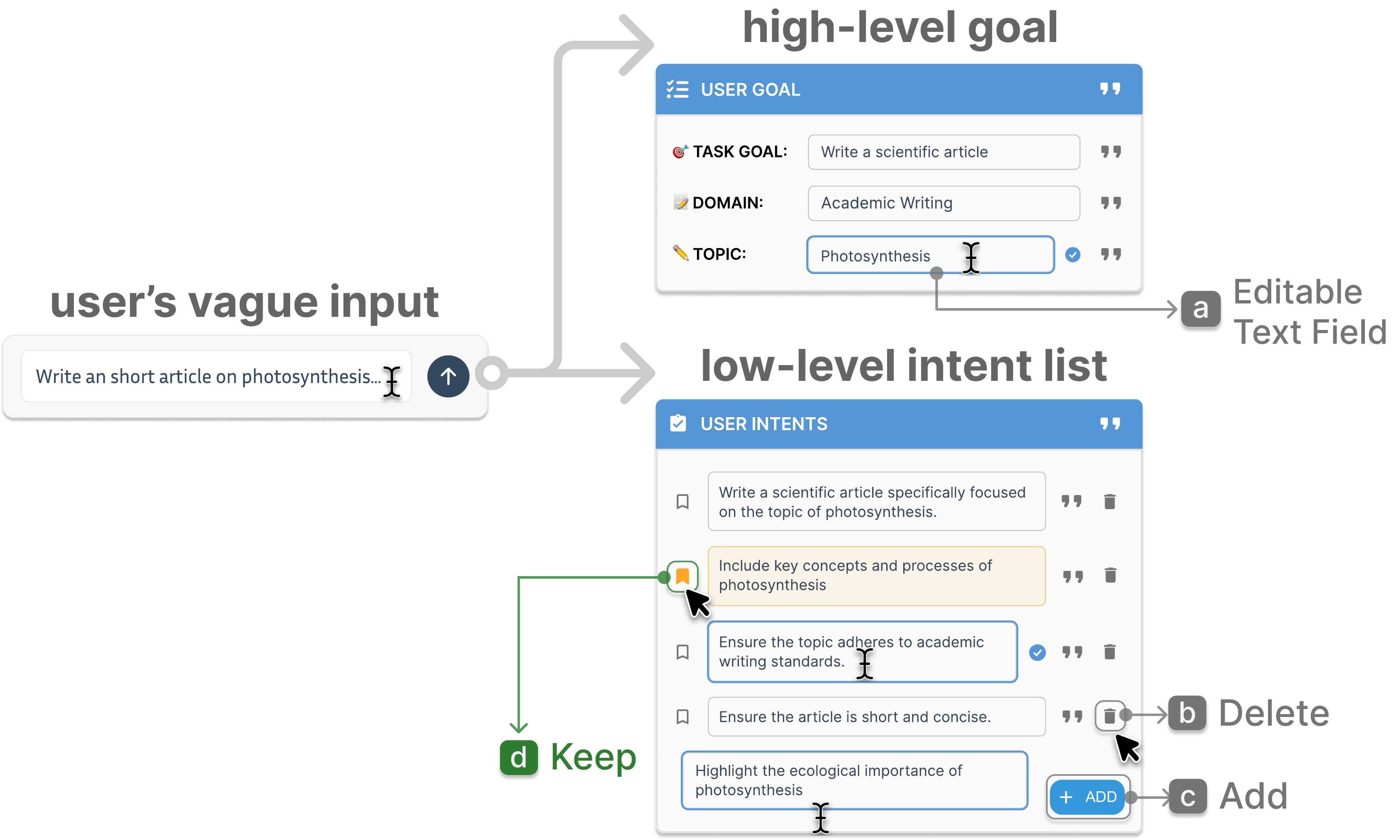}
        \caption{\textbf{Articulation} support in \sysname{}. The system decomposes a user’s vague input into a high-level goal and a set of low-level intents, externalized as editable components. Users can (a) edit, (b) delete, or (c) add intents to refine their intent articulation. (d) Intents can also be pinned to support intent \textbf{management} across iterations.}
        \Description{This is a system diagram illustrating articulation support in \sysname{}. A user’s vague text prompt is decomposed into a high-level goal and a list of low-level intents. The intents are shown as (a) editable text fields, with controls to (b) delete and (c) add intents, supporting articulation by allowing users to refine how their intent is expressed. The figure also includes (d) a pin icon that allows intents to persist across iterations, which supports intent management rather than articulation.}
        \label{fig:articulation}
    \vspace{-\baselineskip}
\end{figure}

For \articulation{}, prior work most commonly supported it by decomposing users' vague input into granular sub-components and enabling direct manipulation to individual intents (A1, A2 in \autoref{tab:slr-feature}). Reflecting this, \sysname{} decomposes a user's chat prompt into a structured representation consisting of a \emph{high-level goal} and a set of \emph{low-level intents}, which are externalized as editable components (\autoref{fig:articulation}). Users can directly revise, add, or remove individual intents, enabling them to progressively articulate and refine their intent beyond an initial vague prompt.

\begin{figure}
    \centering
        \includegraphics[width=\linewidth]{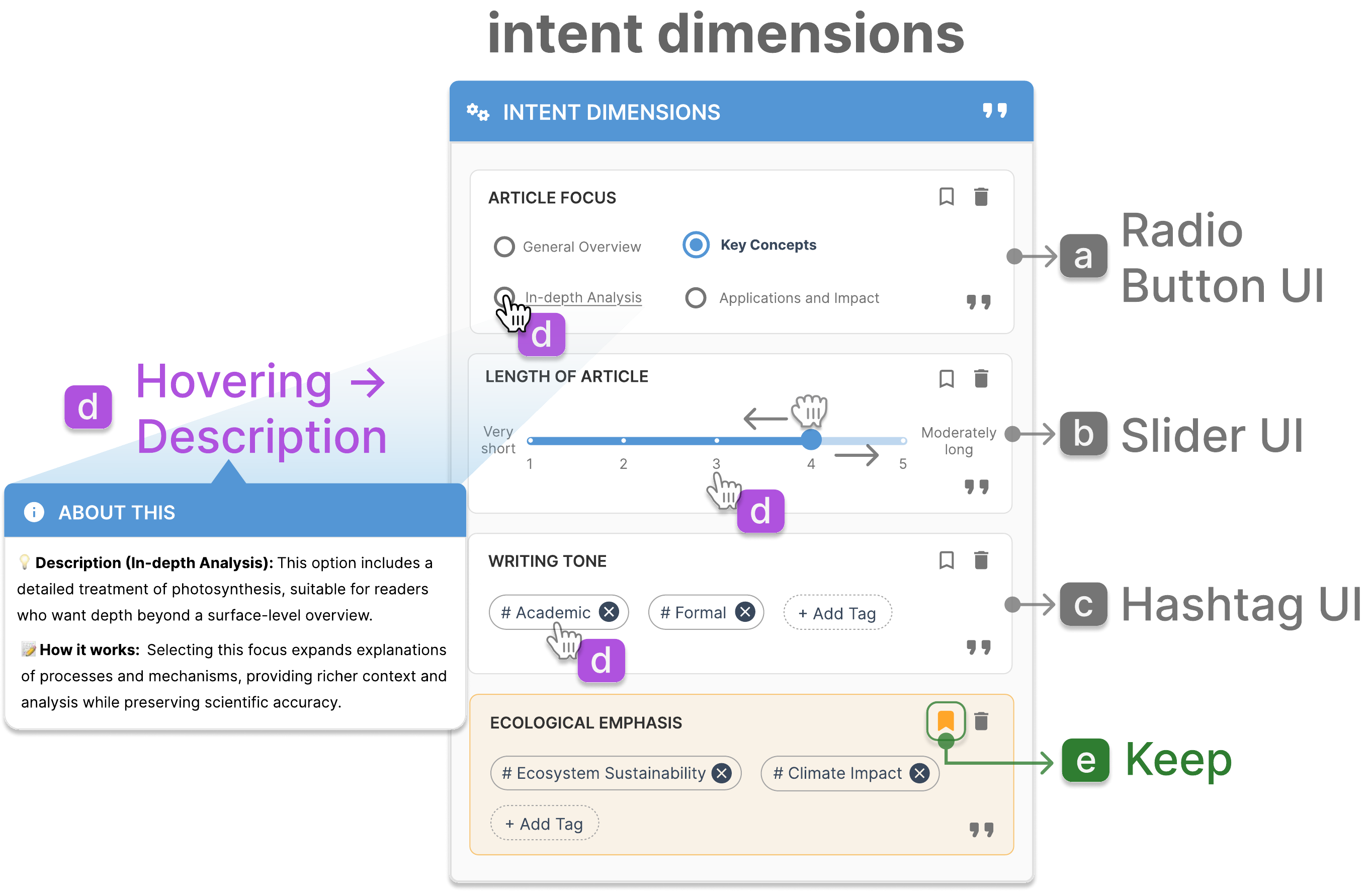}
        \caption{\textbf{Exploration} support in \sysname{}. Intent dimensions are presented through multiple UI controls, including (a) radio buttons, (b) sliders, and (c) tags, enabling users to explore alternative intent configurations. The figure also shows (d) a preview mechanism for \textbf{synchronization} and (e) a pinning option for intent \textbf{management}.}
        \Description{This is a system illustration showing exploration support in \sysname{} through an intent dimension panel. Users can explore alternative intent configurations using multiple UI controls, including (a) radio buttons, (b) sliders, and (c) hashtag-style tags. Hovering over dimension options reveals descriptions that preview how different choices would affect the generated output (d), supporting synchronization. The figure also shows a pin icon (e) that allows selected dimension values to persist across iterations, supporting intent management.}
        \label{fig:exploration}
    \vspace{-\baselineskip}
\end{figure}

To support \exploration{}, we drew on prior systems that enabled users to navigate intent variations and consider alternatives through output spectra or adjustable parameters (E1, E2 in \autoref{tab:slr-feature}). In \sysname{}, exploration is supported through the \emph{intent dimension section},  which exposes adjustable dimensions for each intent using multiple UI controls such as radio buttons, sliders, and tags (\autoref{fig:exploration}). Users can explore how different emphases or alternative options would shape the resulting output by directly manipulating these dimensions.

\begin{figure*}
    \centering
    \includegraphics[width=.85\linewidth]{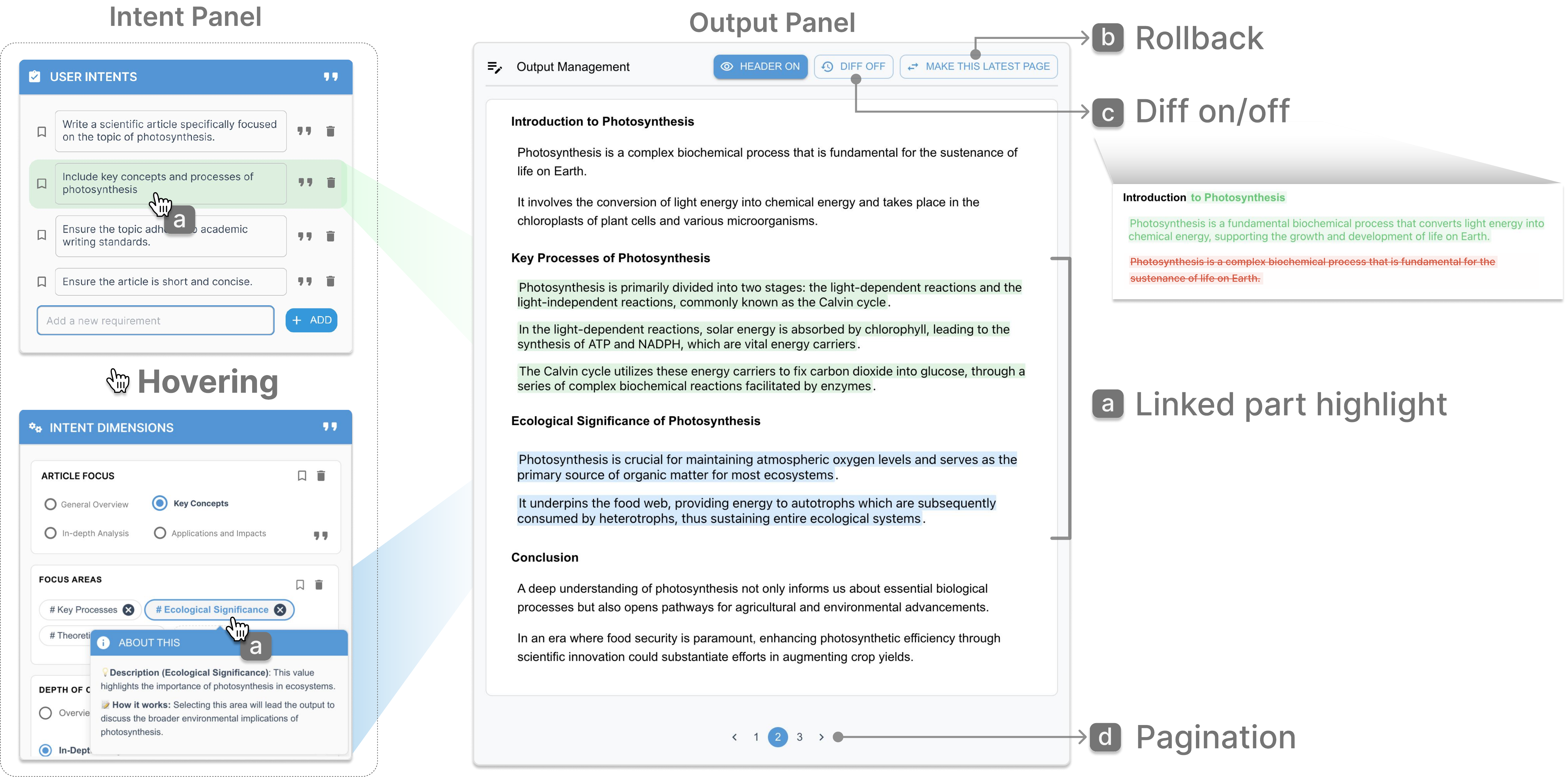}
    \caption{\textbf{Synchronization} support in \sysname{}. (a) Hovering over intents and intent dimension values highlights corresponding linked parts of the generated output in green (intents) or blue (intent dimension values). (b) Users can roll back to any prior version, which brings the selected output and its associated intents to the latest position in the workflow. (c) Diff view compares old and new outputs. (d) Pagination allows users to navigate and manage multiple output versions over time, supporting intent \textbf{management}.}
    \Description{This is a system illustration showing synchronization support in \sysname{}. (a) Hovering over an intent highlights the linked parts of the output in green or blue. (b) There is a ``Rollback'' button, which users can use to move old pages to the back, making it the latest page, restoring the selected output and its intents to the latest position. (c) There is also a ``Diff On/Off'' button that allows users to discern changes from the new output to the old output. To see these differences, users can toggle this button. (d) Pagination at the bottom lets users browse earlier outputs.}
    \label{fig:synchronization}
    \vspace{-\baselineskip}
\end{figure*}

For \management{}, our SLR highlighted features that structure intents into manageable representations and support revisiting and curating past intents over time (M1, M2 in \autoref{tab:slr-feature}). To instantiate these supports, \sysname{} treats intents and their associated dimensions as persistent entities that users can pin or unpin across interaction turns, directly edit, and revisit through versioning (\autoref{fig:articulation}-(d), \autoref{fig:exploration}-(e), \autoref{fig:synchronization}-(d)). This design allows users to curate an evolving set of intents, supporting longer-term intent communication beyond early-stage ideation and enabling us to study how users manage multiple intents across iterative workflows.

Finally, for \synchronization{}, prior work has emphasized helping users preview the effects of intent changes and see how their intents are reflected in the generated output (S1, S2 in \autoref{tab:slr-feature}).
In \sysname{}, synchronization is supported by explicitly linking user intents and intent dimension choices to corresponding segments in the generated output (\autoref{fig:synchronization}) and providing contextual previews for alternative dimension values (\autoref{fig:exploration}-(d)). This design allows users to verify how their intents are reflected in the output and to iteratively adjust this alignment as interaction unfolds.

Together, these design choices allow \sysname{} to function as a research probe for examining how intent-support aspects are enacted, coordinated, and appropriated during intent communication, rather than evaluating individual features in isolation.


\subsection{\sysname{} Interface}
\begin{figure*}[t]
    \centering
    \includegraphics[width=.9\linewidth]{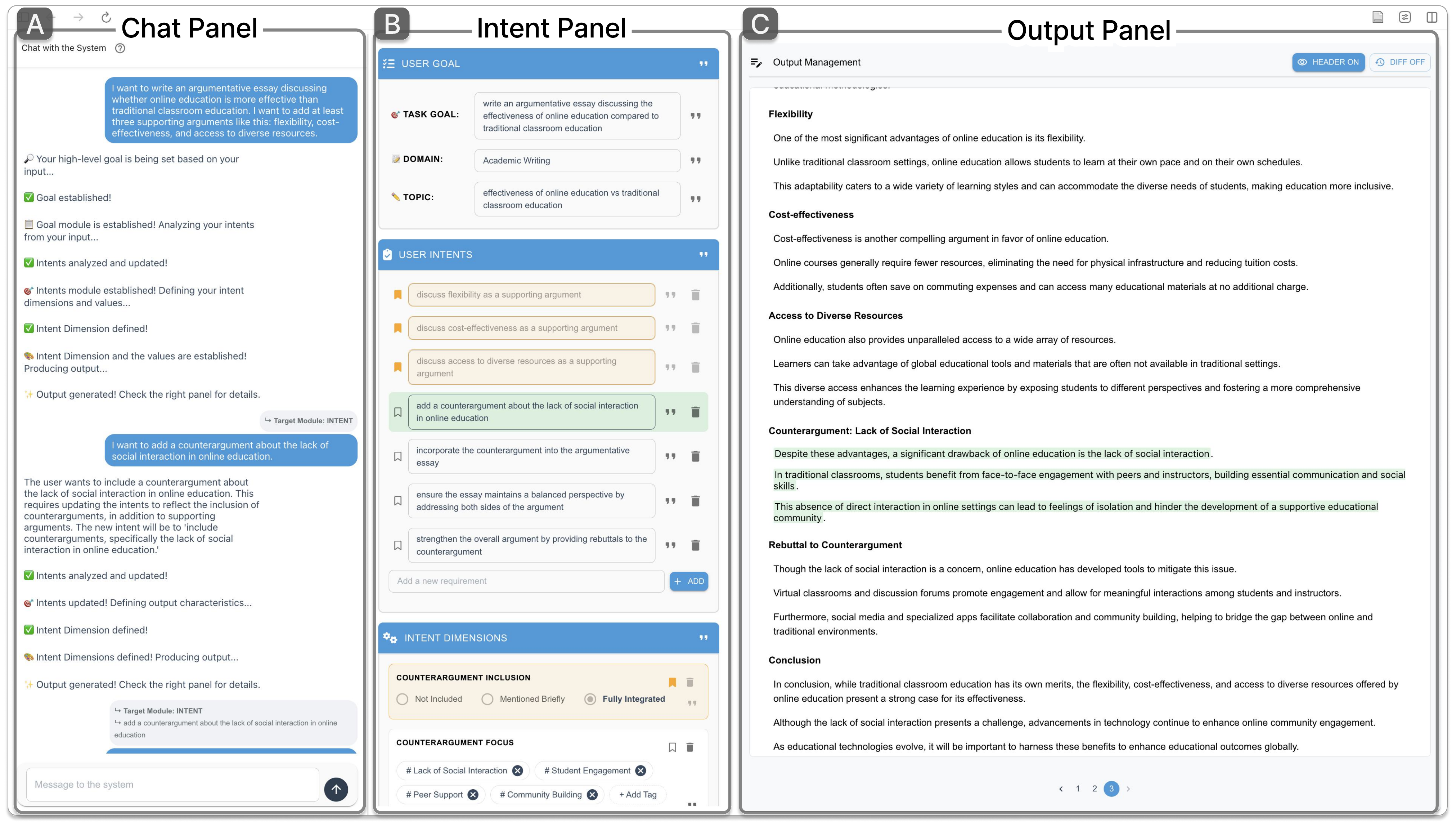}
    \vspace{-1em}
    \caption{Overall interface of \sysname{}. It is split up into three sections: (A) Chat Panel, (B) Intent Panel, and (C) Output Panel.}
    \Description{This figure is a screenshot of the system's overall interface. The screen is split into three sections, with a chat interface on the left (A), an intent panel with several widgets in the middle (B), and the output panel with the generated text output on the right (C). The user initially types a prompt into the chat panel (A), which generates several intent widgets in (B) and an initial output in (C).}
    \label{fig:interface}
    \vspace{-\baselineskip}
\end{figure*} 

We instantiate \sysname{} in the context of LLM-based writing tasks, which provide a suitable testbed for studying intent communication due to their iterative, multi-faceted, and evolving nature~\cite{flower1981cognitive, reza2024abscribe}. Writing tasks often require users to articulate high-level goals, refine multiple interacting intents, explore alternative framings, and iteratively align system outputs with evolving expectations, making them well-suited for examining how intent-support mechanisms operate over time.

As shown in \autoref{fig:interface}, the interface of \sysname{} is organized into three main panels: a \textbf{Chat Panel}, an \textbf{Intent Panel}, and an \textbf{Output Panel}. Together, these panels support a continuous workflow in which users can express intents, refine and manage them, and inspect how they are reflected in the generated output.


The \textbf{Chat Panel} serves as the primary entry point for user input. Users provide an initial, often underspecified prompt describing their writing task. Rather than treating this input as a single-shot instruction, \sysname{} uses it as a starting point for structuring intent communication, triggering the extraction of a high-level goal, low-level intents, and intent dimensions that populate the \textbf{Intent Panel}.

The \textbf{Intent Panel} externalizes users’ intents into structured, editable representations and consists of three sections. 
The \emph{Goal Section} captures stable, high-level aspects of the task, such as the overall writing goal, domain, and topic. Making these elements explicit allows users to verify and revise the system’s understanding of their overarching objective. 
The \emph{Intent List Section} displays a set of low-level, discrete intents inferred from the user’s input. In addition to explicitly stated intents, the system also surfaces implicit intents that are logically required to carry out the task, reflecting how the LLM decomposes an underspecified prompt into fine-grained subtasks. Each intent is presented as an editable item that users can revise, add, or delete, supporting progressive \articulation{} as users clarify and refine what they want to achieve.
The \emph{Intent Dimension Section} further decomposes each intent into adjustable dimensions, exposing parameters such as emphasis, scope, or style through interactive controls. The UI format and initial value of each dimension (e.g., radio buttons, sliders, or tags) is automatically selected based on the characteristics of the associated intent. By directly manipulating these dimensions, users can explore alternative interpretations and configurations of their intents, explicitly supporting \exploration{} during the writing process. In addition, hover-based explanations reveal how different dimension values would affect the generated output, supporting \synchronization{} by helping users anticipate how intent refinements will be reflected in the system’s behavior.
Across both sections, users can pin (\inlineimage{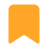}) selected intents and dimension values to persist them across interaction turns, supporting \management{} by allowing them to manage which aspects of their evolving intent should be retained as the writing process unfolds.

The \textbf{Output Panel} displays the text generated by the LLM based on the current configuration of goals, intents, and intent dimensions. Beyond presenting the output, this panel primarily supports \synchronization{} by making the relationship between user intent and generated text explicit. Hovering over an intent or a dimension value highlights the corresponding segments in the output, enabling users to verify how specific intent choices are reflected in the generated content. This panel also supports \management{} by maintaining a version history of each output along with its corresponding intent configuration from the \textbf{Intent Panel}. Users can browse rivisit priot intent-output states, compare changes using a diff view, and roll back to a previous state to continue their work from an earlier configuration. Together, these features allow users to manage, revisit, and refine their intent configurations over time while maintaining alignment between intent and output.

\subsection{Internal Pipeline and Implementation}
\begin{figure*}[t!]
    \includegraphics[width=\linewidth]{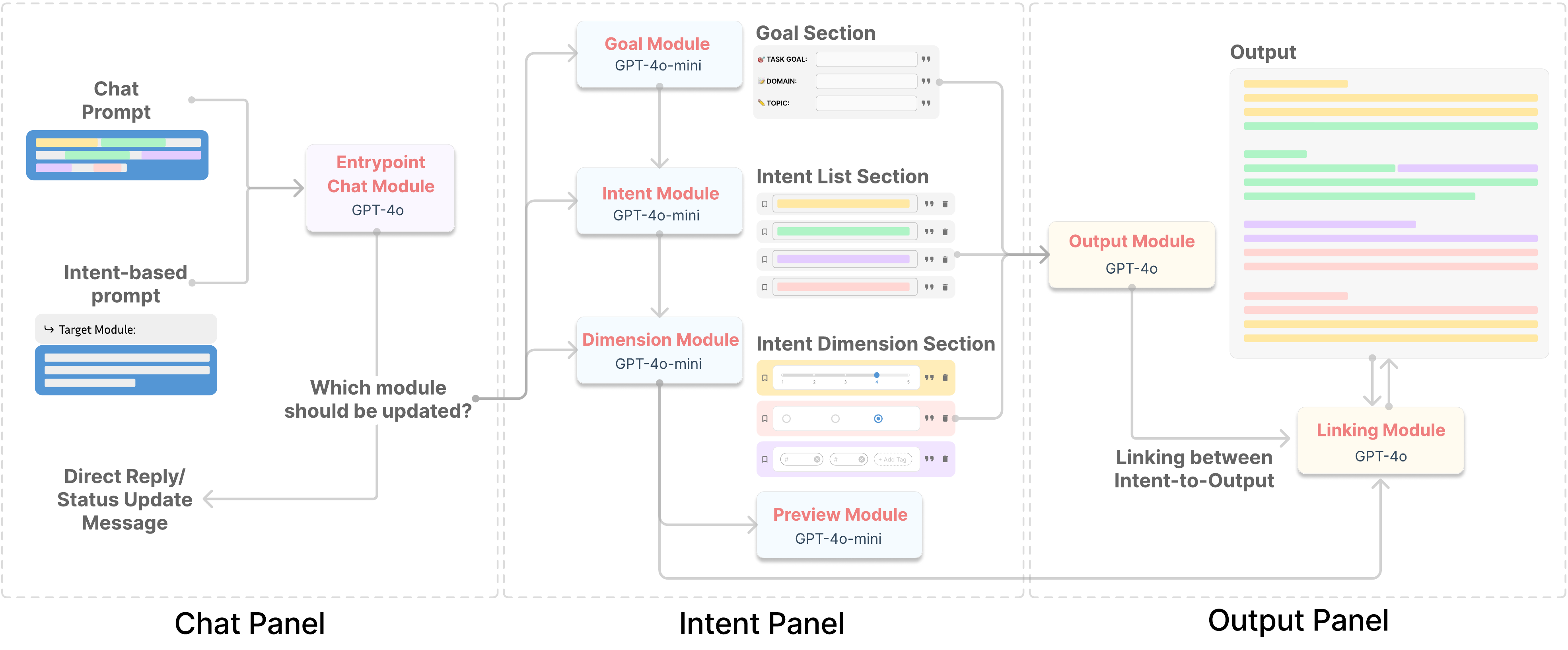}
    \vspace{-1em}
    \caption{Internal pipeline of \sysname{}. The \textit{EntryPoint Chat Module} first interprets the prompt and coordinates three intent-processing modules: the \textit{Goal Module}, \textit{Intent Module}, and \textit{Dimension Module}. These modules extract and structure the user’s goal and intents, which are then used by the \textit{Output Module} to generate text. The \textit{Preview Module} provides brief explanations of dimension values and their potential effects on the output. The \textit{Linking Module} associates each intent with corresponding segments in the generated output. Each module box shows its LLM model name.}
    \Description{This figure shows the backend process of the system's overall pipeline. On the left, the user prompt is processed by the Entrypoint Chat Module. The Entrypoint Chat Module interacts with three modules: the Goal Module, Intent Module, and Dimension Module to form the Intent Panel. After the Dimension Module, the Preview Module generates explanations of dimension values and their effects on the output. These modules are then forwarded to the Output Module, which forms the generated output observed in the Output Panel, as well as the Linking Module, which generates the links between the output and the Intent Panel.}
    \label{fig:system_pipeline}
    \vspace{-\baselineskip}
\end{figure*}

Internally, \sysname{} operates through a modular LLM pipeline that incrementally translates user input and interaction into structured representations of goals, intents, and intent dimensions, which in turn condition output generation. As illustrated in \autoref{fig:system_pipeline}, user inputs and edits are processed through dedicated modules that extract and update intent representations, generate text based on the current intent state, and establish links between intents, dimension values, and corresponding output segments. Rather than generating output solely from the latest prompt, the pipeline is designed to preserve and update intent-related state, enabling \articulation{}, \exploration{}, \management{}, and \synchronization{} to be sustained over time.

We implemented \sysname{} as a web-based application with React for the frontend and Flask for the backend. The backend leveraged the OpenAI API to handle language model functionalities, with each module responsible for a specific role in the pipeline, as illustrated in \autoref{fig:system_pipeline}. To improve responsiveness in real-time interactions, we used a smaller model for the \textit{Goal Module}, \textit{Intent Module}, and \textit{Dimension Module}, observing comparable performance to larger models. For the \textit{Linking Module}, we further optimized it using multiprocessing to reduce latency. Full prompts for each module are provided ~\autoref{appendix:module_prompts} in the Appendix.

\subsection{Pipeline Validation}
Before conducting the user study, we performed a lightweight validation to verify that \sysname{}’s intent-processing pipeline (\autoref{fig:system_pipeline}) operates reliably across diverse writing contexts. The goal of this validation was not to benchmark model performance or compare alternative architectures, but to ensure that the pipeline produces reasonable and stable outputs that can meaningfully support user interaction during the study.

Specifically, we examined whether the pipeline could (1) extract coherent task goals and intents from underspecified prompts, (2) generate relevant and interpretable intent dimensions with appropriate UI representations, and (3) correctly associate intents and dimension values with corresponding parts of the generated output. This validation serves as a sanity check that the system behaves as intended before being deployed as a research probe in the user study.

\subsubsection{Setup}
We selected 12 representative prompts---two from each of the six writing contexts defined by Lee et al.~\cite{lee2024designspace}: academic, creative, journalistic, personal, professional, and technical. For each prompt, we generated a full pipeline output and created a corresponding survey form for evaluation. 
To reduce ambiguity and individual variance, the survey used binary (Yes/No) questions, complemented by an optional free-form section. We recruited a total of 60 evaluators (5 per prompt) via Prolific~\footnote{https://prolific.com}. To ensure evaluation quality, we excluded evaluators who had never used LLMs, lacked experience with the given writing task, or reported unfamiliarity with the given writing topic. Each evaluator was compensated with £5 for completing a 30-minute evaluation task. Detailed information about the prompts, writing tasks, and topics used in the evaluation is provided in the Appendix~\ref{appendix:eval_prompts}.

\subsubsection{Validation Criteria}
We designed a set of questions targeting each pipeline module to assess whether: 
\begin{itemize}
    \item the extracted goal reflected the user’s overall objective (Q1. Goal Alignment),
    \item the set of intents was reasonably complete (Q2. Completeness) , distinct (Q3. Distinctiveness) , and relevant (Q4. Relevance),
    \item intent dimensions were relevant (Q5. Relevance) and paired with appropriate UI controls and values (Q6. UI Appropriateness, Q7. Value Appropriateness), and
    \item highlighted output segments accurately reflected the corresponding intents (Q8. Link Accuracy).
\end{itemize}
Full question wording is provided in Appendix~\ref{appendix:evaluation_questions}.

\subsubsection{Results}
\begin{table}
\vspace{-\baselineskip}
\small
\begin{tabular}{lr}
\toprule
\textbf{Evaluation Criteria} & \textbf{``Agree''(\%)} \\
\midrule
Q1. Goal Alignment & 95.00 \\
Q2. Set of Intents – Completeness & 95.00 \\
Q3. Set of Intents – Distinctiveness & 86.67 \\
Q4. Individual Intents – Relevance & 94.08 \\
Q5. Intent Dimension – Relevance & 86.56 \\
Q6. Intent Dimension – UI Appropriateness & 86.78 \\
Q7. Intent Dimension – Value Appropriateness & 86.14 \\
Q8. Intent-to-Output Linking – Link Accuracy & 94.04 \\
\bottomrule
\end{tabular}%
\caption{Evaluation results for each question in the technical evaluation. Values indicate the percentage of ``Agree'' responses aggregated across all prompts and participants.}
\Description{This table summarizes the results of a technical evaluation based on eight criteria related to goal and intent alignment in the system. Each row lists an evaluation question along with the percentage of participants who answered ``Agree''. The highest ratings (95.00\%) were for Q1 (Goal Alignment) and Q2 (Set of Intents – Completeness). Other notable results include 94.08\% for the relevance of individual intents (Q4) and 94.04\% for the accuracy of intent-to-output linking (Q8). Ratings for various dimensions of intent—such as distinctiveness, relevance, UI appropriateness, and value appropriateness—ranged from approximately 86\% to 87\%, indicating strong overall support for structured and meaningful intent representation in the system.}
\label{table:tech_eval}
\end{table}

Across all prompts and evaluators, the pipeline demonstrated consistently high agreement rates.
As summarized in \autoref{table:tech_eval}, over 85\% of responses were positive across all criteria, suggesting that the pipeline generally produces coherent and interpretable intent representations and links.
We also analyzed qualitative feedback from evaluators. Several participants noted that when intent dimensions were presented without clear descriptions of value meanings, it was difficult to judge their appropriateness.
For example, when the ``Formality Level'' dimension was presented as a slider with an initial value of 4, participants reported that it was hard to judge how formal the value 4 actually was. This likely contributed to the lower agreement in the \textit{Intent Dimension} category since our technical evaluation did not provide a hover-based explanation for each dimension value. This underscores the importance of providing descriptive explanations for dimension values (\autoref{fig:exploration}-d).
\section{User Study}
To understand how the four intent communication supports shape users’ behaviors and experiences compared to current practice, we conducted a within-subjects user study (N=12) comparing \sysname{} with a conventional chat-based LLM interface representative of widely used commercial systems. Rather than evaluating task performance in isolation, we used \sysname{} as a \textbf{\emph{research probe}} to examine how users enact intent communication over time when interacting with systems that differ in their level of explicit intent support. The research questions guiding this investigation are introduced in \autoref{section:intro}. In particular, we centered our analysis on\textbf{ users' action-level intent communication behaviors}---such as adding, adjusting, correcting, and removing intents---captured through participants’ post-hoc annotations of their own inputs. We analyzed these behaviors alongside system logs that recorded \textbf{participants’ use of different intent communication support features} (e.g., articulation, exploration, management, and synchronization), as well as \textbf{self-report surveys} and \textbf{post-study interviews} to capture users’ subjective experiences and interpretations. 
To ground this investigation in users’ existing practices, we implemented a baseline system that reflects common chat-based LLM writing interfaces such as ChatGPT and Claude Artifacts. The baseline represented an ecologically valid status quo for LLM-based writing, in which users primarily communicate intent through linear, conversational turn-taking with free-form prompts. The baseline used the same generation model as \sysname{}, \texttt{gpt-4o-2024-08-06}. The screenshot of the baseline interface can be found in~\autoref{fig:baseline_interface} in the Appendix. 

\subsection{Participants}
We recruited 12 participants (8 male, 4 female; age $\mathit{M}=25.50$, age $\mathit{SD}=2.75$) through online recruitment posting at our university community platforms. Participants were not professional writers; however, all had prior experience with writing tasks and using large language models (LLMs). We intentionally recruited participants with writing and LLM experience to observe how experienced writers manage complex, nuanced, and evolving intents during interaction.
During the recruitment, we administered a pre-survey to collect information about participants' LLM experience, writing background, and familiarity with the task topics (e.g., scientific explanations such as the Doppler Effect). We used this information for pre-screening, excluding those who lacked sufficient topic knowledge or writing experience, as these factors could limit meaningful engagement and evaluation quality.
Participants received 30,000 KRW (around 22 USD) for the 90-minute session, and to encourage careful and motivated engagement, we offered an additional 20,000 KRW (14 USD) performance-based bonus to the top 40\% of participants. Additional details, including participants' LLM experience and writing experiences, are provided in the Appendix~\ref{appendix:participants}.

\subsection{Tasks}
Each participant completed two writing tasks, which were randomly paired with the counterbalanced study conditions. We selected writing tasks commonly used in prior HCI studies~\cite{reza2024abscribe, Gu2024SupportingSOA, kim2023cells} and added short scenarios to each task to introduce more nuanced and multifaceted considerations. 
The first task involved \textbf{writing a social media post} explaining a scientific concept (e.g., the Doppler Effect) for a general audience with little scientific background. The second involved \textbf{writing a professional email} applying for a personal secretary position for a well-known individual outside the participant’s domain. Through multiple rounds of pilot testing, we refined the scenarios to ensure that the two tasks were similar in complexity and involved a comparable level of multifaceted consideration on writing intents---such as tone, structure, and content depth.
To guide the output scope, we set a flexible length guideline of around half an A4 page. We intentionally did not specify an exact word count to prevent users from fixating on meeting a numerical target, which could distract from focusing on the given task scenario and goal.
Importantly, as our study centered on intent communication, participants were encouraged to explore and refine intents rather than produce polished drafts.
Full task descriptions are available in Appendix~\ref{appendix:study_tasks}.

\begin{figure*}
    \centering
    \includegraphics[width=1\textwidth]{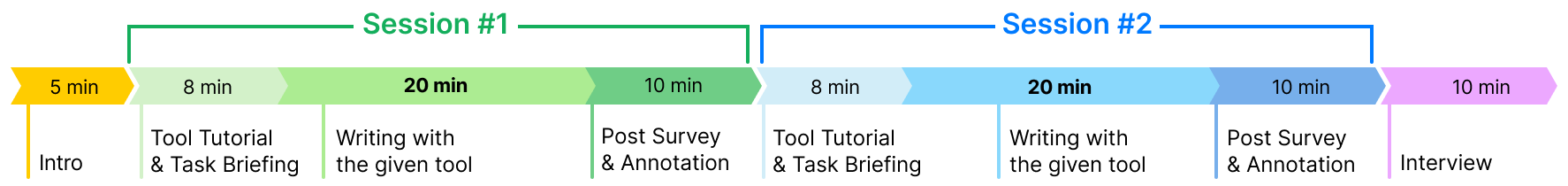}
    \caption{User study procedure. After a brief 5-minute introduction, each participant completed two sessions. The first 8 minutes were for the system tutorial and task briefing. The user had 20 minutes to write using the given system, followed by a 10-minute post-survey and annotation. At the end of both sessions, there was a 10-minute interview.}
    \Description{This figure displays the user study procedure. After a brief 5-minute introduction, two sessions were conducted with each user. The first 8 minutes were for the tool tutorial and task briefing. The user had the next 20 minutes to write with the given tool, and then a 10-minute post survey and annotation were conducted. At the end of both sessions, there was a 10-minute interview.}
    \label{fig:study-process}
\end{figure*}
\subsection{Procedure}
The study was conducted either in person or online via Zoom~\footnote{https://zoom.us/}, depending on participant availability. 
Each study session lasted approximately 90 minutes and followed a fixed protocol~\autoref{fig:study-process}. After a brief introduction and consent process (5 minutes), participants were given a tutorial for the first system (8 minutes). They then completed the first writing task using that system (20 minutes), followed by a post-survey and an annotation activity (10 minutes). 
To gather each participant's intent communication actions, we brought them back to the system after the survey to review their interactions and annotate the purpose of each input. These annotations were used in our analysis of how users engaged in intent communication, as described in more detail in~\autoref{section:measure}.
This process was then repeated for the second system and task. At the end of the sessions, participants completed a semi-structured interview (10 minutes) reflecting on their experience with both systems, including their preferences, their strategies for expressing and adjusting intent, and the usefulness and challenges of each system.

\subsection{Measures}\label{section:measure}
To address our research questions, we collected and analyzed three complementary data sources: (1) annotated intent communication actions derived from user inputs, (2) system logs capturing participants’ use of intent communication support features, and (3) self-report surveys and post-study interviews.

\subsubsection{\textbf{Intent Communicatioin Action Annotation}}
To examine how users enacted intent communication over time (RQ1), we conducted a fine-grained analysis of participants’ intent communication actions throughout the task process. 
We categorized all user inputs into four types of intent communication actions: \add, \delete, \correct, or \adjust. This taxonomy was informed by prior work examining how users express, refine, and repair intent during interactions with LLM-based systems~\cite{Cheng2020ConversationalSPA, bao2023can, kim2023understanding, shin2023planfitting}.
\add\ refers to introducing a new intent not previously expressed. \delete\ refers to removing a previously expressed intent; for instance, a user may have initially requested a summary at the end but later decided to omit it. \correct\ captures cases where a user re-communicates an earlier intent due to a misunderstanding or misalignment by the system; for example, the user asks for a concise style, but the system generates something too verbose, prompting the user to restate the intent more clearly. Finally, \adjust\ involves modifying an existing intent in terms of degree or nuance, for example, slightly increasing the level of formality or adjusting the specificity of a detail without changing the underlying intent altogether.

After completing each task and post-survey, participants were brought back to the system and asked to review their interaction history and annotate eacn of their inputs with an intent communication action. For each input, they indicated which action type best reflected their intent at the time of interaction.
For interface-level interactions in \sysname{} that unambiguously corresponded to a specific action type—such as directly adding or deleting intent items, or adjusting sliders and radio buttons—we automatically labeled the corresponding action and did not require manual annotation.
This post-hoc annotation process allowed us to capture participants’ \emph{intended communicative function} behind each input, enabling analysis of the temporal transitions of intent communication actions over the course of the writing process.

\subsubsection{\textbf{System Logs of Intent Communication Support Feature Usage}}
To examine how different \emph{intent communication support features} mediated users’ intent communication behaviors (RQ2), we analyzed system logs capturing participants’ use of the intent communication support features in \sysname{}. These logs recorded when and how participants engaged with features designed to support intent  \articulation{}, \exploration{}, \management{}, and \synchronization{} during the task process. 
By aligning these feature usage logs with participants’ annotated intent communication actions, we were able to examine how specific intent communication support features shaped, facilitated, or redirected users’ intent communication behaviors over time.

\subsubsection{\textbf{Self-Report Ratings}}
To capture participants’ subjective experiences with each system (RQ3), we collected self-report ratings after each task. Participants responded to a set of statements using 7-point Likert scales (1: Strongly Disagree, 7: Strongly Agree), assessing their subjective perceptions of how each system effectively supported intent communication process. The survey items were adapted from prior work on human–AI interaction, particularly studies on LLM-supported writing systems~\cite{reza2024abscribe, kim2023cells, wu2022aichains}. They measured participants’ perceived \textit{ease} and \textit{clarity} of intent expression, intent \textit{discovery} and \textit{elaboration}, \textit{transparency} and \textit{understanding} of intent–output relationships, and \textit{alignment} between users’ intents and generated outputs. The full list of survey items (M1–M11) is provided in the Appendix \autoref{appendix:survey_items}. In addition, participants completed the NASA Task Load Index (NASA-TLX)~\cite{nasatlx} using a 7-point Likert scale to assess perceived workload.

\subsubsection{Post-study Interviews}
After completing both sessions, participants completed a semi-structured interview to reflect on their experiences with both systems. The goal of these interviews was to gather qualitative data on users’ intent communication processes. Specifically, we asked how they approached expressing, revising, and discarding intents during the task process, how their strategies may have differed across systems, and how they perceived the role of explicit intent representations in supporting or constraining their workflow. In addition, participants were asked to reflect on whether and how the final set of intents they created or kept during the task might be reused in future writing tasks.
The interview data were used to complement our quantitative analysis by providing contextual insights into interaction patterns and feature usage, as well as participants' mental models of intent communication workflows.
\section{result}
For all statistical comparisons, we first conducted Shapiro-Wilk tests to examine the normality of each measure. Based on the results, we used paired t-tests for those that met the normality assumption ($p \ge .05$) and Wilcoxon signed-rank tests otherwise.

\subsection{RQ1. How do users communicate their intents with \sysname{}?}
To examine how intent communication unfolds over time, we analyzed participants’ annotated interaction logs by categorizing each user input into one of four intent communication action types—\add, \delete, \correct, or \adjust—and compared their distributions, temporal sequences, and action-to-action transitions between the Baseline and \sysname{}.
\begin{figure}
    \includegraphics[width=\linewidth]{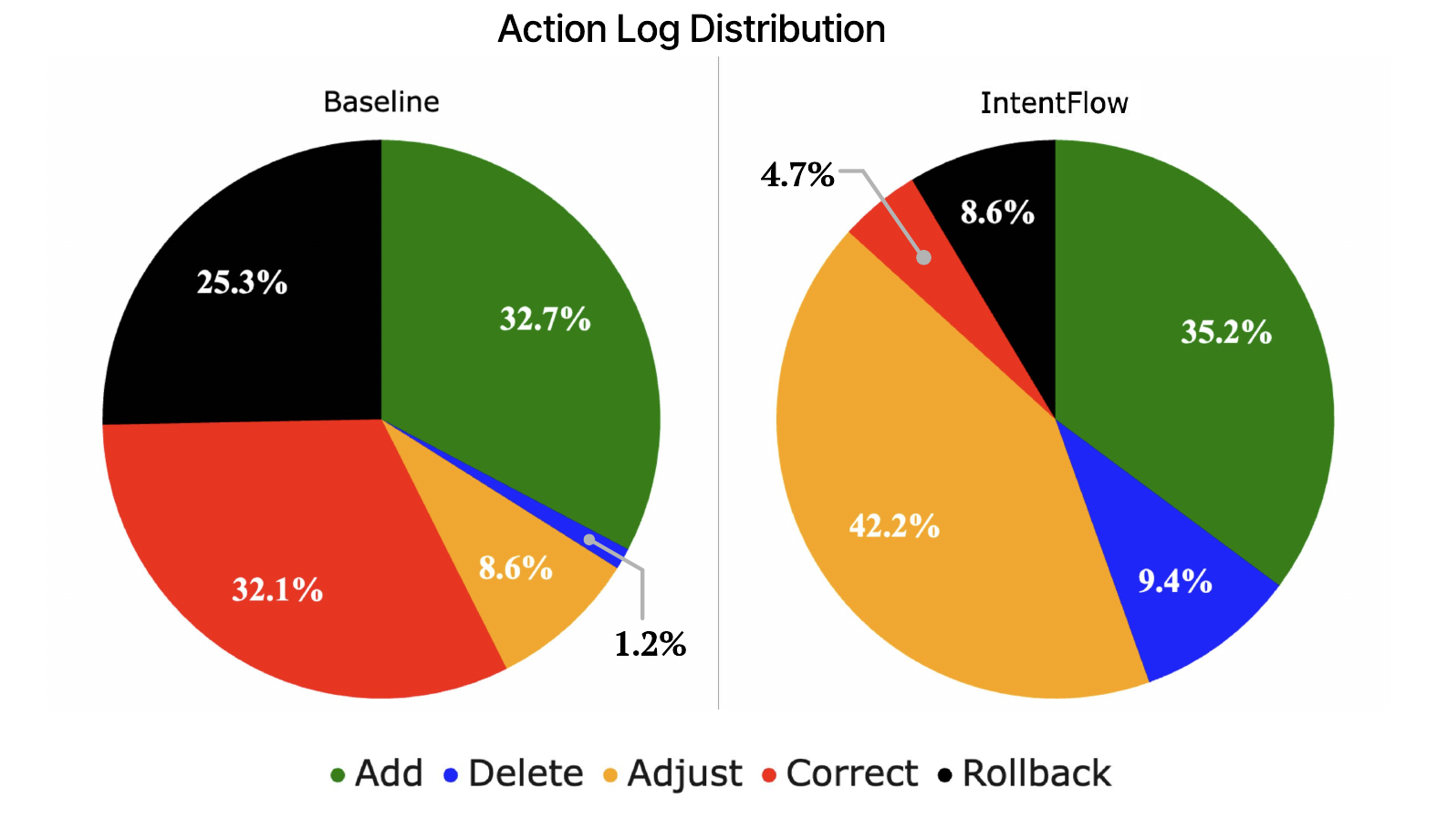}
    \caption{Percentage distribution of action types aggregated across all participants. Each chart reflects the proportion of actions by type in \sysname{} and Baseline.}
    \Description{This figure presents two pie charts comparing the distribution of user action types between the Baseline system and IntentFlow. Each chart shows five action categories: Add (green), Delete (red), Adjust (orange), Correct (blue), and Rollback (black). In the Baseline chart, the most common actions are Add (32.7\%) and Delete (32.1\%), followed by Rollback (25.3\%), Adjust (8.6\%), and Correct (1.2\%). In contrast, the IntentFlow chart shows a higher proportion of Adjust actions (42.2\%) and Add (35.2\%), while Rollback (8.6\%), Correct (4.7\%), and Delete (9.4\%) occur less frequently. The figure highlights how IntentFlow encourages more nuanced intent adjustments and reduces the need for repetitive corrections and rollbacks.}
    \label{fig:action_distribution}
    \vspace{-\baselineskip}
\end{figure}
\begin{figure*}
    \centering
    \includegraphics[width=1\linewidth]{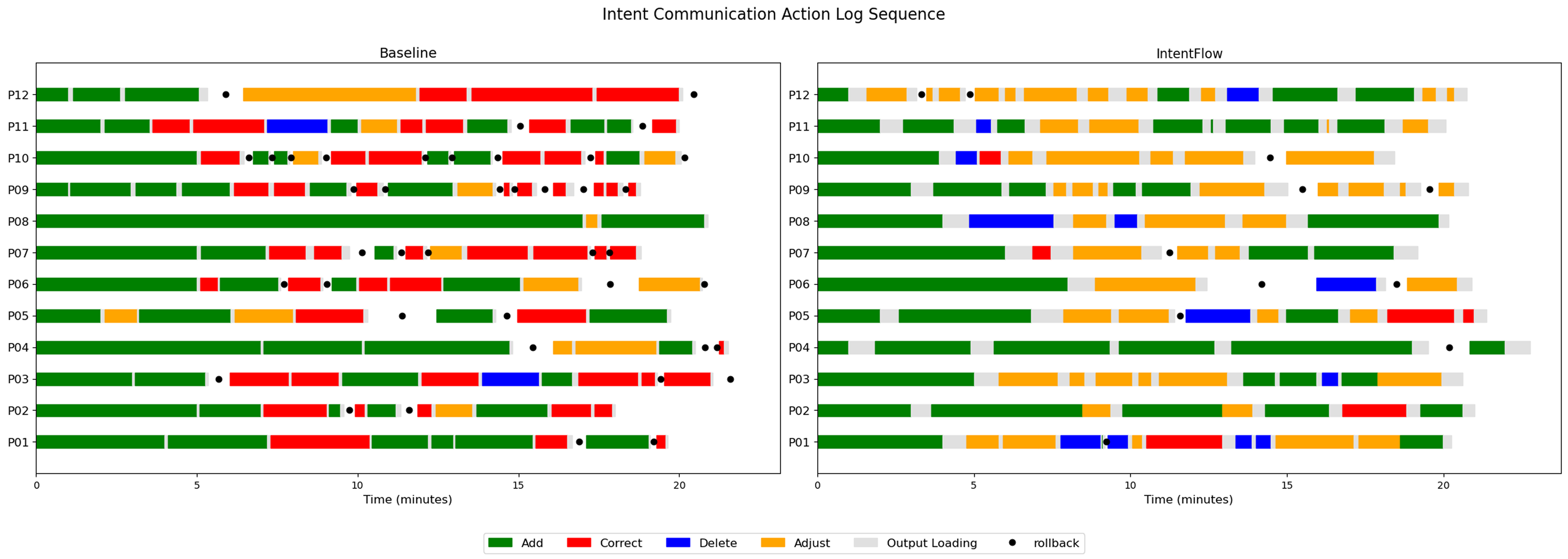}
    \caption{This figure illustrates the action log sequences for 12 participants in each condition: Baseline and \sysname{}.}
    \Description{This figure shows the action log sequences of 12 participants in the Baseline, and 12 participants in \sysname{}. Compared to Baseline, participants in \sysname{} performed significantly fewer Correct actions. Delete and Adjust actions also occurred substantially more.}
    \label{fig:action_log_sequences}
    \vspace{-\baselineskip}
\end{figure*}

\subsubsection{Overall differences in intent communication actions}
As shown in \autoref{fig:action_distribution}, participants’ intent communication behaviors differed substantially across the two systems. Participants performed significantly fewer \correct\ actions when using \sysname{} than in the Baseline (Action Count: \sysname{}: $\mathit{M}=0.50$, $\mathit{SD}=0.67$; Baseline: $\mathit{M}=4.33$, $\mathit{SD}=2.64$; $p<.001$, $t=-4.81$), indicating fewer instances where users had to restate previously expressed intents due to system misalignment. In contrast, the number of \add\ actions was comparable across systems (Action Count: \sysname{}: $\mathit{M}=3.75$, $\mathit{SD}=2.05$; Baseline: $\mathit{M}=4.42$, $\mathit{SD}=1.38$; $p=0.296$, $t=-1.10$), suggesting that users introduced new intents at a similar rate regardless of interface. However, \sysname{} prompted significantly more \adjust\ actions (Action Count: \sysname{}: $\mathit{M}=4.50$, $\mathit{SD}=2.97$; Baseline: $\mathit{M}=1.17$, $\mathit{SD}=0.72$; $p=.005$, $t=3.46$) and more frequent \delete\ actions (Action Count: \sysname{}: $\mathit{M}=1.00$, $\mathit{SD}=1.13$; Baseline: $\mathit{M}=0.17$, $\mathit{SD}=0.39$; $p=.031$, $W=21.00$). These differences indicate a shift from repeatedly correcting misunderstood intents toward actively refining and curating existing intents.

\subsubsection{Temporal evolution of intent communication}
Beyond aggregate counts, the temporal distribution of actions further illustrates how intent communication evolved during the writing process (\autoref{fig:action_log_sequences}). In the Baseline, early stages exhibited diverse transitions such as \add$\rightarrow$\add, \add$\rightarrow$\adjust, and \add$\rightarrow$\correct. Over time, however, interactions increasingly converged on \correct-centered patterns, including repeated \correct$\rightarrow$\correct\ transitions and frequent alternation between \correct\ and \rollback\ . This pattern reflects users repeatedly attempting to reassert previously stated intents or reverting outputs after misalignment, rather than incrementally refining them.
In contrast, \sysname{} showed a different trajectory. Early interactions similarly involved exploratory patterns such as \add$\rightarrow$\adjust\ and \add$\rightarrow$\delete, indicating users externalizing and probing potential intents. As the session progressed, however, intent communication increasingly centered on \adjust, with users repeatedly revisiting and fine-tuning existing intents rather than restating them. This temporal shift suggests that intent communication in \sysname{} stabilized around refinement rather than correction.

\begin{figure*}
    \includegraphics[width=1\linewidth]{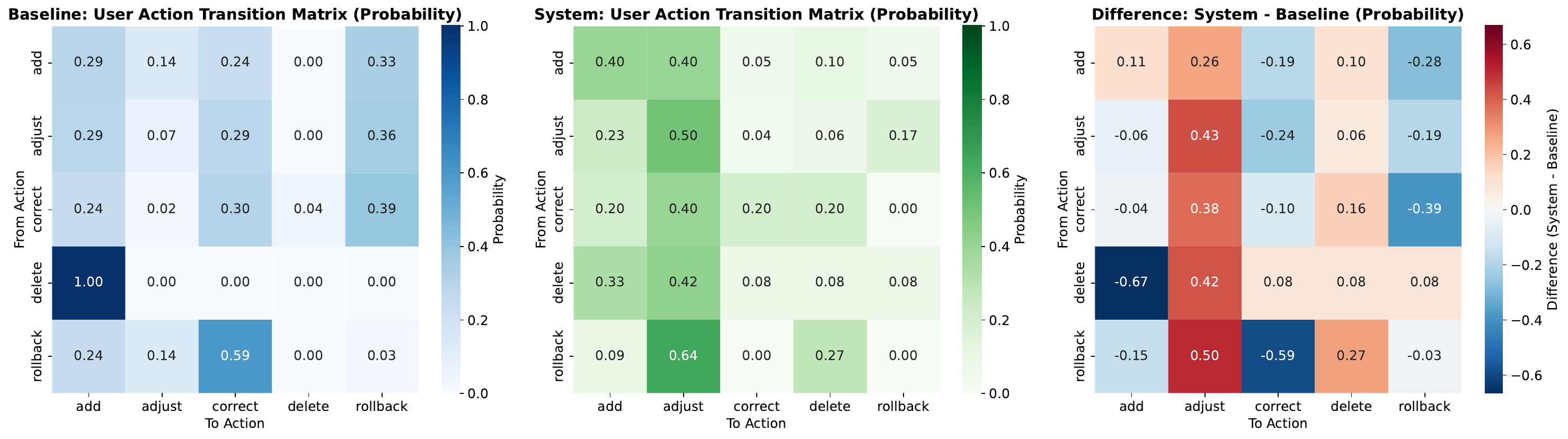}
    \caption{Normalized intent communication action transition matrices. Left: Baseline condition. Middle: \sysname{}. Right: Difference matrix (\sysname{} - Baseline). Each cell represents the conditional probability of transitioning from one action type to another. Compared to the Baseline, \sysname{} shows substantially higher transition probabilities toward \adjust\ and lower probabilities toward \correct\, indicating a shift from correction-driven loops to adjustment-centered refinement.}
    \Description{Three heatmaps showing normalized transition probabilities between intent communication actions. Rows indicate the current action, and columns indicate the subsequent action. In the Baseline heatmap, transitions are concentrated around Correct, including high probabilities for Correct-to-Correct and Rollback-to-Correct. In the IntentFlow heatmap, transitions concentrate around Adjust, with Adjust-to-Adjust as the most prominent self-transition. The difference heatmap highlights higher probabilities for transitions into Adjust and lower probabilities into Correct in IntentFlow compared to the Baseline.}
    \label{fig:action_transition_heatmap}
    \vspace{-\baselineskip}
\end{figure*}

\subsubsection{Transition structure and convergence patterns}
To further characterize these differences, we analyzed normalized action transition matrices for each system (\autoref{fig:action_transition_heatmap}). In the Baseline, transition probabilities were concentrated around \correct\ and \rollback\, indicating recurrent breakdown-recovery cycles. By contrast, \sysname{} exhibited a transition structure centered on \adjust\ action. \adjust$\rightarrow$\adjust\ emerged as the most dominant transition, and transitions to \correct\ or \rollback\ rarely appeared. The difference matrix further highlights this contrast: compared to the Baseline, \sysname{} showed substantially higher transition probabilities into \adjust\ and lower probabilities into \correct. These results quantitatively confirm that intent communication in \sysname{} converged toward adjustment-centered refinement, whereas the Baseline tended to funnel users into correction- and rollback-driven cycles.

\subsubsection{Divergent roles of rollback across systems}
We observed that users' use of \rollback\ differed between the two systems. Rollback occurred significantly more often in the Baseline than in \sysname{} (Action Count: \sysname{}: $\mathit{M}=0.92$, $\mathit{SD}=0.79$; Baseline: $\mathit{M}=3.42$, $\mathit{SD}=2.50$; $p=.003$, $t=-3.80$), but with qualitatively different usage patterns. In the Baseline, \rollback\ was most often paired with \correct\ actions and functioned as a \textit{reset mechanism}: users reverted to earlier outputs after the system failed to respect previously stated intents, then restated or reissued similar instructions. Two participants (P9, P10) explicitly noted frustration, as P9 noted, \textit{``In the B (Baseline), I asked to change only a specific part, but the whole output changed, so I often had to rollback and reissue the same prompt.''} In \sysname{}, \rollback\ more frequently appeared alongside \adjust\ actions, supporting controlled comparison and refinement. Four participants (P7, P9–10, P12) described experimenting with different intent adjustments, comparing resulting outputs, and rolling back to a version that better aligned with their intent. For example, P10 explained, \textit{``In the A (\sysname{}), having the intent dimensions surfaced in the UI made it easier to explore and adjust. I tried different variations and rolled back to the output I liked most.''} This indicates that \rollback\ in \sysname{} supported \textit{exploratory refinement} rather than recovery from breakdowns.

\begin{figure*}
    \includegraphics[width=1\linewidth]{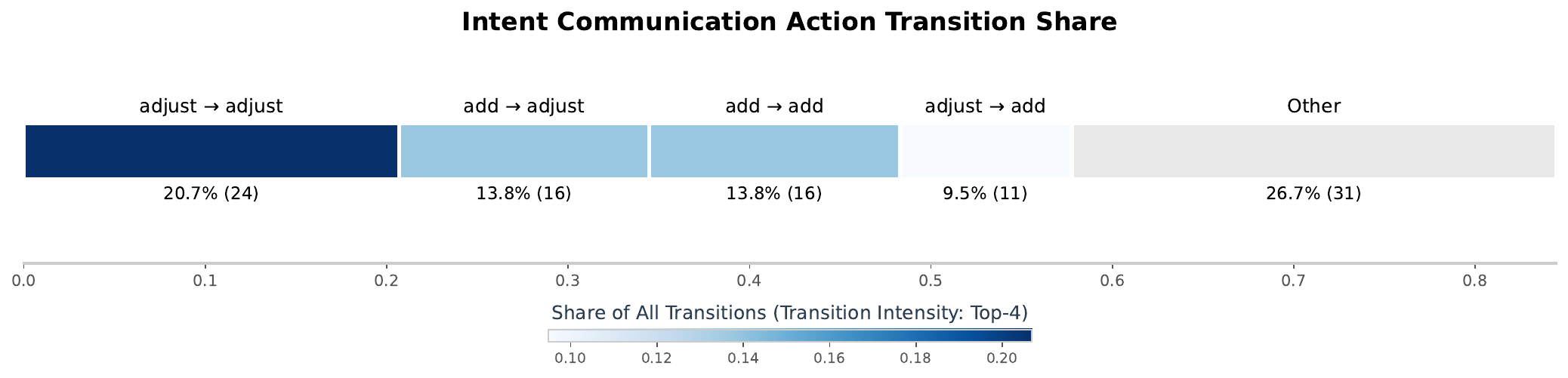}
    \caption{Intent communication action transition shares in \sysname{}. The bar shows the proportion of the most frequent intent communication action transitions observed in \sysname{}, highlighting that \adjust$\rightarrow$\adjust\ is the most dominant transition, followed by transitions involving \add\ and \adjust.}
    \Description{A horizontal bar chart showing the share of the most frequent intent communication action transitions in IntentFlow. Adjust-to-Adjust is the largest segment, followed by Add-to-Adjust, Add-to-Add, and Adjust-to-Add.}
    \label{fig:transition_share}
    \vspace{-\baselineskip}
\end{figure*}

\subsubsection{Dominant transition patterns within \sysname{}}\label{section:dominant_transition}
To further characterize how intent communication unfolded within \sysname{}, we analyzed the distribution of action transitions observed in this condition (Figure~\autoref{fig:transition_share}). The most frequent transition was \adjust$\rightarrow$\adjust, accounting for 20.7\% of all transitions, followed by \add$\rightarrow$\adjust\ (13.8\%), \add$\rightarrow$\add\ (13.8\%), and \adjust$\rightarrow$\add\ (9.5\%). Together, these four transitions accounted for 73.3\% of all transitions, indicating that a small set of transition patterns dominated users’ intent communication in \sysname{}. These patterns reflect a process centered on iterative refinement and structured exploration, rather than repeated correction of misaligned outputs.

\subsection{How do different intent communication support features mediate the intent communication process?}
To understand how different intent communication support features shaped users’ intent communication behaviors, we examined how feature usage sequences mediated the most frequent action transitions observed in \sysname{}. Based on the action transition analysis (Section~\ref{section:dominant_transition}), we focused on the four most prevalent transitions---\adjust$\rightarrow$\adjust, \add$\rightarrow$\adjust, \add$\rightarrow$\add, and \adjust$\rightarrow$\add. For each transition type, we analyzed (1) the most frequent feature sequences and (2) the underlying feature transition flows to uncover recurring interaction patterns.

\begin{figure*}
    \centering
    \includegraphics[width=1\linewidth]{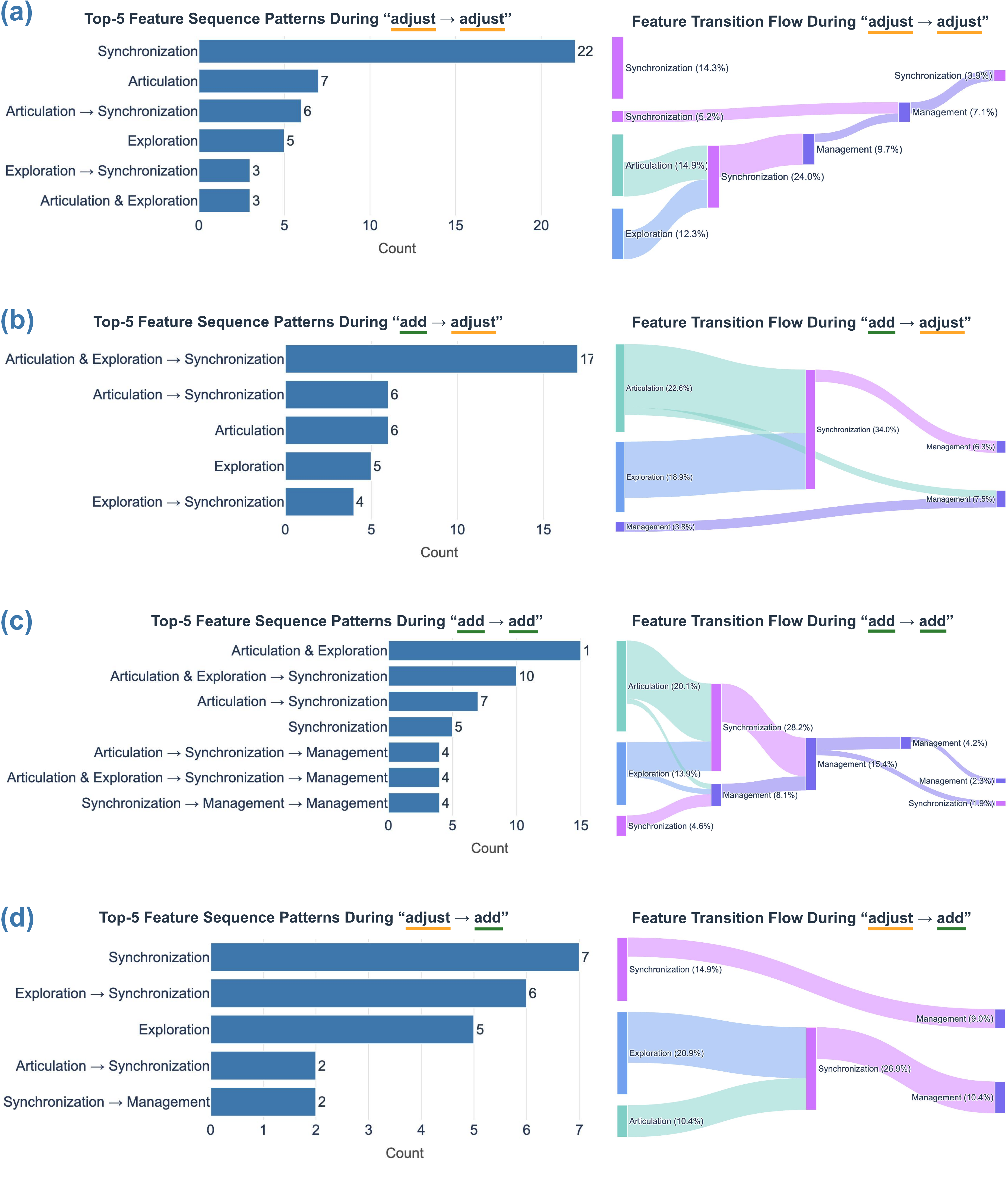}
    \caption{Feature-mediated intent communication patterns in \sysname{}. For each of the four most frequent action transitions---(a) \adjust$\rightarrow$\adjust, (b) \add$\rightarrow$\adjust, (c) \add$\rightarrow$\add, (d) \adjust$\rightarrow$\add---the right column shows the five most frequent feature sequences, and the left column visualizes the corresponding feature transition flows using Sankey diagrams.}
    \Description{A multi-panel figure showing feature usage patterns for four intent communication action transitions in IntentFlow. For each transition, a bar chart summarizes the top five feature sequences, and a Sankey diagram illustrates the flow between articulation, exploration, synchronization, and management. Across all transitions, synchronization appears as a central mediating feature connecting intent articulation, exploration, adjustment, and management.}
    \label{fig:rq2}
\end{figure*}

\subsubsection{Verification-driven refinement and intent curation}
The most frequent transition in \sysname{} was \adjust$\rightarrow$\adjust. As shown in ~\autoref{fig:rq2}-(a), \synchronization{} overwhelmingly dominated both the top feature sequences and the feature transition flow associated with this transition. The Sankey diagram shows that users' interactions were organized around \synchronization{}, with adjustments followed by verification and subsequent management.
In practice, users relied on \synchronization{} to inspect how their current set of intents was reflected in the output, and used this feedback to decide which intents to keep, refine, or deprioritize. Adjustments were often incremental and selective, targeting specific intent dimensions while preserving others, and were interleaved with moments of \management{}, such as curating particular intents. We characterize this pattern as a \textit{\textbf{verification-driven refinement and intent curation}}, in which \synchronization{} serves as the primary mechanism not merely as a verification step, but also as a mechanism for maintaining continuity and supporting intent curation across iterations.

\subsubsection{Exploration-to-alignment transitions}
The second most frequent transition was \add$\rightarrow$\adjust. As shown in ~\autoref{fig:rq2}-(b), the dominant feature sequences combine \articulation{} and \exploration{} followed almost immediately by \synchronization{}. 
The feature transition flow highlights a pattern in which users externalized tentative intents and then relied on \synchronization{} to examine how those intents were reflected in the generated output. Based on this immediate feedback, users transitioned directly into adjustment, refining the scope, emphasis, or specificity of the newly added intent. \management{} appeared only sparingly in this transition, indicating that users were primarily focused on aligning individual intents rather than moving to the stabilizing phase. This pattern reflects an \textbf{\textit{exploration-to-alignment transition}}, where \synchronization{} enables users to quickly assess and refine newly added intents before committing to further expansion.

\subsubsection{Intent space expansion followed by grounding}
Another second most frequent transition was \add$\rightarrow$\add\, which represents moments where users introduced multiple intents in succession. As shown in ~\autoref{fig:rq2}-(c), this transition was strongly associated with combined \articulation{} \& \exploration{} sequences, often followed by \synchronization{} and \management{}. The feature flow reveals a two-phase structure. In the first phase, users expanded the intent space by articulating new intents (\articulation{}) or exploring alternative perspectives (\exploration{}). This \textit{divergence} was then followed by \synchronization{}, allowing users to inspect how the growing set of intents collectively shaped the output. Based on this feedback, users engaged in \management{} by selectively consolidating the expanded intent set, resulting in a \textit{convergence} phase in which the intent space became more stable and structured. This pattern indicates that \add$\rightarrow$\add\ does not reflect unfocused or random intent creation. Instead, it represents a deliberate \textbf{\textit{intent space expansion by grounding pattern}}, in which users first externalize a broad range of intents and then actively organize them once their effects become visible.

\subsubsection{Alignment revealing missing intents}
Finally, the \adjust$\rightarrow$\add\ transition captures moments where refinement surfaced gaps in users’ intent space. As shown in ~\autoref{fig:rq2}-(d), this transition was frequently preceded by \synchronization{} and \exploration{}. The feature transition flow suggests that while users were adjusting an existing intent, \synchronization{} helped reveal limitations in what the current intent configuration could achieve. This realization often prompted the addition of a new, complementary intent. Importantly, this pattern shows that new intents did not emerge only from open-ended exploration; instead, they were often triggered by alignment checks that exposed what was missing. We interpret this as a \textbf{\textit{feedback-driven intent discovery pattern}}, where synchronization during refinement actively supports the emergence of new intents.

\subsection{RQ3.How do intent communication support features affect users’ cognitive effort, sense of control, and understanding of intent--output alignment during interaction?}
To examine how comprehensive intent communication support shaped users’ subjective experiences, we analyzed participants’ self-report ratings (M1–M11) and perceived workload using NASA-TLX.
\subsubsection{Perceived support for intent expression, alignment, and control}
\begin{figure*}
    \centering
    \includegraphics[width=1\linewidth]{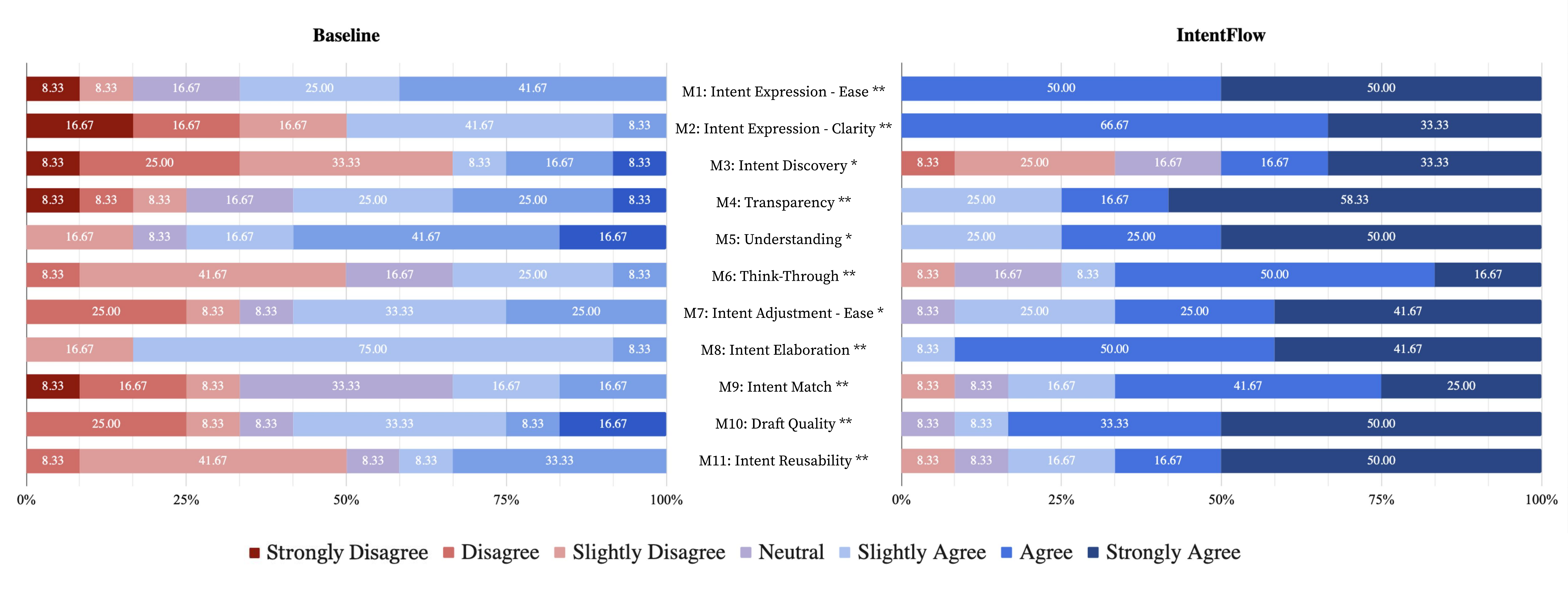}
    \caption{This is a distribution of participants' ratings on intent communication experience. (*$p<.05$, **$p<.01$)}
    \Description{This figure shows two stacked bar charts, one for Baseline and one for \sysname{}, representing participants' ratings on intent communication experience. In each chart, there is a 100\% bar for 11 different dimensions. These are M1: Intent Expression - Ease, M2: Intent Expression - Clarity, M3: Intent Discovery, M4: Transparency, M5: Understanding, M6: Think-Through, M7: Intent Adjustment - Ease, M8: Intent Elaboration, M9: Intent Match, M10: Draft Quality, and M11: Intent Reusability. Compared to Baseline, \sysname{} has significantly more red, which indicates options Strongly Disagree/Disagree/Slightly Disagree. \sysname{} has a substantive amount of blue, with most fields having around 50\% of users rating ``Strongly Agree''.}
    \label{fig:bar-graph}
\end{figure*}

Across all intent communication measures, participants rated \sysname{} significantly higher than the Baseline (M1–M11; all $p < .05$; ~\autoref{fig:bar-graph}). In particular, participants reported substantially greater ease and clarity of intent expression (M1–M2), stronger support for discovering and elaborating intents (M3, M8), and improved understanding of how their intents were reflected in the generated output (M4–M6). Ratings related to intent–output alignment and adjustment were also consistently higher in \sysname{} (M7, M9), indicating that participants felt better able to steer the system toward their intended direction. Notably, ratings associated with perceived control and reusability (M10–M11) suggest that participants viewed intent communication in \sysname{} not merely as a means to complete the current task, but as a transferable strategy for future writing tasks. This aligns with RQ1 and RQ2 findings showing that users converged on stable, well-articulated intent configurations through iterative adjustment rather than repeated correction. Overall, these results indicate that comprehensive intent communication support helped users perceive the interaction as more intentional and controllable, with clearer correspondence between what they wanted to express and what the system produced.

\subsubsection{Reduced cognitive workload and frustration}
\begin{figure*}
    \begin{subfigure}[t]{0.36\textwidth}
        \centering
        \includegraphics[width=\linewidth]{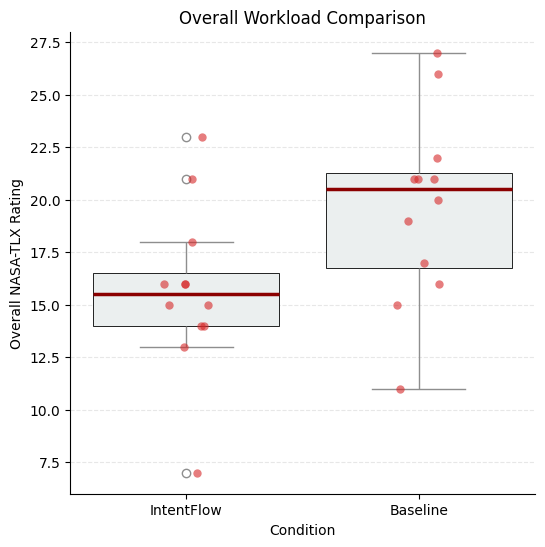}
        \caption{Comparison of overall NASA-TLX ratings between \sysname{} and Baseline. The dark red line indicates the median score per condition.}
        \Description{This figure is a boxplot comparing the overall NASA-TLX ratings between IntentFlow and Baseline. The boxplot indicates that IntentFlow's median overall NASA-TLX rating of 15.67 is significantly lower than Baseline's median overall NASA-TLX rating of 19.67.}
        \label{fig:overall_nasatlx_boxplot}
    \end{subfigure}
    \hspace{0.05\textwidth}
    \begin{subfigure}[t]{0.5\textwidth}
        \centering
        \includegraphics[width=\linewidth]{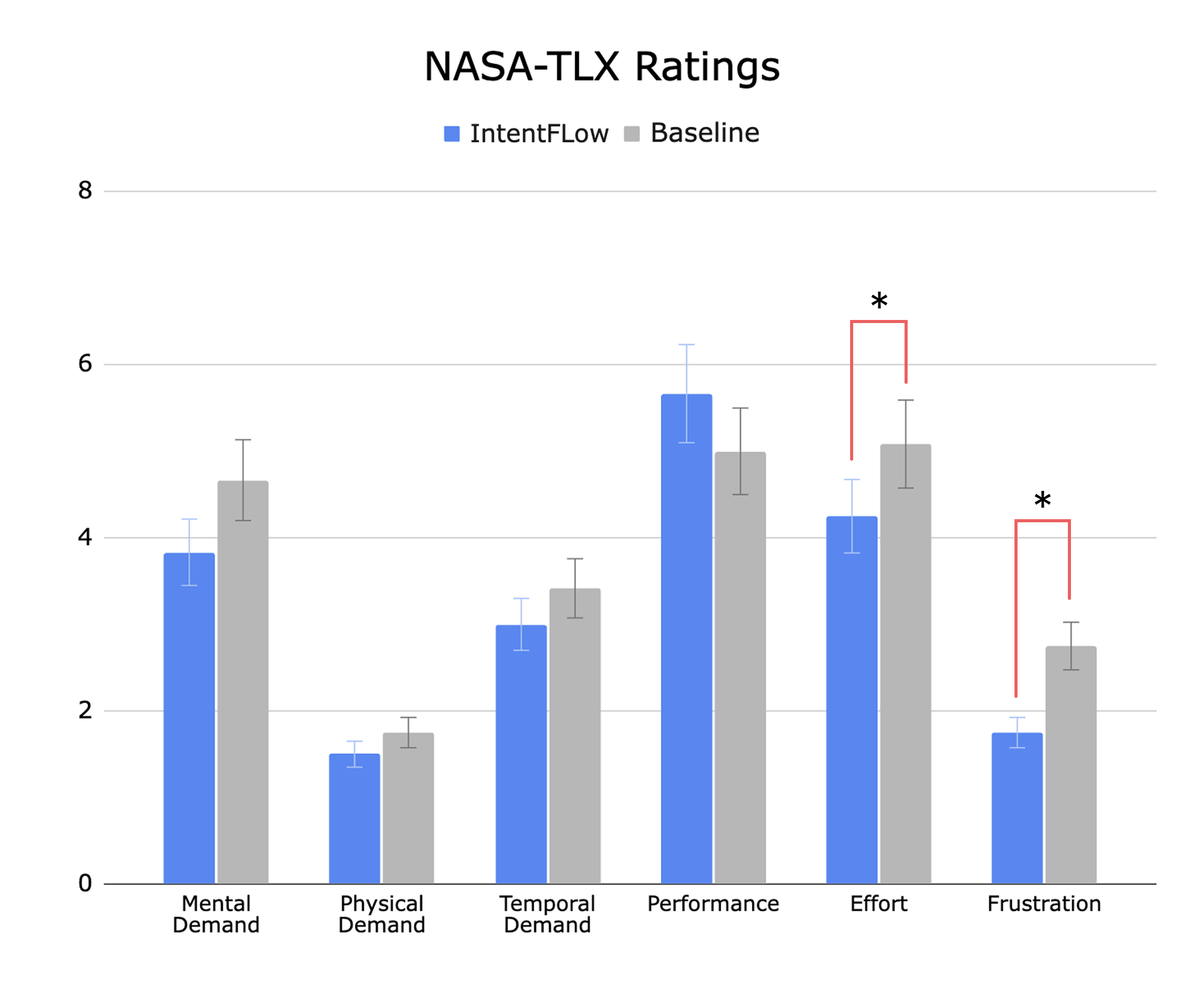}
        \caption{Comparison of NASA-TLX ratings between \sysname{} and Baseline. (*$p<.05$).}
        \label{fig:nasa-tlx-bar}
        \Description{A grouped bar chart comparing NASA-TLX ratings between IntentFlow (blue bars) and the Baseline (gray bars) across six dimensions: Mental Demand, Physical Demand, Temporal Demand, Performance, Effort, and Frustration. Notably, Effort and Frustration show significant differences, with IntentFlow receiving lower ratings, indicated by red brackets and asterisks (*p < .05).}
    \end{subfigure}
    \caption{NASA-TLX ratings between \sysname{} and Baseline.}
    \vspace{-\baselineskip}
\end{figure*}

NASA-TLX results further show that these perceived benefits were accompanied by reduced cognitive effort. Participants reported significantly lower overall workload when using \sysname{} compared to the Baseline (\sysname{}: $\mathit{M}=15.67$, $\mathit{SD}=4.01$; Baseline: $\mathit{M}=19.67$, $\mathit{SD}=4.50$; $p=.004$, $t=2.02$; ~\autoref{fig:overall_nasatlx_boxplot}). A breakdown of individual NASA-TLX dimensions revealed that this reduction was driven primarily by lower perceived \textbf{Effort} (\sysname{}: $\mathit{M}=4.25$, $\mathit{SD}=1.22$; Baseline: $\mathit{M}=5.08$, $\mathit{SD}=0.90$; $p=.048$) and \textbf{Frustration} (\sysname{}: $\mathit{M}=2.75$, $\mathit{SD}=1.49$; Baseline: $\mathit{M}=3.83$, $\mathit{SD}=1.75$; $p=.034$), while differences in mental, temporal, and physical demand were less pronounced (\autoref{fig:nasa-tlx-bar}). These findings suggest that \sysname{} did not reduce workload by simplifying the task itself, but by reducing the cognitive overhead associated with maintaining, restating, and repairing intent. Instead of repeatedly correcting misaligned outputs or managing intent implicitly through chat history, participants could externalize, inspect, and incrementally adjust their intents with clearer feedback.

\section{Discussion}
Our study reveals that supporting intent communication is not simply a matter of improving how users express prompts, but of shaping how users interact with their own evolving intentions over time. By synthesizing findings across RQ1–RQ3, we argue that effective intent communication emerges when four forms of support---\articulation, \exploration, \synchronization, and \management---work together as a cycle rather than as isolated features.

\subsection{Intent Communication as Ongoing Alignment: A Cyclical Process}
Across our analyses, breakdowns in intent communication did not primarily arise from users’ inability to express their intentions. Instead, they emerged when users lacked support for 1) verifying whether their intents were being reflected, 2) retaining previously stated constraints, or 3) selectively revising parts of their intent without destabilizing others. In such situations, users were forced into corrective strategies---repeating instructions, rolling back outputs, or restating intents---treating intent communication as a fragile, error-driven recovery process. 

In contrast, when intent articulation, exploration, synchronization, and management were jointly supported, intent communication became a form of ongoing alignment work. Users could externalize tentative intents, inspect how they were realized, and refine or remove them without losing contextual continuity. This shift fundamentally changed how users engaged with the system: instead of compensating for misalignment after the fact, they actively calibrated the system's behavior by operating directly on intents as they emerged and evolved.

Our findings reveal that these four aspects function as a cyclical process, manifesting in four distinct interaction patterns that trace intent communication's evolution from initial externalization to sustained refinement. Intent communication typically begins with \articulation{} and \exploration{}, where users externalize multiple tentative intents. The \textit{\textbf{intent space expansion by grounding pattern}} demonstrated this: users diverged by articulating multiple candidate intents, then converged by using \synchronization{} and \management{} to stabilize and structure the expanded set.
As users inspect how their intents are realized, \synchronization{} enables rapid alignment. The \textit{\textbf{exploration-to-alignment pattern}} showed users externalizing intents through \articulation{} \& \exploration{}, then immediately using \synchronization{} to examine their realization and refine scope or emphasis based on concrete feedback.
\management{} stabilizes which intents persist, creating a coherent context where synchronization reveals gaps. The \textit{\textbf{feedback-driven intent discovery pattern}} showed how \synchronization{} during refinement exposed limitations in the current configuration, prompting addition of complementary intents to address newly surfaced needs.
As the intent set matures, the cycle converges toward sustained refinement. The \textbf{\textit{verification-driven refinement and intent curation pattern}} demonstrated this.
Breaking any link forces manual compensation. Without management, exploration fragments; without synchronization, articulation becomes guesswork; without exploration, articulation stagnates; without articulation, nothing synchronizes. The Baseline's correction-driven loops demonstrate this: absent persistent management and explicit synchronization, users repeatedly restated intents. Supporting all four aspects allowed sustained, iterative process, reflected in \sysname{}'s adjustment-centered patterns and significantly higher ratings on intent expression, alignment, and control (M1–M11, all p<.05), with reduced effort and frustration.

This perspective reframes intent communication not as a single act of specification, but as a sustained interactional process in which understanding is progressively negotiated between the user and the system. Effective intent communication emerges not from perfecting any single moment of expression, but from supporting the continuous cycle through which users and systems come to shared understanding.

\subsection{Design Implications for Supporting Cyclical Intent Communication}
Our findings suggest that supporting intent communication requires attention not only to individual interaction mechanisms, but to how they are connected across time. Building on this insight, we present four design implications that articulate how intent communication support should be composed and coordinated in human-GenAI interaction.

\smallskip
\noindent
\textbf{DI1: Enable immediate, bidirectional traceability between articulation and synchronization.}
To support verification-driven refinement, systems should tightly couple intent articulation with immediate and interpretable feedback about how those intents are realized in the output. When users articulate or adjust an intent, the system should promptly reveal which parts of the output were affected, making the system’s interpretation visible and inspectable. Conversely, selecting or inspecting parts of the output should surface the intents that contributed to those segments, enabling reverse tracing from realization back to intention. This bidirectional traceability transforms articulation from a one-way instruction into an interactive dialogue. Rather than articulating intents and waiting to infer outcomes retrospectively, users can incrementally calibrate their expressions based on concrete evidence of how the system interpreted them.

\smallskip
\noindent
\textbf{DI2: Use synchronized views as scaffolds for exploration.}
Exploration should not be treated as a separate phase or mode that users must explicitly enter. Instead, synchronized views---where users inspect how their current intents are reflected---should act as natural launch points for exploratory refinement. Systems should allow users to immediately try alternatives from within these views, supported by lightweight parameter adjustments, instant previews, and clear visual diffs that communicate the effects of each change. Embedding exploration directly into synchronization lowers the cost of asking ``what if'' and encourages users to probe variations without abandoning their current mental context. Our analysis of \add$\rightarrow$\adjust\ and \adjust$\rightarrow$\adjust\ transitions suggests that when users could immediately explore after inspecting alignment, exploration flowed seamlessly into refinement, rather than triggering disruptive cycles of re-articulation.

\smallskip
\noindent
\textbf{DI3: Support progressive commitment between exploration and management.}
As users explore intent variations, systems should allow tentative ideas to coexist with more stable, committed intents. Rather than forcing immediate decisions about whether an explored variation should replace the current configuration, systems should support progressive commitment: keeping exploratory branches available for comparison while enabling users to selectively promote, merge, or discard them. This requires treating exploration history as a first-class structure, such as comparable alternatives or branches, rather than ephemeral trials. Transitioning an explored intent into the managed set should be a lightweight operation that preserves continuity with other intents.

\smallskip
\noindent
\textbf{DI4: Contextualize new articulation through visible and structured intent spaces.}
Articulation is most effective when it is informed by awareness of what has already been specified, stabilized, or deprioritized. Systems should therefore make the current intent configuration explicitly visible whenever users add or modify intents. This includes surfacing the managed intent set alongside articulation interfaces, highlighting relationships such as overlap or tension among intents, and revealing gaps that may guide what to articulate next. By structuring the intent space, through categories, dimensions, or summaries, systems can transform articulation from isolated expression into contextualized refinement of an evolving configuration.

\smallskip
\noindent
Taken together, these implications emphasize that supporting intent communication is not a matter of adding more controls or making individual features more powerful. Rather, it requires designing the transitions between articulation, synchronization, exploration, and management so that each step naturally invites the next. When these aspects are orchestrated to form a self-reinforcing cycle, intent communication becomes less about recovering from misalignment and more about progressively shaping, validating, and reusing what users want to express.

\subsection{Generalizability of Design Implications Across Domains}
Although \sysname{} was instantiated in a writing context, the intent communication patterns and design implications discussed above are not specific to writing. Many generative domains---such as data analysis, design, image editing, and code generation---similarly involve communicating multiple, evolving intents and iteratively aligning them with system outputs over time~\cite{son2024genquery, xie2024waitgpt, lee2025veriplan, feng2025cocoa}. Across these domains, users face recurring challenges: maintaining constraints across iterations, understanding how partial specifications are interpreted, and refining intent configurations without repeatedly starting from scratch. Thus, the four aspects of intent communication identified in this work (\articulation{}, \exploration{}, \management{}, and \synchronization{} constitute a domain-agnostic interaction lens. By adapting how these aspects are instantiated in domain-specific representations and interfaces, future systems can support sustained intent communication in a broad range of human–GenAI workflows.

\subsection{Limitations and Future Work}

Our work highlights several directions for future research on intent communication in generative AI systems.

First, while our discussion conceptually addressed generalizability beyond writing tasks, we did not empirically examine how intent communication dynamics manifest across different domains. Future work should investigate how users articulate, refine, and coordinate intents in domains such as design, data analysis, programming, and visual content creation. Such studies could reveal which intent communication dynamics are consistent across domains and which emerge from domain-specific constraints and practices, helping to identify more general principles of intent-centered interaction.

Second, our study focused on short-term interactions within a single session, capturing fine-grained, moment-to-moment intent communication dynamics. However, this could not reflect how users’ interaction strategies evolve over longer periods of real-world use. Longitudinal studies are needed to examine how patterns of intent communication change as users gain experience with a system, including how cycles of articulation, exploration, synchronization, and management stabilize, diversify, or transform over time at a coarser temporal granularity.

Third, our work treats intents as structured representations that exist during an interaction session, but does not address how such representations persist beyond it. Future research could explore intents as long-term interaction artifacts: how intent representations can be stored, reused, merged, or adapted across tasks, how histories of intents can support reflection and personalization, and how conflicts or redundancies among accumulated intents are detected and resolved. This direction concerns the lifecycle of intent representations themselves, rather than the interaction behaviors through which they are produced.

Finally, our research probe currently treats user intent primarily from explicit inputs and in-system actions. In practice, however, user intent is often shaped by broader task context and implicit behavioral signals. External context---such as information gathered through web browsing, documents consulted in other tools, ongoing projects, or activities outside the AI system---can substantially influence how intents form and shift. Likewise, users communicate aspects of their intent implicitly through behavioral cues such as hesitation, repeated revisions, interaction tempo, patterns of tool switching, or localized focus on specific parts of an artifact. Future systems should investigate how such external and implicit signals can be integrated to construct richer models of evolving user intent, and how intent communication interfaces should adapt when intent is inferred not only from what users explicitly state or manipulate, but also from what they do across tools and over time.

\section{Conclusion}
In this work, we investigated intent communication as a dynamic process and how generative AI systems can effectively support it. Based on a systematic literature review of 46 HCI papers on generative AI systems, we synthesized four key aspects of intent communication support, \articulation{}, \exploration{}, \management{}, and \synchronization{}, along with common features used to implement them. While prior systems individually supported these aspects, it has remained unclear how they interact in practice and how their combination shapes users’ intent communication behaviors over time. To address this gap, we instantiated all four aspects together in a research probe, enabling fine-grained, action-level analysis of how users articulate, refine, and stabilize evolving intents through interaction.

Through a within-subjects user study, we showed that supporting and connecting these aspects changes how users communicate intent, shifting interaction patterns from repetitive correction toward deliberate, verification-driven refinement. Our analyses revealed how users iteratively add, adjust, inspect, and manage intents, and how different forms of intent support mediate these behaviors and users’ perceived effort, control, and intent–output alignment. Building on these findings, we articulated design implications that emphasize orchestrating intent communication supports across time rather than treating them as isolated features. Although instantiated in a writing context, the interaction patterns and design implications generalize to other generative tasks in which users iteratively align evolving intents with system outputs. By foregrounding intent communication as a dynamic process and using a research probe to study it holistically, this work contributes an interaction-centered foundation for designing more transparent, controllable, and sustained human–AI collaboration.

\bibliographystyle{ACM-Reference-Format}
\bibliography{references}

@String{Computing = "Computing" }

@misc{AIFirstN88:online,
author = {Jakob Nielsen},
title = {AI: First New UI Paradigm in 60 Years},
howpublished = {\url{https://www.nngroup.com/articles/ai-paradigm/}},
month = {},
year = {2023},
note = {(Accessed on 02/18/2025)}
}

@article{kim2023understanding,
  title={Understanding Users' Dissatisfaction with ChatGPT Responses: Types, Resolving Tactics, and the Effect of Knowledge Level},
  author={Kim, Yoonsu and Lee, Jueon and Kim, Seoyoung and Park, Jaehyuk and Kim, Juho},
  journal={arXiv preprint arXiv:2311.07434},
  year={2023}
}

@article{qian2024tell,
  title={Tell me more! towards implicit user intention understanding of language model driven agents},
  author={Qian, Cheng and He, Bingxiang and Zhuang, Zhong and Deng, Jia and Qin, Yujia and Cong, Xin and Lin, Yankai and Zhang, Zhong and Liu, Zhiyuan and Sun, Maosong},
  journal={arXiv preprint arXiv:2402.09205},
  year={2024}
}

@inproceedings{wu2022aichains,
author = {Wu, Tongshuang and Terry, Michael and Cai, Carrie Jun},
title = {AI Chains: Transparent and Controllable Human-AI Interaction by Chaining Large Language Model Prompts},
year = {2022},
isbn = {9781450391573},
publisher = {Association for Computing Machinery},
address = {New York, NY, USA},
url = {https://doi.org/10.1145/3491102.3517582},
doi = {10.1145/3491102.3517582},
booktitle = {Proceedings of the 2022 CHI Conference on Human Factors in Computing Systems},
articleno = {385},
numpages = {22},
location = {New Orleans, LA, USA},
series = {CHI '22}
}

@misc{deng2023surveyproactivedialoguesystems,
      title={A Survey on Proactive Dialogue Systems: Problems, Methods, and Prospects}, 
      author={Yang Deng and Wenqiang Lei and Wai Lam and Tat-Seng Chua},
      year={2023},
      eprint={2305.02750},
      archivePrefix={arXiv},
      primaryClass={cs.CL},
      url={https://arxiv.org/abs/2305.02750}, 
}

@misc{kaddour2023challengesapplicationslargelanguage,
      title={Challenges and Applications of Large Language Models}, 
      author={Jean Kaddour and Joshua Harris and Maximilian Mozes and Herbie Bradley and Roberta Raileanu and Robert McHardy},
      year={2023},
      eprint={2307.10169},
      archivePrefix={arXiv},
      primaryClass={cs.CL},
      url={https://arxiv.org/abs/2307.10169}, 
}

@misc{anwar2024foundationalchallengesassuringalignment,
      title={Foundational Challenges in Assuring Alignment and Safety of Large Language Models}, 
      author={Usman Anwar and Abulhair Saparov and Javier Rando and Daniel Paleka and Miles Turpin and Peter Hase and Ekdeep Singh Lubana and Erik Jenner and Stephen Casper and Oliver Sourbut and Benjamin L. Edelman and Zhaowei Zhang and Mario Günther and Anton Korinek and Jose Hernandez-Orallo and Lewis Hammond and Eric Bigelow and Alexander Pan and Lauro Langosco and Tomasz Korbak and Heidi Zhang and Ruiqi Zhong and Seán Ó hÉigeartaigh and Gabriel Recchia and Giulio Corsi and Alan Chan and Markus Anderljung and Lilian Edwards and Aleksandar Petrov and Christian Schroeder de Witt and Sumeet Ramesh Motwan and Yoshua Bengio and Danqi Chen and Philip H. S. Torr and Samuel Albanie and Tegan Maharaj and Jakob Foerster and Florian Tramer and He He and Atoosa Kasirzadeh and Yejin Choi and David Krueger},
      year={2024},
      eprint={2404.09932},
      archivePrefix={arXiv},
      primaryClass={cs.LG},
      url={https://arxiv.org/abs/2404.09932}, 
}

@inproceedings{subramonyam2024bridging,
author = {Subramonyam, Hari and Pea, Roy and Pondoc, Christopher and Agrawala, Maneesh and Seifert, Colleen},
title = {Bridging the Gulf of Envisioning: Cognitive Challenges in Prompt Based Interactions with LLMs},
year = {2024},
isbn = {9798400703300},
publisher = {Association for Computing Machinery},
address = {New York, NY, USA},
url = {https://doi.org/10.1145/3613904.3642754},
doi = {10.1145/3613904.3642754},
booktitle = {Proceedings of the 2024 CHI Conference on Human Factors in Computing Systems},
articleno = {1039},
numpages = {19},
location = {Honolulu, HI, USA},
series = {CHI '24}
}

@misc{rodriguez2024intentgptfewshotintentdiscovery,
      title={IntentGPT: Few-shot Intent Discovery with Large Language Models}, 
      author={Juan A. Rodriguez and Nicholas Botzer and David Vazquez and Christopher Pal and Marco Pedersoli and Issam Laradji},
      year={2024},
      eprint={2411.10670},
      archivePrefix={arXiv},
      primaryClass={cs.CL},
      url={https://arxiv.org/abs/2411.10670}, 
}

@misc{wu2025collabllmpassiverespondersactive,
      title={CollabLLM: From Passive Responders to Active Collaborators}, 
      author={Shirley Wu and Michel Galley and Baolin Peng and Hao Cheng and Gavin Li and Yao Dou and Weixin Cai and James Zou and Jure Leskovec and Jianfeng Gao},
      year={2025},
      eprint={2502.00640},
      archivePrefix={arXiv},
      primaryClass={cs.AI},
      url={https://arxiv.org/abs/2502.00640}, 
}

@inproceedings{peng2025navigating,
author = {Peng, Yingzhe and Qin, Xiaoting and Zhang, Zhiyang and Zhang, Jue and Lin, Qingwei and Yang, Xu and Zhang, Dongmei and Rajmohan, Saravan and Zhang, Qi},
title = {Navigating the Unknown: A Chat-Based Collaborative Interface for Personalized Exploratory Tasks},
year = {2025},
isbn = {9798400713064},
publisher = {Association for Computing Machinery},
address = {New York, NY, USA},
url = {https://doi.org/10.1145/3708359.3712093},
doi = {10.1145/3708359.3712093},
booktitle = {Proceedings of the 30th International Conference on Intelligent User Interfaces},
pages = {1048–1063},
numpages = {16},
location = {
},
series = {IUI '25}
}

@incollection{clark2008cognitive,
  title={Cognitive task analysis},
  author={Clark, Richard E and Feldon, David F and Van Merrienboer, Jeroen JG and Yates, Kenneth A and Early, Sean},
  booktitle={Handbook of research on educational communications and technology},
  pages={577--593},
  year={2008},
  publisher={Routledge}
}

@inproceedings{Kazemitabaar2024Improving,
author = {Kazemitabaar, Majeed and Williams, Jack and Drosos, Ian and Grossman, Tovi and Henley, Austin Zachary and Negreanu, Carina and Sarkar, Advait},
title = {Improving Steering and Verification in AI-Assisted Data Analysis with Interactive Task Decomposition},
year = {2024},
isbn = {9798400706288},
publisher = {Association for Computing Machinery},
address = {New York, NY, USA},
url = {https://doi.org/10.1145/3654777.3676345},
doi = {10.1145/3654777.3676345},
booktitle = {Proceedings of the 37th Annual ACM Symposium on User Interface Software and Technology},
articleno = {92},
numpages = {19},
location = {Pittsburgh, PA, USA},
series = {UIST '24}
}

@misc{feng2025cocoa,
      title={Cocoa: Co-Planning and Co-Execution with AI Agents}, 
      author={K. J. Kevin Feng and Kevin Pu and Matt Latzke and Tal August and Pao Siangliulue and Jonathan Bragg and Daniel S. Weld and Amy X. Zhang and Joseph Chee Chang},
      year={2025},
      eprint={2412.10999},
      archivePrefix={arXiv},
      primaryClass={cs.HC},
      url={https://arxiv.org/abs/2412.10999}, 
}

@inproceedings{lee2024designspace,
author = {Lee, Mina and Gero, Katy Ilonka and Chung, John Joon Young and Shum, Simon Buckingham and Raheja, Vipul and Shen, Hua and Venugopalan, Subhashini and Wambsganss, Thiemo and Zhou, David and Alghamdi, Emad A. and August, Tal and Bhat, Avinash and Choksi, Madiha Zahrah and Dutta, Senjuti and Guo, Jin L.C. and Hoque, Md Naimul and Kim, Yewon and Knight, Simon and Neshaei, Seyed Parsa and Shibani, Antonette and Shrivastava, Disha and Shroff, Lila and Sergeyuk, Agnia and Stark, Jessi and Sterman, Sarah and Wang, Sitong and Bosselut, Antoine and Buschek, Daniel and Chang, Joseph Chee and Chen, Sherol and Kreminski, Max and Park, Joonsuk and Pea, Roy and Rho, Eugenia Ha Rim and Shen, Zejiang and Siangliulue, Pao},
title = {A Design Space for Intelligent and Interactive Writing Assistants},
year = {2024},
isbn = {9798400703300},
publisher = {Association for Computing Machinery},
address = {New York, NY, USA},
url = {https://doi.org/10.1145/3613904.3642697},
doi = {10.1145/3613904.3642697},
booktitle = {Proceedings of the 2024 CHI Conference on Human Factors in Computing Systems},
articleno = {1054},
numpages = {35},
location = {Honolulu, HI, USA},
series = {CHI '24}
}

@book{knapp2011sage,
  title={The SAGE handbook of interpersonal communication},
  author={Knapp, Mark L and Daly, John A},
  isbn={9781412974745},
  lccn={2011024470},
  year={2011},
  publisher={Sage Publications}
}

@article{stamp1990construct,
  title={The construct of intent in interpersonal communication},
  author={Stamp, Glen H and Knapp, Mark L},
  journal={Quarterly journal of speech},
  volume={76},
  number={3},
  pages={282--299},
  year={1990},
  publisher={Taylor \& Francis}
}

@inproceedings{kim2024evallm,
author = {Kim, Tae Soo and Lee, Yoonjoo and Shin, Jamin and Kim, Young-Ho and Kim, Juho},
title = {EvalLM: Interactive Evaluation of Large Language Model Prompts on User-Defined Criteria},
year = {2024},
isbn = {9798400703300},
publisher = {Association for Computing Machinery},
address = {New York, NY, USA},
url = {https://doi.org/10.1145/3613904.3642216},
doi = {10.1145/3613904.3642216},
booktitle = {Proceedings of the 2024 CHI Conference on Human Factors in Computing Systems},
articleno = {306},
numpages = {21},
location = {Honolulu, HI, USA},
series = {CHI '24}
}

@inproceedings{kim2023cells,
author = {Kim, Tae Soo and Lee, Yoonjoo and Chang, Minsuk and Kim, Juho},
title = {Cells, Generators, and Lenses: Design Framework for Object-Oriented Interaction with Large Language Models},
year = {2023},
isbn = {9798400701320},
publisher = {Association for Computing Machinery},
address = {New York, NY, USA},
url = {https://doi.org/10.1145/3586183.3606833},
doi = {10.1145/3586183.3606833},
booktitle = {Proceedings of the 36th Annual ACM Symposium on User Interface Software and Technology},
articleno = {4},
numpages = {18},
location = {San Francisco, CA, USA},
series = {UIST '23}
}

@inproceedings{laban2024beyondthechat,
author = {Laban, Philippe and Vig, Jesse and Hearst, Marti and Xiong, Caiming and Wu, Chien-Sheng},
title = {Beyond the Chat: Executable and Verifiable Text-Editing with LLMs},
year = {2024},
isbn = {9798400706288},
publisher = {Association for Computing Machinery},
address = {New York, NY, USA},
url = {https://doi.org/10.1145/3654777.3676419},
doi = {10.1145/3654777.3676419},
booktitle = {Proceedings of the 37th Annual ACM Symposium on User Interface Software and Technology},
articleno = {20},
numpages = {23},
location = {Pittsburgh, PA, USA},
series = {UIST '24}
}

@inproceedings{suh2024Luminate,
author = {Suh, Sangho and Chen, Meng and Min, Bryan and Li, Toby Jia-Jun and Xia, Haijun},
title = {Luminate: Structured Generation and Exploration of Design Space with Large Language Models for Human-AI Co-Creation},
year = {2024},
isbn = {9798400703300},
publisher = {Association for Computing Machinery},
address = {New York, NY, USA},
url = {https://doi.org/10.1145/3613904.3642400},
doi = {10.1145/3613904.3642400},
booktitle = {Proceedings of the 2024 CHI Conference on Human Factors in Computing Systems},
articleno = {644},
numpages = {26},
location = {Honolulu, HI, USA},
series = {CHI '24}
}

@inproceedings{Zamfirescu2023whyjohnny,
author = {Zamfirescu-Pereira, J.D. and Wong, Richmond Y. and Hartmann, Bjoern and Yang, Qian},
title = {Why Johnny Can’t Prompt: How Non-AI Experts Try (and Fail) to Design LLM Prompts},
year = {2023},
isbn = {9781450394215},
publisher = {Association for Computing Machinery},
address = {New York, NY, USA},
url = {https://doi.org/10.1145/3544548.3581388},
doi = {10.1145/3544548.3581388},
booktitle = {Proceedings of the 2023 CHI Conference on Human Factors in Computing Systems},
articleno = {437},
numpages = {21},
location = {Hamburg, Germany},
series = {CHI '23}
}

@book{buxton2010sketching,
  title={Sketching user experiences: getting the design right and the right design},
  author={Buxton, Bill},
  year={2010},
  publisher={Morgan kaufmann}
}

@article{schon1992designing,
title = {Designing as reflective conversation with the materials of a design situation},
journal = {Knowledge-Based Systems},
volume = {5},
number = {1},
pages = {3-14},
year = {1992},
note = {Artificial Intelligence in Design Conference 1991 Special Issue},
issn = {0950-7051},
doi = {https://doi.org/10.1016/0950-7051(92)90020-G},
url = {https://www.sciencedirect.com/science/article/pii/095070519290020G},
author = {D.A. Schön}
}

@inproceedings{Tankelevitch2024metacognitivedemands,
author = {Tankelevitch, Lev and Kewenig, Viktor and Simkute, Auste and Scott, Ava Elizabeth and Sarkar, Advait and Sellen, Abigail and Rintel, Sean},
title = {The Metacognitive Demands and Opportunities of Generative AI},
year = {2024},
isbn = {9798400703300},
publisher = {Association for Computing Machinery},
address = {New York, NY, USA},
url = {https://doi.org/10.1145/3613904.3642902},
doi = {10.1145/3613904.3642902},
booktitle = {Proceedings of the 2024 CHI Conference on Human Factors in Computing Systems},
articleno = {680},
numpages = {24},
location = {Honolulu, HI, USA},
series = {CHI '24}
}

@article{Zhao2024Explainability,
author = {Zhao, Haiyan and Chen, Hanjie and Yang, Fan and Liu, Ninghao and Deng, Huiqi and Cai, Hengyi and Wang, Shuaiqiang and Yin, Dawei and Du, Mengnan},
title = {Explainability for Large Language Models: A Survey},
year = {2024},
issue_date = {April 2024},
publisher = {Association for Computing Machinery},
address = {New York, NY, USA},
volume = {15},
number = {2},
issn = {2157-6904},
url = {https://doi.org/10.1145/3639372},
doi = {10.1145/3639372},
journal = {ACM Trans. Intell. Syst. Technol.},
month = feb,
articleno = {20},
numpages = {38}
}

@incollection{nasatlx,
title = {Development of NASA-TLX (Task Load Index): Results of Empirical and Theoretical Research},
editor = {Peter A. Hancock and Najmedin Meshkati},
series = {Advances in Psychology},
publisher = {North-Holland},
volume = {52},
pages = {139-183},
year = {1988},
booktitle = {Human Mental Workload},
issn = {0166-4115},
doi = {https://doi.org/10.1016/S0166-4115(08)62386-9},
url = {https://www.sciencedirect.com/science/article/pii/S0166411508623869},
author = {Sandra G. Hart and Lowell E. Staveland}
}

@article{shneiderman2007creativity,
  title={Creativity support tools: accelerating discovery and innovation},
  author={Shneiderman, Ben},
  journal={Communications of the ACM},
  volume={50},
  number={12},
  pages={20--32},
  year={2007},
  publisher={ACM New York, NY, USA}
}

@inproceedings{terry2002recognizing,
  title={Recognizing creative needs in user interface design},
  author={Terry, Michael and Mynatt, Elizabeth D},
  booktitle={Proceedings of the 4th Conference on Creativity \& Cognition},
  pages={38--44},
  year={2002}
}

@inproceedings{brade2023promptify,
  title={Promptify: Text-to-image generation through interactive prompt exploration with large language models},
  author={Brade, Stephen and Wang, Bryan and Sousa, Mauricio and Oore, Sageev and Grossman, Tovi},
  booktitle={Proceedings of the 36th Annual ACM Symposium on User Interface Software and Technology},
  pages={1--14},
  year={2023}
}

@article{yeh2024ghostwriter,
  title={Ghostwriter: Augmenting collaborative human-ai writing experiences through personalization and agency},
  author={Yeh, Catherine and Ramos, Gonzalo and Ng, Rachel and Huntington, Andy and Banks, Richard},
  journal={arXiv preprint arXiv:2402.08855},
  year={2024}
}

@inproceedings{arawjo2024chainforge,
  title={Chainforge: A visual toolkit for prompt engineering and llm hypothesis testing},
  author={Arawjo, Ian and Swoopes, Chelse and Vaithilingam, Priyan and Wattenberg, Martin and Glassman, Elena L},
  booktitle={Proceedings of the 2024 CHI Conference on Human Factors in Computing Systems},
  pages={1--18},
  year={2024}
}

@inproceedings{zhang2023visar,
  title={Visar: A human-ai argumentative writing assistant with visual programming and rapid draft prototyping},
  author={Zhang, Zheng and Gao, Jie and Dhaliwal, Ranjodh Singh and Li, Toby Jia-Jun},
  booktitle={Proceedings of the 36th annual ACM symposium on user interface software and technology},
  pages={1--30},
  year={2023}
}

@inproceedings{Gu2024SupportingSOA,
author = {Gero, Katy Ilonka and Swoopes, Chelse and Gu, Ziwei and Kummerfeld, Jonathan K. and Glassman, Elena L.},
title = {Supporting Sensemaking of Large Language Model Outputs at Scale},
year = {2024},
isbn = {9798400703300},
publisher = {Association for Computing Machinery},
address = {New York, NY, USA},
url = {https://doi.org/10.1145/3613904.3642139},
doi = {10.1145/3613904.3642139},
booktitle = {Proceedings of the 2024 CHI Conference on Human Factors in Computing Systems},
articleno = {838},
numpages = {21},
location = {Honolulu, HI, USA},
series = {CHI '24}
}

@article{Cheng2020ConversationalSPA,
  title={Conversational semantic parsing for dialog state tracking},
  author={Cheng, Jianpeng and Agrawal, Devang and Alonso, H{\'e}ctor Mart{\'\i}nez and Bhargava, Shruti and Driesen, Joris and Flego, Federico and Ghosh, Shaona and Kaplan, Dain and Kartsaklis, Dimitri and Li, Lin and others},
  journal={arXiv preprint arXiv:2010.12770},
  year={2020}
}

@article{bao2023can,
  title={Can Foundation Models Watch, Talk and Guide You Step by Step to Make a Cake?},
  author={Bao, Yuwei and Yu, Keunwoo Peter and Zhang, Yichi and Storks, Shane and Bar-Yossef, Itamar and De La Iglesia, Alexander and Su, Megan and Zheng, Xiao Lin and Chai, Joyce},
  journal={arXiv preprint arXiv:2311.00738},
  year={2023}
}

@article{shin2023planfitting,
  title={PlanFitting: Tailoring Personalized Exercise Plans with Large Language Models},
  author={Shin, Donghoon and Hsieh, Gary and Kim, Young-Ho},
  journal={arXiv preprint arXiv:2309.12555},
  year={2023}
}

@inproceedings{son2024genquery,
author = {Son, Kihoon and Choi, DaEun and Kim, Tae Soo and Kim, Young-Ho and Kim, Juho},
title = {GenQuery: Supporting Expressive Visual Search with Generative Models},
year = {2024},
isbn = {9798400703300},
publisher = {Association for Computing Machinery},
address = {New York, NY, USA},
url = {https://doi.org/10.1145/3613904.3642847},
doi = {10.1145/3613904.3642847},
booktitle = {Proceedings of the 2024 CHI Conference on Human Factors in Computing Systems},
articleno = {180},
numpages = {19},
location = {Honolulu, HI, USA},
series = {CHI '24}
}

@inproceedings{xie2024waitgpt,
author = {Xie, Liwenhan and Zheng, Chengbo and Xia, Haijun and Qu, Huamin and Zhu-Tian, Chen},
title = {WaitGPT: Monitoring and Steering Conversational LLM Agent in Data Analysis with On-the-Fly Code Visualization},
year = {2024},
isbn = {9798400706288},
publisher = {Association for Computing Machinery},
address = {New York, NY, USA},
url = {https://doi.org/10.1145/3654777.3676374},
doi = {10.1145/3654777.3676374},
booktitle = {Proceedings of the 37th Annual ACM Symposium on User Interface Software and Technology},
articleno = {119},
numpages = {14},
location = {Pittsburgh, PA, USA},
series = {UIST '24}
}

@inproceedings{Khurana2024whyandwhen,
author = {Khurana, Anjali and Subramonyam, Hariharan and Chilana, Parmit K},
title = {Why and When LLM-Based Assistants Can Go Wrong: Investigating the Effectiveness of Prompt-Based Interactions for Software Help-Seeking},
year = {2024},
isbn = {9798400705083},
publisher = {Association for Computing Machinery},
address = {New York, NY, USA},
url = {https://doi.org/10.1145/3640543.3645200},
doi = {10.1145/3640543.3645200},
booktitle = {Proceedings of the 29th International Conference on Intelligent User Interfaces},
pages = {288–303},
numpages = {16},
location = {Greenville, SC, USA},
series = {IUI '24}
}

@inproceedings{goldi2024intelligentwriters,
author = {G\"{o}ldi, Andreas and Wambsganss, Thiemo and Neshaei, Seyed Parsa and Rietsche, Roman},
title = {Intelligent Support Engages Writers Through Relevant Cognitive Processes},
year = {2024},
isbn = {9798400703300},
publisher = {Association for Computing Machinery},
address = {New York, NY, USA},
url = {https://doi.org/10.1145/3613904.3642549},
doi = {10.1145/3613904.3642549},
booktitle = {Proceedings of the 2024 CHI Conference on Human Factors in Computing Systems},
articleno = {1047},
numpages = {12},
location = {Honolulu, HI, USA},
series = {CHI '24}
}

@inproceedings{zamfirescu2023herding,
author = {Zamfirescu-Pereira, J.D. and Wei, Heather and Xiao, Amy and Gu, Kitty and Jung, Grace and Lee, Matthew G and Hartmann, Bjoern and Yang, Qian},
title = {Herding AI Cats: Lessons from Designing a Chatbot by Prompting GPT-3},
year = {2023},
isbn = {9781450398930},
publisher = {Association for Computing Machinery},
address = {New York, NY, USA},
url = {https://doi.org/10.1145/3563657.3596138},
doi = {10.1145/3563657.3596138},
booktitle = {Proceedings of the 2023 ACM Designing Interactive Systems Conference},
pages = {2206–2220},
numpages = {15},
location = {Pittsburgh, PA, USA},
series = {DIS '23}
}

@inproceedings{liang2024largescale,
author = {Liang, Jenny T. and Yang, Chenyang and Myers, Brad A.},
title = {A Large-Scale Survey on the Usability of AI Programming Assistants: Successes and Challenges},
year = {2024},
isbn = {9798400702174},
publisher = {Association for Computing Machinery},
address = {New York, NY, USA},
url = {https://doi.org/10.1145/3597503.3608128},
doi = {10.1145/3597503.3608128},
booktitle = {Proceedings of the IEEE/ACM 46th International Conference on Software Engineering},
articleno = {52},
numpages = {13},
location = {Lisbon, Portugal},
series = {ICSE '24}
}

@incollection{gussow2023language,
  title={Language production under message uncertainty: When, how, and why we speak before we think},
  author={Gussow, Arella E},
  booktitle={Psychology of Learning and Motivation},
  volume={78},
  pages={83--117},
  year={2023},
  publisher={Elsevier}
}

@article{Kraljic2024Collaborating,
author = {Kraljic, Tanya and Lahav, Michal},
title = {From Prompt Engineering to Collaborating: A Human-Centered Approach to AI Interfaces},
year = {2024},
issue_date = {May - June 2024},
publisher = {Association for Computing Machinery},
address = {New York, NY, USA},
volume = {31},
number = {3},
issn = {1072-5520},
url = {https://doi.org/10.1145/3652622},
doi = {10.1145/3652622},
journal = {Interactions},
month = may,
pages = {30–35},
numpages = {6}
}

@article{page2021prisma,
  title={The PRISMA 2020 statement: an updated guideline for reporting systematic reviews},
  author={Page, Matthew J and McKenzie, Joanne E and Bossuyt, Patrick M and Boutron, Isabelle and Hoffmann, Tammy C and Mulrow, Cynthia D and Shamseer, Larissa and Tetzlaff, Jennifer M and Akl, Elie A and Brennan, Sue E and others},
  journal={bmj},
  volume={372},
  year={2021},
  publisher={British Medical Journal Publishing Group}
}

@inproceedings{Liu2024WeNeedStructured,
author = {Liu, Michael Xieyang and Liu, Frederick and Fiannaca, Alexander J. and Koo, Terry and Dixon, Lucas and Terry, Michael and Cai, Carrie J.},
title = {"We Need Structured Output": Towards User-centered Constraints on Large Language Model Output},
year = {2024},
isbn = {9798400703317},
publisher = {Association for Computing Machinery},
address = {New York, NY, USA},
url = {https://doi.org/10.1145/3613905.3650756},
doi = {10.1145/3613905.3650756},
booktitle = {Extended Abstracts of the CHI Conference on Human Factors in Computing Systems},
articleno = {10},
numpages = {9},
location = {Honolulu, HI, USA},
series = {CHI EA '24}
}

@inproceedings{liu2023whatitwantsmetosay,
author = {Liu, Michael Xieyang and Sarkar, Advait and Negreanu, Carina and Zorn, Benjamin and Williams, Jack and Toronto, Neil and Gordon, Andrew D.},
title = {“What It Wants Me To Say”: Bridging the Abstraction Gap Between End-User Programmers and Code-Generating Large Language Models},
year = {2023},
isbn = {9781450394215},
publisher = {Association for Computing Machinery},
address = {New York, NY, USA},
url = {https://doi.org/10.1145/3544548.3580817},
doi = {10.1145/3544548.3580817},
booktitle = {Proceedings of the 2023 CHI Conference on Human Factors in Computing Systems},
articleno = {598},
numpages = {31},
location = {Hamburg, Germany},
series = {CHI '23}
}

@inproceedings{reza2024abscribe,
author = {Reza, Mohi and Laundry, Nathan M and Musabirov, Ilya and Dushniku, Peter and Yu, Zhi Yuan “Michael” and Mittal, Kashish and Grossman, Tovi and Liut, Michael and Kuzminykh, Anastasia and Williams, Joseph Jay},
title = {ABScribe: Rapid Exploration \& Organization of Multiple Writing Variations in Human-AI Co-Writing Tasks using Large Language Models},
year = {2024},
isbn = {9798400703300},
publisher = {Association for Computing Machinery},
address = {New York, NY, USA},
url = {https://doi.org/10.1145/3613904.3641899},
doi = {10.1145/3613904.3641899},
booktitle = {Proceedings of the 2024 CHI Conference on Human Factors in Computing Systems},
articleno = {1042},
numpages = {18},
location = {Honolulu, HI, USA},
series = {CHI '24}
}

@inproceedings{riche2025ai-instruments,
author = {Riche, Nathalie and Offenwanger, Anna and Gmeiner, Frederic and Brown, David and Romat, Hugo and Pahud, Michel and Marquardt, Nicolai and Inkpen, Kori and Hinckley, Ken},
title = {AI-Instruments: Embodying Prompts as Instruments to Abstract \& Reflect Graphical Interface Commands as General-Purpose Tools},
year = {2025},
isbn = {9798400713941},
publisher = {Association for Computing Machinery},
address = {New York, NY, USA},
url = {https://doi.org/10.1145/3706598.3714259},
doi = {10.1145/3706598.3714259},
booktitle = {Proceedings of the 2025 CHI Conference on Human Factors in Computing Systems},
articleno = {1104},
numpages = {18},
location = {
},
series = {CHI '25}
}

@inproceedings{Gmeiner2024Evidence-based,
author = {Gmeiner, Frederic and Conlin, Jamie Lynn and Tang, Eric Handa and Martelaro, Nikolas and Holstein, Kenneth},
title = {An Evidence-based Workflow for Studying and Designing Learning Supports for Human-AI Co-creation},
year = {2024},
isbn = {9798400703317},
publisher = {Association for Computing Machinery},
address = {New York, NY, USA},
url = {https://doi.org/10.1145/3613905.3650763},
doi = {10.1145/3613905.3650763},
booktitle = {Extended Abstracts of the CHI Conference on Human Factors in Computing Systems},
articleno = {42},
numpages = {15},
location = {Honolulu, HI, USA},
series = {CHI EA '24}
}

@inproceedings{kim2025applying,
author = {Kim, Yoonsu and Chin, Brandon and Son, Kihoon and Kim, Seoyoung and Kim, Juho},
title = {Applying the Gricean Maxims to a Human-LLM Interaction Cycle: Design Insights from a Participatory Approach},
year = {2025},
isbn = {9798400713958},
publisher = {Association for Computing Machinery},
address = {New York, NY, USA},
url = {https://doi.org/10.1145/3706599.3719759},
doi = {10.1145/3706599.3719759},
booktitle = {Proceedings of the Extended Abstracts of the CHI Conference on Human Factors in Computing Systems},
articleno = {72},
numpages = {8},
location = {
},
series = {CHI EA '25}
}

@inproceedings{chen2025dango,
author = {Chen, Wei-Hao and Tong, Weixi and Case, Amanda, Ph.D. and Zhang, Tianyi},
title = {Dango: A Mixed-Initiative Data Wrangling System using Large Language Model},
year = {2025},
isbn = {9798400713941},
publisher = {Association for Computing Machinery},
address = {New York, NY, USA},
url = {https://doi.org/10.1145/3706598.3714135},
doi = {10.1145/3706598.3714135},
booktitle = {Proceedings of the 2025 CHI Conference on Human Factors in Computing Systems},
articleno = {389},
numpages = {28},
location = {
},
series = {CHI '25}
}

@inproceedings{kim2024DiaryMate,
author = {Kim, Taewan and Shin, Donghoon and Kim, Young-Ho and Hong, Hwajung},
title = {DiaryMate: Understanding User Perceptions and Experience in Human-AI Collaboration for Personal Journaling},
year = {2024},
isbn = {9798400703300},
publisher = {Association for Computing Machinery},
address = {New York, NY, USA},
url = {https://doi.org/10.1145/3613904.3642693},
doi = {10.1145/3613904.3642693},
booktitle = {Proceedings of the 2024 CHI Conference on Human Factors in Computing Systems},
articleno = {1046},
numpages = {15},
location = {Honolulu, HI, USA},
series = {CHI '24}
}

@inproceedings{masson2024directgpt,
author = {Masson, Damien and Malacria, Sylvain and Casiez, G\'{e}ry and Vogel, Daniel},
title = {DirectGPT: A Direct Manipulation Interface to Interact with Large Language Models},
year = {2024},
isbn = {9798400703300},
publisher = {Association for Computing Machinery},
address = {New York, NY, USA},
url = {https://doi.org/10.1145/3613904.3642462},
doi = {10.1145/3613904.3642462},
booktitle = {Proceedings of the 2024 CHI Conference on Human Factors in Computing Systems},
articleno = {975},
numpages = {16},
location = {Honolulu, HI, USA},
series = {CHI '24}
}

@inproceedings{choi2025Expandora,
author = {Choi, DaEun and Son, Kihoon and Jung, HyunJoon and Kim, Juho},
title = {Expandora: Broadening Design Exploration with Text-to-Image Model},
year = {2025},
isbn = {9798400713958},
publisher = {Association for Computing Machinery},
address = {New York, NY, USA},
url = {https://doi.org/10.1145/3706599.3720189},
doi = {10.1145/3706599.3720189},
booktitle = {Proceedings of the Extended Abstracts of the CHI Conference on Human Factors in Computing Systems},
articleno = {232},
numpages = {10},
location = {
},
series = {CHI EA '25}
}

@inproceedings{wang2025harmonycut,
author = {Wang, Huanchen and Qiu, Tianrun and Li, Jiaping and Lu, Zhicong and Ma, Yuxin},
title = {HarmonyCut: Supporting Creative Chinese Paper-cutting Design with Form and Connotation Harmony},
year = {2025},
isbn = {9798400713941},
publisher = {Association for Computing Machinery},
address = {New York, NY, USA},
url = {https://doi.org/10.1145/3706598.3714159},
doi = {10.1145/3706598.3714159},
booktitle = {Proceedings of the 2025 CHI Conference on Human Factors in Computing Systems},
articleno = {661},
numpages = {22},
location = {
},
series = {CHI '25}
}

@inproceedings{gmeiner2025intenttagging,
author = {Gmeiner, Frederic and Marquardt, Nicolai and Bentley, Michael and Romat, Hugo and Pahud, Michel and Brown, David and Roseway, Asta and Martelaro, Nikolas and Holstein, Kenneth and Hinckley, Ken and Riche, Nathalie},
title = {Intent Tagging: Exploring Micro-Prompting Interactions for Supporting Granular Human-GenAI Co-Creation Workflows},
year = {2025},
isbn = {9798400713941},
publisher = {Association for Computing Machinery},
address = {New York, NY, USA},
url = {https://doi.org/10.1145/3706598.3713861},
doi = {10.1145/3706598.3713861},
booktitle = {Proceedings of the 2025 CHI Conference on Human Factors in Computing Systems},
articleno = {531},
numpages = {31},
location = {
},
series = {CHI '25}
}

@inproceedings{Wang2025IntentPrism,
author = {Wang, Zehuan and Xiao, Jiaqi and Sun, Jingwei and Liu, Can},
title = {IntentPrism: Human-AI Intent Manifestation for Web Information Foraging},
year = {2025},
isbn = {9798400713958},
publisher = {Association for Computing Machinery},
address = {New York, NY, USA},
url = {https://doi.org/10.1145/3706599.3719744},
doi = {10.1145/3706599.3719744},
booktitle = {Proceedings of the Extended Abstracts of the CHI Conference on Human Factors in Computing Systems},
articleno = {345},
numpages = {11},
location = {
},
series = {CHI EA '25}
}

@inproceedings{Shanmugarasa2025Privacy,
author = {Shanmugarasa, Yashothara and Pan, Shidong and Ding, Ming and Zhao, Dehai and Rakotoarivelo, Thierry},
title = {Privacy Meets Explainability: Managing Confidential Data and Transparency Policies in LLM-Empowered Science},
year = {2025},
isbn = {9798400713958},
publisher = {Association for Computing Machinery},
address = {New York, NY, USA},
url = {https://doi.org/10.1145/3706599.3720099},
doi = {10.1145/3706599.3720099},
booktitle = {Proceedings of the Extended Abstracts of the CHI Conference on Human Factors in Computing Systems},
articleno = {448},
numpages = {8},
location = {
},
series = {CHI EA '25}
}

@inproceedings{Wang2024PromptCharm,
author = {Wang, Zhijie and Huang, Yuheng and Song, Da and Ma, Lei and Zhang, Tianyi},
title = {PromptCharm: Text-to-Image Generation through Multi-modal Prompting and Refinement},
year = {2024},
isbn = {9798400703300},
publisher = {Association for Computing Machinery},
address = {New York, NY, USA},
url = {https://doi.org/10.1145/3613904.3642803},
doi = {10.1145/3613904.3642803},
booktitle = {Proceedings of the 2024 CHI Conference on Human Factors in Computing Systems},
articleno = {185},
numpages = {21},
location = {Honolulu, HI, USA},
series = {CHI '24}
}

@inproceedings{Siddiqui2025ScriptShift,
author = {Siddiqui, Momin N and Pea, Roy D and Subramonyam, Hari},
title = {Script\&Shift: A Layered Interface Paradigm for Integrating Content Development and Rhetorical Strategy with LLM Writing Assistants},
year = {2025},
isbn = {9798400713941},
publisher = {Association for Computing Machinery},
address = {New York, NY, USA},
url = {https://doi.org/10.1145/3706598.3714119},
doi = {10.1145/3706598.3714119},
booktitle = {Proceedings of the 2025 CHI Conference on Human Factors in Computing Systems},
articleno = {532},
numpages = {19},
location = {
},
series = {CHI '25}
}

@inproceedings{Kim2025ShoeGenAI,
author = {Kim, Hui-Jun and Kim, Jeongho and Jeong, Sohyun and Lee, Minbong and Choo, Jaegul and Kim, Sung-Hee},
title = {ShoeGenAI: A Creativity Support Tool for High-Feasible Shoe Product Design},
year = {2025},
isbn = {9798400713958},
publisher = {Association for Computing Machinery},
address = {New York, NY, USA},
url = {https://doi.org/10.1145/3706599.3721204},
doi = {10.1145/3706599.3721204},
booktitle = {Proceedings of the Extended Abstracts of the CHI Conference on Human Factors in Computing Systems},
articleno = {478},
numpages = {11},
location = {
},
series = {CHI EA '25}
}

@inproceedings{lee2025veriplan,
author = {Lee, Christine P. and Porfirio, David and Wang, Xinyu Jessica and Zhao, Kevin Chenkai and Mutlu, Bilge},
title = {VeriPlan: Integrating Formal Verification and LLMs into End-User Planning},
year = {2025},
isbn = {9798400713941},
publisher = {Association for Computing Machinery},
address = {New York, NY, USA},
url = {https://doi.org/10.1145/3706598.3714113},
doi = {10.1145/3706598.3714113},
booktitle = {Proceedings of the 2025 CHI Conference on Human Factors in Computing Systems},
articleno = {247},
numpages = {19},
location = {
},
series = {CHI '25}
}

@inproceedings{drosos2024dynamicpromptmiddlewarecontextual,
author = {Drosos, Ian and Williams, Jack and Sarkar, Advait and Wilson, Nicholas and Rintel, Sean and Panda, Payod},
title = {Dynamic Prompt Middleware: Contextual Prompt Refinement Controls for Comprehension Tasks},
year = {2025},
isbn = {9798400713842},
publisher = {Association for Computing Machinery},
address = {New York, NY, USA},
url = {https://doi.org/10.1145/3729176.3729203},
doi = {10.1145/3729176.3729203},
booktitle = {Proceedings of the 4th Annual Symposium on Human-Computer Interaction for Work},
articleno = {24},
numpages = {23},
location = {
},
series = {CHIWORK '25}
}

@inproceedings{Lee2024Closer,
author = {Lee, Cassandra and Mindel, Jessica R},
title = {Closer and Closer Worlds: Using LLMs to Surface Personal Stories in World-building Conversation Games},
year = {2024},
isbn = {9798400706325},
publisher = {Association for Computing Machinery},
address = {New York, NY, USA},
url = {https://doi.org/10.1145/3656156.3665430},
doi = {10.1145/3656156.3665430},
booktitle = {Companion Publication of the 2024 ACM Designing Interactive Systems Conference},
pages = {289–293},
numpages = {5},
location = {IT University of Copenhagen, Denmark},
series = {DIS '24 Companion}
}

@inproceedings{Lim2024Co-Creating,
author = {Lim, Hyunseung and Cho, Ji Yong and Kim, Taewan and Park, Jeongeon and Shin, Hyungyu and Choi, Seulgi and Park, Sunghyun and Lee, Kyungjae and Kim, Juho and Lee, Moontae and Hong, Hwajung},
title = {Co-Creating Question-and-Answer Style Articles with Large Language Models for Research Promotion},
year = {2024},
isbn = {9798400705830},
publisher = {Association for Computing Machinery},
address = {New York, NY, USA},
url = {https://doi.org/10.1145/3643834.3660705},
doi = {10.1145/3643834.3660705},
booktitle = {Proceedings of the 2024 ACM Designing Interactive Systems Conference},
pages = {975–994},
numpages = {20},
location = {Copenhagen, Denmark},
series = {DIS '24}
}

@inproceedings{Peng2024DesignPrompt,
author = {Peng, Xiaohan and Koch, Janin and Mackay, Wendy E.},
title = {DesignPrompt: Using Multimodal Interaction for Design Exploration with Generative AI},
year = {2024},
isbn = {9798400705830},
publisher = {Association for Computing Machinery},
address = {New York, NY, USA},
url = {https://doi.org/10.1145/3643834.3661588},
doi = {10.1145/3643834.3661588},
booktitle = {Proceedings of the 2024 ACM Designing Interactive Systems Conference},
pages = {804–818},
numpages = {15},
location = {Copenhagen, Denmark},
series = {DIS '24}
}

@inproceedings{Zhou2024GlassMail,
author = {Zhou, Chen and Yan, Zihan and Ram, Ashwin and Gu, Yue and Xiang, Yan and Liu, Can and Huang, Yun and Ooi, Wei Tsang and Zhao, Shengdong},
title = {GlassMail: Towards Personalised Wearable Assistant for On-the-Go Email Creation on Smart Glasses},
year = {2024},
isbn = {9798400705830},
publisher = {Association for Computing Machinery},
address = {New York, NY, USA},
url = {https://doi.org/10.1145/3643834.3660683},
doi = {10.1145/3643834.3660683},
booktitle = {Proceedings of the 2024 ACM Designing Interactive Systems Conference},
pages = {372–390},
numpages = {19},
location = {Copenhagen, Denmark},
series = {DIS '24}
}

@inproceedings{Tilekbay2024ExpressEdit,
author = {Tilekbay, Bekzat and Yang, Saelyne and Lewkowicz, Michal Adam and Suryapranata, Alex and Kim, Juho},
title = {ExpressEdit: Video Editing with Natural Language and Sketching},
year = {2024},
isbn = {9798400705083},
publisher = {Association for Computing Machinery},
address = {New York, NY, USA},
url = {https://doi.org/10.1145/3640543.3645164},
doi = {10.1145/3640543.3645164},
booktitle = {Proceedings of the 29th International Conference on Intelligent User Interfaces},
pages = {515–536},
numpages = {22},
location = {Greenville, SC, USA},
series = {IUI '24}
}

@inproceedings{Chen2024AutoSpark,
author = {Chen, Liuqing and Jing, Qianzhi and Tsang, Yixin and Wang, Qianyi and Liu, Ruocong and Xia, Duowei and Zhou, Yunzhan and Sun, Lingyun},
title = {AutoSpark: Supporting Automobile Appearance Design Ideation with Kansei Engineering and Generative AI},
year = {2024},
isbn = {9798400706288},
publisher = {Association for Computing Machinery},
address = {New York, NY, USA},
url = {https://doi.org/10.1145/3654777.3676337},
doi = {10.1145/3654777.3676337},
booktitle = {Proceedings of the 37th Annual ACM Symposium on User Interface Software and Technology},
articleno = {108},
numpages = {19},
location = {Pittsburgh, PA, USA},
series = {UIST '24}
}

@inproceedings{Chung2024Patchview,
author = {Chung, John Joon Young and Kreminski, Max},
title = {Patchview: LLM-powered Worldbuilding with Generative Dust and Magnet Visualization},
year = {2024},
isbn = {9798400706288},
publisher = {Association for Computing Machinery},
address = {New York, NY, USA},
url = {https://doi.org/10.1145/3654777.3676352},
doi = {10.1145/3654777.3676352},
booktitle = {Proceedings of the 37th Annual ACM Symposium on User Interface Software and Technology},
articleno = {77},
numpages = {19},
location = {Pittsburgh, PA, USA},
series = {UIST '24}
}

@inproceedings{Zhang2024ProtoDreamer,
author = {Zhang, Hongbo and Chen, Pei and Xie, Xuelong and Lin, Chaoyi and Liu, Lianyan and Li, Zhuoshu and You, Weitao and Sun, Lingyun},
title = {ProtoDreamer: A Mixed-prototype Tool Combining Physical Model and Generative AI to Support Conceptual Design},
year = {2024},
isbn = {9798400706288},
publisher = {Association for Computing Machinery},
address = {New York, NY, USA},
url = {https://doi.org/10.1145/3654777.3676399},
doi = {10.1145/3654777.3676399},
booktitle = {Proceedings of the 37th Annual ACM Symposium on User Interface Software and Technology},
articleno = {97},
numpages = {18},
location = {Pittsburgh, PA, USA},
series = {UIST '24}
}

@article{flower1981cognitive,
  title={A cognitive process theory of writing},
  author={Flower, Linda and Hayes, John R},
  journal={College Composition \& Communication},
  volume={32},
  number={4},
  pages={365--387},
  year={1981},
  publisher={NCTE}
}

@inproceedings{Vaithilingam2024Imagining,
author = {Vaithilingam, Priyan and Arawjo, Ian and Glassman, Elena L.},
title = {Imagining a Future of Designing with AI: Dynamic Grounding, Constructive Negotiation, and Sustainable Motivation},
year = {2024},
isbn = {9798400705830},
publisher = {Association for Computing Machinery},
address = {New York, NY, USA},
url = {https://doi.org/10.1145/3643834.3661525},
doi = {10.1145/3643834.3661525},
booktitle = {Proceedings of the 2024 ACM Designing Interactive Systems Conference},
pages = {289–300},
numpages = {12},
location = {Copenhagen, Denmark},
series = {DIS '24}
}

@inproceedings{Vaithilingam2025Semantic,
author = {Vaithilingam, Priyan and Kim, Munyeong and Acosta-Parenteau, Frida-Cecilia and Lee, Daniel and Mhedhbi, Amine and Glassman, Elena L. and Arawjo, Ian},
title = {Semantic Commit: Helping Users Update Intent Specifications for AI Memory at Scale},
year = {2025},
isbn = {9798400720376},
publisher = {Association for Computing Machinery},
address = {New York, NY, USA},
url = {https://doi.org/10.1145/3746059.3747778},
doi = {10.1145/3746059.3747778},
booktitle = {Proceedings of the 38th Annual ACM Symposium on User Interface Software and Technology},
articleno = {137},
numpages = {18},
location = {
},
series = {UIST '25}
}

@inproceedings{Ding2025Towards,
author = {Ding, Zijian and Li, Fenghai and Yu, Haofei and Chan, Joel},
title = {Towards Direct Intent Manipulation: Drag-Based Research Ideation, Evaluation and Evolution},
year = {2025},
isbn = {9798400720369},
publisher = {Association for Computing Machinery},
address = {New York, NY, USA},
url = {https://doi.org/10.1145/3746058.3758453},
doi = {10.1145/3746058.3758453},
booktitle = {Adjunct Proceedings of the 38th Annual ACM Symposium on User Interface Software and Technology},
articleno = {171},
numpages = {3},
location = {
},
series = {UIST Adjunct '25}
}

@inproceedings{Zhang2025NeuroSync,
author = {Zhang, Wenshuo and Shen, Leixian and Xu, Shuchang and Wang, Jindu and Zhao, Jian and Qu, Huamin and Yuan, Lin-Ping},
title = {NeuroSync: Intent-Aware Code-Based Problem Solving via Direct LLM Understanding Modification},
year = {2025},
isbn = {9798400720376},
publisher = {Association for Computing Machinery},
address = {New York, NY, USA},
url = {https://doi.org/10.1145/3746059.3747668},
doi = {10.1145/3746059.3747668},
booktitle = {Proceedings of the 38th Annual ACM Symposium on User Interface Software and Technology},
articleno = {30},
numpages = {19},
location = {
},
series = {UIST '25}
}

@inproceedings{Huang2025SketchGPT,
author = {Huang, Zeyuan and Gao, Cangjun and Shan, Yaxian and Hu, Haoxiang and Li, Qingkun and Deng, Xiaoming and Ma, Cuixia and Lai, Yu-Kun and Liu, Yong-Jin and Tian, Feng and Dai, Guozhong and Wang, Hongan},
title = {SketchGPT: A Sketch-based Multimodal Interface for Application-Agnostic LLM Interaction},
year = {2025},
isbn = {9798400720376},
publisher = {Association for Computing Machinery},
address = {New York, NY, USA},
url = {https://doi.org/10.1145/3746059.3747598},
doi = {10.1145/3746059.3747598},
booktitle = {Proceedings of the 38th Annual ACM Symposium on User Interface Software and Technology},
articleno = {157},
numpages = {18},
location = {
},
series = {UIST '25}
}

@inproceedings{Yen2024CoLadder,
author = {Yen, Ryan and Zhu, Jiawen Stefanie and Suh, Sangho and Xia, Haijun and Zhao, Jian},
title = {CoLadder: Manipulating Code Generation via Multi-Level Blocks},
year = {2024},
isbn = {9798400706288},
publisher = {Association for Computing Machinery},
address = {New York, NY, USA},
url = {https://doi.org/10.1145/3654777.3676357},
doi = {10.1145/3654777.3676357},
booktitle = {Proceedings of the 37th Annual ACM Symposium on User Interface Software and Technology},
articleno = {11},
numpages = {20},
location = {Pittsburgh, PA, USA},
series = {UIST '24}
}

@inproceedings{Yen2024Memolet,
author = {Yen, Ryan and Zhao, Jian},
title = {Memolet: Reifying the Reuse of User-AI Conversational Memories},
year = {2024},
isbn = {9798400706288},
publisher = {Association for Computing Machinery},
address = {New York, NY, USA},
url = {https://doi.org/10.1145/3654777.3676388},
doi = {10.1145/3654777.3676388},
booktitle = {Proceedings of the 37th Annual ACM Symposium on User Interface Software and Technology},
articleno = {58},
numpages = {22},
location = {Pittsburgh, PA, USA},
series = {UIST '24}
}

@inproceedings{Leung2025SQUIRE,
author = {Leung, Alan and Cheng, Ruijia and Wu, Jason and Nichols, Jeffrey and Barik, Titus},
title = {SQUIRE: Interactive UI Authoring via Slot QUery Intermediate REpresentations},
year = {2025},
isbn = {9798400720376},
publisher = {Association for Computing Machinery},
address = {New York, NY, USA},
url = {https://doi.org/10.1145/3746059.3747672},
doi = {10.1145/3746059.3747672},
booktitle = {Proceedings of the 38th Annual ACM Symposium on User Interface Software and Technology},
articleno = {199},
numpages = {17},
location = {
},
series = {UIST '25}
}

@inproceedings{Sun2025Creative,
author = {Sun, Zhida and Zhang, Zhenyao and Zhang, Yue and Lu, Min and Lischinski, Dani and Cohen-Or, Daniel and Huang, Hui},
title = {Creative Blends of Visual Concepts},
year = {2025},
isbn = {9798400713941},
publisher = {Association for Computing Machinery},
address = {New York, NY, USA},
url = {https://doi.org/10.1145/3706598.3713683},
doi = {10.1145/3706598.3713683},
booktitle = {Proceedings of the 2025 CHI Conference on Human Factors in Computing Systems},
articleno = {542},
numpages = {17},
location = {
},
series = {CHI '25}
}

@inproceedings{Zhang2025ChainBuddy,
author = {Zhang, Jingyue and Arawjo, Ian},
title = {ChainBuddy: An AI-assisted Agent System for Generating LLM Pipelines},
year = {2025},
isbn = {9798400713941},
publisher = {Association for Computing Machinery},
address = {New York, NY, USA},
url = {https://doi.org/10.1145/3706598.3714085},
doi = {10.1145/3706598.3714085},
booktitle = {Proceedings of the 2025 CHI Conference on Human Factors in Computing Systems},
articleno = {241},
numpages = {21},
location = {
},
series = {CHI '25}
}

@inproceedings{Marquardt2025ImaginationVellum,
author = {Marquardt, Nicolai and Roseway, Asta and Romat, Hugo and Panda, Payod and Pahud, Michel and Ramos, Gonzalo and Drucker, Steven M. and Wilson, Andrew D. and Hinckley, Ken and Riche, Nathalie},
title = {ImaginationVellum: Generative-AI Ideation Canvas with Spatial Prompts, Generative Strokes, and Ideation History},
year = {2025},
isbn = {9798400720376},
publisher = {Association for Computing Machinery},
address = {New York, NY, USA},
url = {https://doi.org/10.1145/3746059.3747631},
doi = {10.1145/3746059.3747631},
booktitle = {Proceedings of the 38th Annual ACM Symposium on User Interface Software and Technology},
articleno = {159},
numpages = {19},
location = {
},
series = {UIST '25}
}

@inproceedings{Choi2025IdeaBlocks,
author = {Choi, DaEun and Son, Kihoon and Yu, Jaesang and Jung, HyunJoon and Kim, Juho},
title = {IdeaBlocks: Expressing and Reusing Exploratory Intents for Design Exploration with Generative AI},
year = {2025},
isbn = {9798400720369},
publisher = {Association for Computing Machinery},
address = {New York, NY, USA},
url = {https://doi.org/10.1145/3746058.3759001},
doi = {10.1145/3746058.3759001},
booktitle = {Adjunct Proceedings of the 38th Annual ACM Symposium on User Interface Software and Technology},
articleno = {43},
numpages = {4},
location = {
},
series = {UIST Adjunct '25}
}

@inproceedings{Lee2025ThematicPlane,
author = {Lee, Daniel and Sharma, Nikhil and Shin, Donghoon and Choi, DaEun and Sharma, Harsh and Kim, Jeonghwan and Ji, Heng},
title = {ThematicPlane: Bridging Tacit User Intent and Latent Spaces for Image Generation},
year = {2025},
isbn = {9798400720369},
publisher = {Association for Computing Machinery},
address = {New York, NY, USA},
url = {https://doi.org/10.1145/3746058.3758376},
doi = {10.1145/3746058.3758376},
booktitle = {Adjunct Proceedings of the 38th Annual ACM Symposium on User Interface Software and Technology},
articleno = {120},
numpages = {3},
location = {
},
series = {UIST Adjunct '25}
}

@inproceedings{boehner2007probes,
author = {Boehner, Kirsten and Vertesi, Janet and Sengers, Phoebe and Dourish, Paul},
title = {How HCI interprets the probes},
year = {2007},
isbn = {9781595935939},
publisher = {Association for Computing Machinery},
address = {New York, NY, USA},
url = {https://doi.org/10.1145/1240624.1240789},
doi = {10.1145/1240624.1240789},
booktitle = {Proceedings of the SIGCHI Conference on Human Factors in Computing Systems},
pages = {1077–1086},
numpages = {10},
location = {San Jose, California, USA},
series = {CHI '07}
}
\clearpage

\appendix
\section{Appendices}
\subsection{Pipeline Module Prompts} \label{appendix:module_prompts}
\begin{promptbox}
\textbf{Entrypoint Chat Module}
\tcblower
\textbf{System Prompt}
\begin{Verbatim}[breaklines, fontsize=\fontsize{7}{8}\selectfont]
You are a **highly intelligent AI assistant** designed to **analyze user queries and determine how to update or refine their task-related information**. Your primary role is **not** to directly respond to user queries, but to **decide which module(s) should be updated** and explain why.

## Your Role:
- Your main responsibility is to **analyze user queries** and determine how they impact the **Goal** and **Intent** modules.
- **Do not directly answer user queries** unless they are explicitly asking about **why the Goal and Intent modules were set in a certain way** (e.g., "Why is my goal set this way?" or "Why are these intents selected?").
- By default, for the **user's first query**, always return **both the Goal and Intent modules** as updated. 
- If a selected module is provided, the module must be set to be updated.

## Inputs You Will Receive:
1. **User Query:** The latest user input.
2. Selected Module:
   - This indicates a specific module (Goal or Intents or Intent Dimensions) the user is currently focusing on.
   - If the selected Module is not null, you must always include this module as updated.
   - Additionally, analyze how the user query affects this selected module specifically.
3. **Chat History:** Previous interactions with the user.
4. **Current Module States:** The latest information from:
   - **Goal Module**: Contains the user's task objective, topic, and domain.
   - **Intent Module**: Contains the user's specific **requirements, preferences, and strategies** for achieving their task objective.
   - **User Intent Dimensions**: Represents the **dimensions of the user's intents as UI components**, storing these dimensions and their corresponding values.

## Your Tasks:
1. **Determine whether the query requires updating the Goal or Intent module.**
    - **By default, always return both the Goal and Intent modules as updated for the user's first query.**
    - If a selected module is provided, always include that module as updated.
    - Carefully analyze how the user's query is intended to refine or update the selected module.
    - Provide a clear explanation of how the user query affects the selected module's information.
    - The **Goal module** should remain largely unchanged unless the user presents an entirely new task.
    - If the query does not require updating the user's Goal or Intent, and is instead a meta-question (e.g., "Why is my goal set this way?"), then provide a direct response instead of updating any modules.

2.**If a module needs updating, return the recommended module(s) along with a clear explanation.**
    - If multiple modules require updates, list all relevant ones.
    - Ensure there are no duplicate modules in the updated modules list. Each module (goal, intents, intent dimensions) should appear at most once.
    - Ensure your reasoning is clear, well-structured, and directly tied to the user's task.

\#\# When to Directly Answer the User's Query:
- **Only** respond directly if the query is about the **reasoning behind the Goal or Intent modules' configuration**.
- Example queries that should be answered directly:
  - "Why was my Goal set to this topic?"
  - "How were my intents determined?"
- In all other cases, **focus on module updates rather than answering the query directly**.

Return your response in the following JSON format EXACTLY:
```json
{
    "response": "Direct response to the user's query (if applicable).",
    "updated_modules": [
        {
            "module": "goal || intents || intent_dimensions",
            "reason": "Why this module needs updating."
        }
    ]
}
\end{Verbatim}


\end{promptbox}

\begin{promptbox}
\textbf{Goal Module}
\tcblower
\textbf{System Prompt}
\begin{Verbatim}[breaklines, fontsize=\fontsize{7}{8}\selectfont]
You are a helpful and analytical assistant tasked with analyzing the user's query and extracting their task, domain, and topic. The user provides a writing task as a query through the chat interface in the system. Analyze the provided query to identify: 
    - What the writing task (`task` in the output) is asking for. 
    - Which domain (`domain` in the output) the writing task belongs to (e.g., Journalism Writing, Academic Writing, Creative Writing, Technical Writing, etc.). 
    - What the topic (`topic` in the output) of the writing task is. 

## Input You Will Receive: 
1. **User Query:** The user input.
2. **Interaction History:** Previous user query and goal output history.

## Your task 
First, you need to carefully review the user's query and reasoning the user's request deeply. 
Second, you need to extract the task, domain, and topic from the user's query.
Lastly, you need to provide the extracted information in the JSON format, like the example below. 

Return your response in the following JSON format EXACTLY:
{
    "query": "user provided query", 
    "task": { 
        "value": "task/objective of the user query", 
    },  
    "domain": { 
        "value": "domain of the user query", 
    }, 
    "topic": { 
        "value": "topic of the user query", 
    }  
}

\end{Verbatim}


\end{promptbox}

\begin{promptbox}
\textbf{Intent Module}
\tcblower
\textbf{System Prompt}
\begin{Verbatim}[breaklines, fontsize=\fontsize{7}{8}\selectfont]
You are a helpful and analytical assistant tasked with analyzing the user's `query` along with its `context` (such as the provided `task`, `domain`, and `topic`), and then extracting specific and actionable `intent`(s) from the user `query` and `context`.

The user has requested a writing task as a `query` through the chat interface in the current system. You are provided with four pieces of information: `query`, `task`, `domain`, and `topic`. Your task is extracting concrete and actionable intents based on these four information.
`query` is the user's input query. `task` is the objective of the user query. `domain` is the writing task's domain (e.g., Journalism Writing, Academic Writing, Creative Writing, Technical Writing, etc.). `topic` is the writing task's topic.

## Input You Will Receive: 
1. **User Query:** The user input.
2. **Current Goal Context:** The current user context (task goal, domain, topic).
3. **Interaction History:** Previous user query, context, and intent list output history.

## Your task 
First, you need to carefully review the given input information. In this process, you must deeply reason about what the user truly wants.
Second, you should extract the task, domain, and topic from the user's query.


## You should extract both: 
- Explicit intents: Clearly and directly stated intentions in the user's query and context. 
- Implicit intents: Essential, reasonable, or logically required steps, processes, or goals that are not directly mentioned but are necessary to accomplish the user's writing task successfully.
 

## The extracted intents must: 
- Be specific, explicit, and actionable, so that the output can be generated immediately based on them. 
- Include all relevant implicit intents inferred from the task, domain, and topic, even if not directly stated by the user. 
- Do not contain duplicates. 
- Do not include the task itself as part of the user intents. 

Return your response in the following JSON format EXACTLY:
{
    "intents": [
        { 
            "intent": <specific, explicit, or implicit actionable intent>, 
        }
        ... ,
        {
            "intent": <specific, explicit, or implicit actionable intent>,
        } 
    ] 
}   
\end{Verbatim}


\end{promptbox}

\begin{promptbox}
\textbf{Intent Dimension Module}
\tcblower
\textbf{System Prompt}
\begin{Verbatim}[breaklines, fontsize=\fontsize{7}{8}\selectfont]
You are an analytical and precise assistant tasked with defining appropriate intent dimensions for the given user's intent and selecting the most suitable UI layout to clearly represent key aspects of their task-related needs.

Your role is to analyze the user's intents (requirements, preferences, and strategies) and determine appropriate **intent dimensions** that capture key aspects of their task-related needs. You will then assign the most suitable UI layout for each dimension and set an **initial value** based on the user's query, task goal, and current intents.

There are three possible UI layouts for each intent dimension. You can use as many or as few of each layout as you want:

1. Likert Scale Layout:
    - When appropriate: Used for dimensions with discrete, ordered options
    - Output requirements: title and array of options from left to right
    - Example: Writing Stage (options: ["Idea Generation", "Planning", "Drafting", "Revision"])

2. Sliding Scale Layout:
    - When appropriate: Used for dimensions with continuous numeric values, up to 5 values (min=1, max=5)
    - Output requirements: title, left label, right label, min value, max value
    **RESTRICTION**:
        - Avoid overly granular or excessively narrow scales (e.g., avoid using min=0 and max=200 just because the user mentions 200 words. Instead, group it in practical ranges like 50-100, 100-300, etc.).
        - Example: Specificity (left: "General Overview", right: "Detailed Requirements", range: 1-5)
       
3. Hashtag Layout:
    - When appropriate: Used for dimensions with multiple selectable tags
    - Output requirements: title and array of possible tags
    - Example: Writing Context (tags: ["Academic", "Creative", "Technical", "Professional"])

## Input You Will Receive: 
1. **User Query:** The user input.
2. **Current Goal Context:** The current user context (task goal, domain, topic).
3. **Current Intent List:** The current user intent list.
4. **Interaction History:** Previous user query, context, intent list, and intent dimension output history.

## Your Task
Based on the user's query, task goal, and context, and user intents, determine at least three dimensions that are relevant and which UI layout is most appropriate for each. Examples of dimensions are: Writing Stage, Writing Context, Purpose, Specificity, Audience, and Background Knowledge.

These are only examples!! Please come up with at least one new dimension that isn't mentioned in the examples, and try not to use these examples as a direct reference. You also don't have to use all three layouts; you can use as many or as few of each layout as you want.
Return your response with dimensions in the following JSON format. If there are n dimensions, there should be n elements in the dimensions array:

Likert Scale:
    {
        "dimensions": [ 
            { 
                "type": "likert", 
                "title": "dimension title", 
                "options": ["option1", "option2", "option3"], 
                "selected": "currently selected option", 
            }, 
        ]
    } 
Sliding Scale: 
    { 
        "dimensions": [ 
            { 
                "type": "slider", 
                "title": "dimension title", 
                "leftlabel": "minimum description", 
                "rightlabel": "maximum description", 
                "min": minimum number (1), 
                "max": maximum number (5), 
                "value": current value, 
            }, 
        ]
    } 
Hashtag: 
    { 
        "dimensions": [ 
                "type": "hashtags", 
                "title": "dimension title", 
                "tags": [ 
                    { 
                        "tag": "tag text", 
                        "selected": true/false, 
                    }  
                ], 
            } 
        ] 
    }}

\end{Verbatim}


\end{promptbox}

\begin{promptbox}
\textbf{Preview Module}
\tcblower
\textbf{System Prompt}
\begin{Verbatim}[breaklines, fontsize=\fontsize{7}{8}\selectfont]
You are the Intent Dimension Value Preview Assistant. Your role is to provide a clear, user-friendly explanation of each Intent Dimension Value, describing what it means and how including the value affects the final output.

## Input You Will Receive: 
1. **User Query:** The user input.
2. **Confirmed Intent Dimensions:** Intent dimensions and each confirmed value.
3. **Interaction History:** Previous user query, intent dimensions and confirmed values, and preview output history.

For each Intent Dimension Value, provide the following fields:
1. **intentDimensionValue**: The name of the intent dimension value.
2. **description**: A concise description explaining what this value represents to the user. Should be one sentence.
3. **effectExplanation**: A clear explanation of how including this value will influence or shape the LLM's output, written in a way that helps the user understand its purpose. Should be one sentence.
4. **isSelected**: A boolean indicating if the value is currently selected (True or False). Only include this field for previews involving likert scales or sliding scales.

Additionally, you may also receive:
- **Specific change**
If specific change is provided:
1. Focus only on the intent dimension value related to the specific change.
2. Do not regenerate all for unchanged intent dimensions.

For likert scales, provide previews for each option in the scale. 
For sliding scales, provide previews for each numerical value in the scale (e.g., from 1 to 10).

The name of each preview should be the name of the dimension in lowercase with underscores instead of spaces, for example, preview style.
Return your response in the following EXACT format (for as many dimensions as given):
{
    "preview_dimension1": [{
        "intentDimensionValue": "each intent dimension value",
        "description": "what this value means to the user",
        "effectExplanation": "how including this value affects the output",
        "isSelected": "True or False (if applicable)"
    }],
    "preview_dimension2": [{
        "intentDimensionValue": "each intent dimension value",
        "description": "what this value means to the user",
        "effectExplanation": "how including this value affects the output",
        "isSelected": "True or False (if applicable)"
    }]
}
\end{Verbatim}


\end{promptbox}

\begin{promptbox}
\textbf{Output Module}
\tcblower
\textbf{System Prompt}
\begin{Verbatim}[breaklines, fontsize=\fontsize{7}{8}\selectfont]
You are an advanced LLM assistant designed to generate coherent and well-structured outputs based on information provided by the user.
Your task is to generate the final output text, composed of logically flowing sentences that fulfill the task goal and user intents.

## Input You Will Receive: 
1. **User Query:** The user input.
2. **Current Goal Context:** The current user context (task goal, domain, topic).
3. **Current Intent List:** The current user intent list.
4. **Current Intent Dimensions and Previews:** The current intent dimensions and corresponding previews.
5. **Interaction History:** Previous user query, context, intent list, intent dimensions, previews, and output history.

Additionally, you may also receive: 
- **Specific changes**
 
When specific changes are provided: 
1. Carefully analyze the changes and how they will change the output.
2. Modify only the necessary parts of the output to reflect these updates, while keeping unaffected parts consistent.
3. Ensure that the overall output remains coherent and aligned with the updated requirements.

## Key Rules:
1. Divide the output into clear sections using subheaders.
2. Within each section, include sentences that logically build toward fulfilling the user's task goal and intents.
3. If specific changes are provided, reflect them accurately and revise relevant sections as needed.

Return your response in the following JSON format EXACTLY:
{
    "generatedoutput": [ 
    { 
        "subheader": "subtask title", 
        "content": [
            { 
                "sentence": "sentence", 
            } 
        ] 
    }]
}  

\end{Verbatim}


\end{promptbox}

\begin{promptbox}
\textbf{Linking Module}
\tcblower
\textbf{System Prompt}
\begin{Verbatim}[breaklines, fontsize=\fontsize{7}{8}\selectfont]
You are an accurate and capable assistant tasked with creating links between each given **intent** or **intent dimension** and the specific phrases in the output. The intents are extracted from the user's writing task request query, reflecting the user's goals. The intent dimensions represent the controllable aspects of the intents, expressed in UI elements (one of: slider, Likert scale, or hashtags), allowing the user to adjust specific parts of their intent. The output is the writing result generated based on the user's intents and intent dimensions.
For this task, you are provided with the following input information:

## Input You Will Receive: 
1. **Current Intent List:** The current user intent list.
2. **Current Intent Dimensions and Their Selected Value:** The current intent dimensions and their selected value.
3. **Output Text:** Output text is provided as a list of phrases (each phrase separated as an individual item).
4. **`Specific Change`**: Specific change contains modifications made by the user regarding the intents and intent dimensions. For example, the user may delete, edit, or add intents. The results of these changes are provided in specific change (where `from` refers to the pre-modified state and `to` refers to the post-modified state). Similarly, for intent dimensions, if the user modifies UI elements, such as adding hashtags, changing slider values, or updating Likert scale selections, those changes will also be reflected here (`from` indicates before modification, `to` indicates after modification). If the `Specific Change` input is None, it means the user has made no changes, and you can ignore this information. 
5. **`Total Number of Links Required`**: An integer specifying exactly how many **linking entries** must be returned. 

## Your Task: 
First, thoroughly review the provided input information and `identify the relationships between the output and each **intent**/**intent dimension**.  
Second, for each given **intent** or **intent dimension**, identify all phrases in the output that are possibly related to it. 
Third, create links between every phrase and every **intent**/** intent dimension** it may be related to, even if the connection is weak or indirect. Be exhaustive and avoid missing potentially relevant connections. 
    - For every intent in the intents list, link **each intent** to the relevant phrases in the output that fulfill or address that intent. You should identify every links that are relevant to the intent, even if it is indirectly related. 
    - For every intent dimension in the intent dimensions list, also create links connecting each intent dimension to specific phrases in the output. You should identify every links that are relevant to the intent, even if it is indirectly related. 

## Special Guidelines for Intent Dimensions: 
In particular, for intent dimensions, follow these specific rules based on the UI element type: 
- For the **likert scale format**, link the selected intent dimension value to the relevant phrases in the output.
- For the **slider scale format**, link the selected intent dimension value to the relevant phrases in the output. 
- For **hashtags**, **process each individual hashtag separately**: 
  - **Create a separate link entry for each hashtag.** 
  - For each hashtag, link it only to the specific phrases that are relevant to that particular hashtag. 
  - Do **NOT group multiple hashtags together** in a single entry. 
  
## Important Note: 
If an intent or intent dimension primarily affects the overall structure, flow, style, or tone of the output rather than specific individual phrases, you may link it to the entire output. 
In such cases, set the given entire output to the `linkingphrases` to indicate its pervasive influence throughout the entire output. 

**RESTRICTION**: 
- **Every intent and every intent dimension value provided as input MUST be linked to at least one relevant phrase in the output.** 
- You must return *exactly* as many link entries as specified by `Total Number of Links Required`. 
- Do not skip or omit any input intent or intent dimension value. Even if the link seems minor, it must be explicitly included. 

Output Structure (Ensure the response is valid JSON without any comments or trailing commas) 
You must generate the exact number of links as instructed. 
Return your response in the following JSON format:

{ 
    "links": [ 
    { 
      "intent": { 
        "title": "Intent Text" 
      }, 
      "linkingphrases": ["exact phrase 1", "exact phrase 2"] 
    }, 
    { 
      "intentDimensionValue": { 
        "type": "likert", 
        "title": "intent dimension title", 
        "specificValue": "selected value" 
      }, 
      "linkingphrases": ["exact phrase 1", "exact phrase 2"] 
    }, 
    {  
      "intentDimensionValue": { 
        "type": "slider", 
        "title": "intent dimension title", 
        "specificValue": "selected value" 
      }, 
      "linkingphrases": ["exact phrase 1", "exact phrase 2"] 
    }, 
    { 
      "intentDimensionValue": {  
        "type": "hashtags", 
        "title": "intent dimension title", 
        "specificValue": "tag1" 
      }, 
      "linkingphrases": ["exact phrase 1", "exact phrase 2"] 
    }, 
    { 
      "intentDimensionValue": { 
        "type": "hashtags", 
        "title": "intent dimension title", 
        "specificValue": "tag2" 
      }, 
      "linkingphrases": ["exact phrase 1", "exact phrase 2"] 
    }
  ]
}
\end{Verbatim}


\end{promptbox}

\clearpage
\subsection{Pipeline Validation Details}

\subsubsection{Question list}\label{appendix:evaluation_questions}
The full phrase for pipeline validation questions is provided below:

\smallskip\noindent 
\textbf{[\textit{Goal Module} --- Q1. Goal Alignment]}: ``Do you think the task goal, domain, and topic described below appropriately reflect the user's high-level and overall goal?''

\smallskip\noindent 
\textbf{[\textit{Intent Module}]} 
\begin{itemize}
    \item \textbf{[Set of Intents --- Q2. Completeness]}: `` Do you think the set of intents cover all key aspects of the user prompt without missing anything important?''
    \item \textbf{[Set of Intents --- Q3. Distinctiveness]}: ``Do you think the intents are meaningfully distinct from each other without redundancy?''
    \item \textbf{[Individual Intents --- Q4. Relevance]}: ``Do you think this intent is relevant to the user prompt?''
\end{itemize}

\smallskip\noindent 
\textbf{[\textit{Dimension Module}]}
\begin{itemize}
    \item \textbf{[Q5. Relevance]}: ``Do you think this intent dimension is relevant to the user prompt?''
    \item \textbf{[Q6. UI Appropriateness]}: ``Do you think the UI component (e.g., hashtags, slider, radio buttons) is appropriate to control this intent dimension's value?''
    \item \textbf{[Q7. Value Appropriateness]}: ``Do you think this intent dimension value in this UI component is appropriate to the user's prompt?''
\end{itemize}

\smallskip\noindent 
\textbf{[\textit{Linking Module} --- Q8. Link Accuracy]}: ``Does the highlighted part correspond to the intent?''

\subsubsection{Task and prompts table}\label{appendix:eval_prompts}
\autoref{table:eval_prompts_table} presents the 12 writing tasks and their corresponding user prompts used in the technical evaluation. These tasks span six representative writing contexts---academic, creative, journalistic, personal, professional, and technical. Each prompt is designed to reflect realistic writing goals within its context, providing concrete task instructions that the system must interpret and respond to. This curated set enables a comprehensive assessment of the system’s ability to support intent understanding and generation across a wide range of writing scenarios.

\begin{table*}[ht]
\caption{Task contexts, topics, and user prompts used in the technical evaluation.}
\Description{This table lists the task contexts, topics, and user prompts used in the technical evaluation of the system. It includes 12 entries spanning six different writing contexts: Academic, Creative, Journalistic, Technical, Personal, and Professional. Each entry specifies a writing task (e.g., argumentative essay, poetry, technical report), a topic (e.g., online education, solitude in nature, smartphone battery life), and a corresponding user prompt that provides detailed instructions for the task. The prompts vary in tone and structure depending on the context—for instance, academic prompts emphasize structure and formality, while creative prompts focus on emotional expression and vivid imagery. The table ensures diversity in task types and writing goals, serving as a comprehensive benchmark for system evaluation.}
\label{table:eval_prompts_table}
\centering
\small
\begin{tabular}{p{0.1\textwidth}|p{0.12\textwidth}|p{0.2\textwidth}|p{0.58\textwidth}}

\toprule
\textbf{Writing Context} & 
\textbf{Task} & 
\textbf{Topic} & 
\textbf{User Prompt} \\

\midrule
Academic & 
Argumentative essay writing & 
The effectiveness of online education compared to traditional classroom & 
Write an argumentative essay discussing whether online education is more effective than traditional classroom education. Include a clear thesis statement, at least three supporting arguments with evidence, and address one counterargument. Use a formal academic tone throughout. \\
\hline

Academic & 
Research proposal writing & 
Investigating the impact of social media usage on student academic performance & 
Write a research proposal exploring how social media usage affects student academic performance. Your proposal should include the research objective, a brief review of potential related factors, proposed methodology, and expected outcomes. Use a formal academic tone and structure. \\
\hline

Creative & 
Poetry writing & 
The feeling of solitude in nature & 
Write a free verse poem that captures the feeling of solitude experienced while walking alone in a dense forest. Use vivid sensory imagery and metaphors to evoke the atmosphere and emotion. \\
\hline

Creative & 
Fiction writing &  
A mysterious letter arrives without a sender & 
Write a short fiction story about a character who receives a mysterious letter with no return address. The letter contains a cryptic message that leads them on an unexpected journey. Focus on building suspense, the character’s emotional response, and detailed scene descriptions. \\
\hline

Journalistic & 
Article writing & 
Local community launching a zero-waste initiative & 
Write a news article covering a local community’s launch of a zero-waste initiative. Include a clear headline, an engaging lead, factual details about the initiative, and quotes from key people involved. Adopt an objective, informative journalistic style. \\
\hline

Journalistic & 
Opinion column writing & 
The feeling of solitude in nature & 
Write a free verse poem that captures the feeling of solitude experienced while walking alone in a dense forest. Use vivid sensory imagery and metaphors to evoke the atmosphere and emotion. \\
\hline

Technical & 
Science explanation writing & 
How photosynthesis works &
Explain how photosynthesis works in a way that is accessible to high school students. Break down the key steps and components involved, using clear language and relatable analogies where helpful. \\
\hline

Technical & 
Technical report writing & 
Smartphone battery life test report & 
Write a technical report evaluating the battery life of your smartphone under different usage conditions (e.g., watching videos, browsing, idle). Include sections for the objective, testing methodology, key findings (such as average battery drain rate), identified issues, and suggestions to optimize battery usage. Use formal technical language and organize the report clearly. \\
\hline

Personal & 
Letter writing & 
Letter to a childhood friend after years apart & 
Write a personal letter to a childhood friend you haven’t spoken to in years. Reflect on a fond memory you shared, share how your life has been, and express your interest in reconnecting. Keep the tone warm and genuine. \\
\hline

Personal & 
Social media post writing &  
Sharing a recent personal achievement & 
Write a social media post sharing a recent personal achievement. Make it engaging and authentic, and include a positive or motivational message for your audience. It should be under 200 words. \\
\hline

Professional & 
Elevator pitch writing & 
Introducing a new productivity app & 
Write a 60-second elevator pitch introducing a new productivity app designed to help remote teams collaborate efficiently. Highlight the key features and the specific problem the app solves, keeping the pitch confident and compelling. \\
\hline

Professional & 
Business email writing & 
Requesting a meeting to discuss a potential partnership & 
Write a professional email to a potential partner organization, requesting a meeting to explore collaboration opportunities. Politely introduce yourself and your organization, explain the reason for reaching out, propose a meeting time, and close with a courteous sign-off. \\
\bottomrule
\end{tabular}
\end{table*}

\clearpage
\subsection{User Study Details}
\subsubsection{\textbf{Study Task Descriptions }}\label{appendix:study_tasks}

The full task descriptions provided to participants in the user study are presented below.

\noindent\hrulefill

\noindent
\textbf{Task A: Social Media Post Writing}

\noindent
\textsf{
\textbf{User Scenario.} Imagine you are working as a content creator for an online educational platform that aims to make complex scientific concepts engaging and relatable for a general audience. Your task is to write a short, compelling social media post about a scientific phenomenon, the Doppler Effect, that connects to everyday life.}

\noindent
\textsf{\textbf{Your goal is to:}
\begin{itemize}
    \item Grab the reader's attention with a relatable or thought-provoking message.
    \item Explain the scientific concept clearly and concisely, making it accessible to people with little or no background in the subject.
    \item Use examples or scenarios from daily life to help readers connect with the topic.
    \item Encourage reader interaction by posing a question or prompt that invites them to share their observations or thoughts.
\end{itemize}
Your audience consists of curious individuals who enjoy learning through social media but may not have a scientific background. The tone should be engaging, conversational, and easy to understand. Your challenge is to make the concept as clear, relatable, and thought-provoking as possible while keeping the post concise. The length would be suitable if it fits about half an A4 page.
}

\noindent\hrulefill

\noindent
\textbf{Task B: Job Application Email Writing}

\noindent
\textsf{
\textbf{User Scenario.} Imagine you are applying for a personal secretary position for a well-known professional in a field unrelated to your expertise (e.g., an artist, entrepreneur, scientist, or athlete). The employer is looking for a secretary with strong organizational skills, communication ability, and adaptability rather than specialized knowledge in the employer’s domain. Your task is to write a compelling job application email introducing yourself and demonstrating why you would be a great fit for this role.}

\noindent
\textsf{\textbf{Your goal is to:}
\begin{itemize}
    \item Leave a strong impression to the professional.
    \item Clearly express your motivation for applying, emphasizing skills that make you a strong candidate.
    \item Share a relevant personal experience that highlights your ability to adapt, learn quickly, or support a busy professional.
    \item Show your genuine interest in working closely with someone whose work may be outside your area of expertise.
\end{itemize}
Your audience is a busy professional who likely receives many applications. Your challenge is to stand out by being clear, professional, and persuasive while keeping your email concise. The length would be suitable if it fits about half an A4 page.
}

\noindent\hrulefill

\subsubsection{\textbf{Self-Report Survey Items}} \label{appendix:survey_items}
The full statement for self-report survey items is provided below:

\begin{enumerate}
    \item \textbf{M1. Intent Expression --- Ease}: \textit{``I could easily express my intent to the system.''}
    \item \textbf{M2. Intent Expression --- Clarity}: \textit{``The system helped me express my intent clearly.''}
    \item \textbf{M3. Intent Discovery}: \textit{``The system helped me recognize or discover additional intents that I had not explicitly considered at the start.''}
    \item \textbf{M4. Transparency}: \textit{``The system helped me see how each of my intents influenced the output generation.''}
    \item \textbf{M5. Understanding}: \textit{``I understood how each of my intents was reflected in the output.''}
    \item \textbf{M6. Think-Through}: \textit{``The system helped me think what kinds of intents I would want to complete the task goal, and how to complete the task.''}
    \item \textbf{M7. Intent Adjustment --- Ease}: \textit{``I was able to adjust my intent to achieve the output aligned with my task goal.''}
    \item \textbf{M8. Intent Elaboration}: \textit{``The intents I created/kept were specific, detailed, and well-articulated.''}
    \item \textbf{M9. Intent Match}: \textit{``The system helped me obtain intents and an output that better matched what I wanted.''}
    \item \textbf{M10. Draft Quality}: \textit{``I felt like communicating intents using this workflow will help me have a better final draft.''}
    \item \textbf{M11. Intent Reusability}: \textit{``I would reuse the intents I created/kept in future similar tasks.''}
\end{enumerate}

\subsubsection{\textbf{Baseline Interface}}
\begin{figure}[h!]
    \centering
    \includegraphics[width=\linewidth]{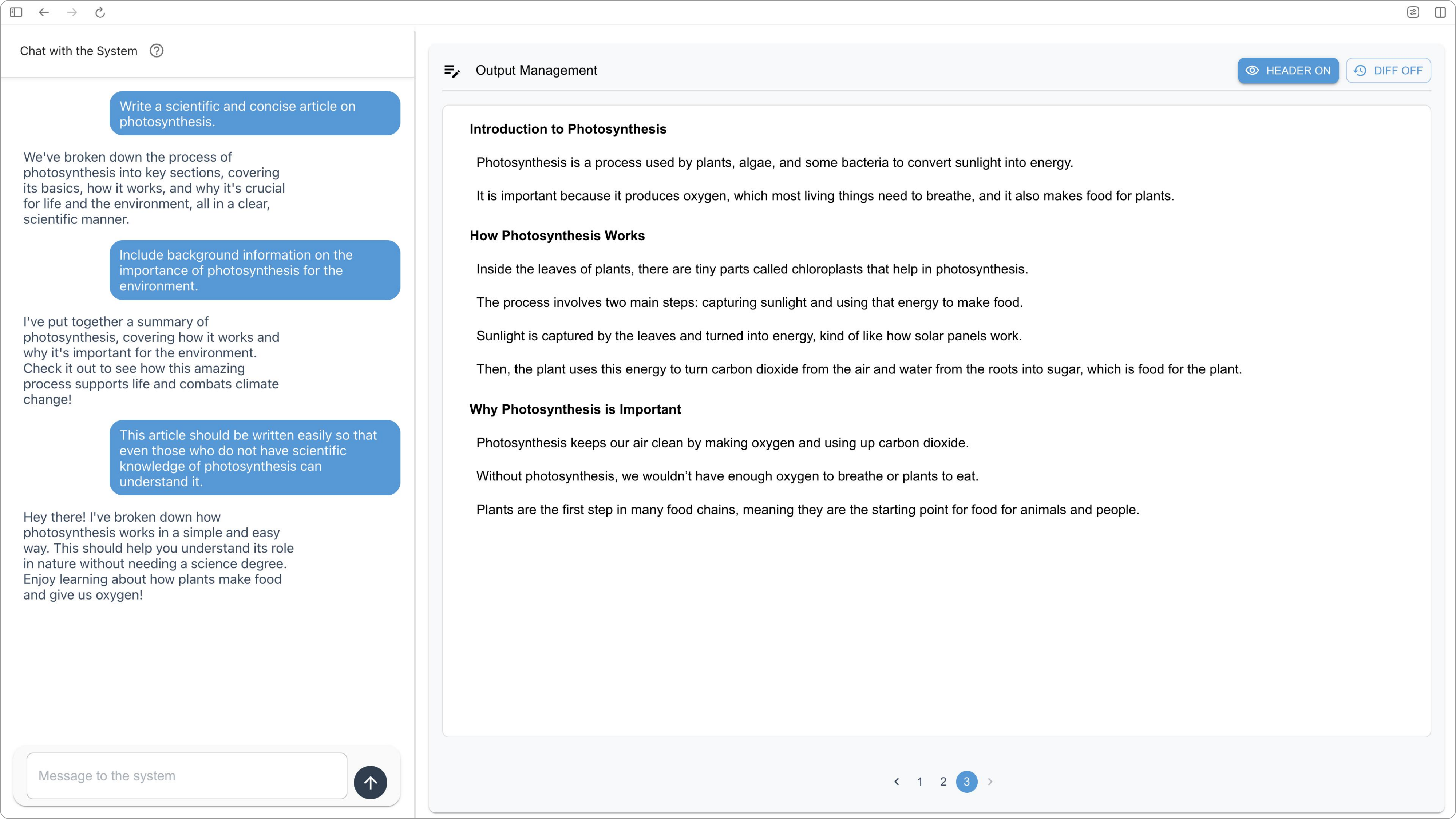}
    \caption{A screenshot of the Baseline Interface}
    \label{fig:baseline_interface}
    \Description{This figure is a screenshot of the baseline interface. The baseline interface is divided into two sections: a chat panel on the left where users interact through free-form prompts and an output panel on the right where the system generates and displays the resulting text in response to those prompts.}
\end{figure}
\autoref{fig:baseline_interface} shows the baseline interface, which consists of a chat panel for user prompts and a separate panel displaying the generated writing output.

\subsubsection{\textbf{User Study Participants Details}}\label{appendix:participants}
\autoref{table:participant_background} presents detailed demographic and background information about the user study participants, including their LLM usage frequency, self-rated English writing experience, and qualitative descriptions of their writing backgrounds.

\begin{table*}[ht]
\caption{Self-reported background information of user study participants, including their frequency of LLM usage, English writing experience levels, and free-form descriptions of their English writing backgrounds (1: No experience, 7: Extensive experience).}
\Description{This table presents detailed background information for 12 user study participants. It includes six columns: participant ID, age, gender, LLM usage frequency over the past 6 months (e.g., ``Every day'', ``2–5 times a week''), self-rated English writing experience on a 7-point Likert scale (1 = No experience, 7 = Extensive experience), and a free-text description of their English writing background. The descriptions reflect diverse writing experiences, such as academic writing, email communication, technical reports, and use of LLMs for writing assistance.}
\label{table:participant_background}
\centering
\small
\resizebox{\textwidth}{!}{%
\begin{tabularx}{\textwidth}{p{0.04\textwidth}|p{0.03\textwidth}|p{0.06\textwidth}|p{0.14\textwidth}|p{0.14\textwidth}|X}
\toprule
\textbf{ID} & \textbf{Age} & \textbf{Gender} & \textbf{LLM Usage Experience (Past 6 Months)} & \textbf{English Writing Experience (Self-rating)} & \textbf{English Writing Experience Description (Free Response)} \\
\midrule
P01 & 26 & Female & 2–5 times a week & 7 (Extensive experience) & Has extensive experience writing in English, including research papers. Frequently uses LLMs to improve grammar and overall writing quality. \\ \hline
P02 & 28 & Female & 2–5 times a week & 7 (Extensive experience) & Currently works as an HCI researcher and writes in English daily for research papers, grant proposals and reports, and professional emails.\\ \hline
P03 & 26 & Female & Every day & 6 (High experience) & Primarily focused on academic writing. Occasionally uses LLMs to paraphrase for better wording and sentence structure, but prefers to independently craft the outline and logical flow. \\ \hline
P04 & 21 & Male   & Every day & 5 (Moderate experience) & Has written numerous papers during high school.\\ \hline
P05 & 25 & Male   & 2–5 times a week & 5 (Moderate experience) & Regularly writes assignments in English and took English writing courses during undergraduate studies. \\ \hline
P06 & 23 & Male   & Every day & 6 (High experience) & Has experience writing in English for various purposes including TOEFL preparation, reports, research papers, and CVs. \\ \hline
P07 & 23 & Female & 2–5 times a week & 7 (Extensive experience) & As a biology major, wrote weekly experimental reports in English for several years. Currently works as a freelancer producing English reports using prompts they created, leveraging ChatGPT for the task. \\ \hline
P08 & 30 & Male   & Every day & 6 (High experience) & Has written papers, emails, reviews, and reports in English. However, has no experience with emotional or narrative writing in English, and reading experience is limited to nonfiction, academic texts, and news articles. \\ \hline
P09 & 28 & Male   & Every day & 7 (Extensive experience) & Completed a master’s thesis written in English. \\ \hline
P10 & 27 & Male   & Once every 2–3 months & 7 (Extensive experience) & Has experience writing both academic research papers and English-language newspaper articles. \\ \hline
P11 & 22 & Male   & 2–5 times a week & 5 (Moderate experience) & English writing experience is limited to general education course assignments. \\ \hline
P12 & 27 & Male   & 2–5 times a week & 7 (Extensive experience) & Attended an international school from kindergarten through 12th grade and served as Editor-in-Chief of a university English-language student newspaper. \\
\bottomrule
\end{tabularx}
}
\end{table*}

\end{document}